\documentclass[10pt,twocolumn,letterpaper]{article}

\usepackage[pagenumbers]{iccv}      %

\usepackage{times}
\usepackage{epsfig}
\usepackage{graphicx}
\usepackage{amsmath}
\usepackage{amssymb}
\usepackage{cuted}

\usepackage{stmaryrd}

\usepackage{overpic}
\usepackage[dvipsnames]{xcolor}
\usepackage{times}
\usepackage{epsfig}
\usepackage{graphicx}
\usepackage{amsmath}
\usepackage{amssymb}
\usepackage{amsthm}
\usepackage{mathtools}
\usepackage{bm}
\makeatletter
\@namedef{ver@everyshi.sty}{}
\makeatother
\usepackage[mode=buildnew]{standalone}
\usepackage{tikz}
\usetikzlibrary{positioning}
\usetikzlibrary{shapes} 
\usetikzlibrary{matrix}
\usetikzlibrary{calc} 
\usetikzlibrary{intersections}
\usepackage[accsupp]{axessibility}  %
\usepackage{tikzscale}
\newcommand{\cir}[2][red,fill=red]{
\hspace{-0.1cm}\tikz[baseline=-0.5ex]\draw[#1,radius=#2] (0,0) circle;\hspace{-0.1cm}
}
\usepackage{pgfplots}
\usepackage[eulergreek]{sansmath}
\usepackage{caption,subcaption}
\usepackage{bbm}
\usepackage{comment}
\usepackage{acro}
\usepackage{listings}
\usepackage{multirow}
\usepackage{booktabs}
\usepackage{wrapfig}
\usepackage{algorithm}
\usepackage{algpseudocode}
\usepackage{float}
\usepackage{overpic}
\usepackage{cuted}
\usepackage{pbox}
\usepackage{tabularx}
\setkeys{Gin}{keepaspectratio}
\usepackage{adjustbox}
\usepackage{enumitem}

\newcommand{\verts}{\mathcal{V}}
\newcommand{\faces}{\mathcal{F}}
\newcommand{\edges}{\mathcal{E}}

\newcommand{\contour}{\mathcal{C}}
\newcommand{\contourVerts}{\verts_\contour}

\newcommand{\contourEdgs}{\edges_\contour}
\newcommand{\meshX}{\mathcal{X}}
\newcommand{\meshY}{\mathcal{Y}}

\newcommand{\prodGraph}{\mathcal{P}}

\newcommand{\hyperProdGraph}{\mathcal{H}}

\newcommand{\graph}{\mathcal{G}}

\newcommand{\twovector}[2]{\bigl(\begin{smallmatrix}#1 \\#2\end{smallmatrix}\bigr)}

\newcommand{\linking}{\mathcal{L}}
\newcommand{\injectivity}{\mathcal{S}}

\newcommand{\surfacecycles}{{surface cycles}}
\newcommand{\surfacecycle}{{surface cycle}}

\newcommand{\windheuser}{Windheuser~\etal}
\newcommand{\ours}{Ours}

\newcommand{\faust}{FAUST}

\newcommand{\caoetal}{ULRSSM}
\newcommand{\smcomb}[0]{SM-Comb}
\newcommand{\discreteopt}[0]{DiscrOpt}
\newcommand{\smooths}[0]{SmoothShells}
\newcommand{\discomatch}[0]{DiscoMatch}
\newcommand{\spidermatch}[0]{SpiderMatch}
\newcommand{\hyperedge}[2]{\left\llbracket{#1}\rightarrowtriangle {#2}\right\rrbracket}

\definecolor{cRED}{HTML}{F94144}
\definecolor{cORANGE}{HTML}{F3722C}
\definecolor{cYELLOW}{HTML}{F9C74F}
\definecolor{cLIGTHGREEN}{HTML}{90BE6D}
\definecolor{cGREEN}{HTML}{43AA8B}
\definecolor{cDARKGREEN}{HTML}{4D908E}
\definecolor{cDARKBLUE}{HTML}{577590}
\definecolor{cBLUE}{HTML}{277DA1}
\definecolor{cPINK}{HTML}{F1C8DB}

\definecolor{scLIGHTBLUE}{HTML}{A1C4E2}
\definecolor{scBLUE}{HTML}{5393CA}
\definecolor{scYELLOW}{HTML}{D6B656}
\definecolor{scPURPLE}{HTML}{A680B8}
\definecolor{scGREEN}{HTML}{67AB9F}
\definecolor{scLIGHTYELLOW}{HTML}{E7D49C}
\definecolor{scLIGHTPURPLE}{HTML}{E6D0DE}
\definecolor{scLIGHTGREEN}{HTML}{9AC7BF}
\definecolor{scLIGHTDARKBLUE}{HTML}{73A6D4}
\definecolor{scLIGHTGREEN}{HTML}{D5E8D4}

\definecolor{cSYELLOW}{HTML}{FFBE0B}
\definecolor{cSORANGE}{HTML}{FB5607}
\definecolor{cSPINK}{HTML}{FF006E}
\definecolor{cSPURPLE}{HTML}{8338EC}
\definecolor{cSBLUE}{HTML}{3A86FF}
\definecolor{cSGREEN}{HTML}{20BF55}

\colorlet{ulrssmcolour}{cYELLOW}
\colorlet{discocolor}{cSBLUE}
\colorlet{windheusercolor}{cSPINK}
\colorlet{laehnercolor}{cSPINK}
\colorlet{roetzercolor}{cSBLUE}
\colorlet{ourcolor}{cSGREEN}

\definecolor{cSGPUDARKPINK}{HTML}{9D0C4B}
\definecolor{cSGPUDARKBLUE}{HTML}{4479CE}
\definecolor{cSGPUYELLOW}{HTML}{E8AB04}
\colorlet{mccgpunormalcolor}{cSGPUDARKPINK}
\colorlet{mccgpucolor}{cSGPUYELLOW}

\definecolor{cPLOT0}{HTML}{4D908E}%
\definecolor{cPLOT2}{HTML}{90E39A}%
\definecolor{cPLOT3}{HTML}{F9844A}%
\colorlet{cPLOT5}{cSPURPLE}%
\definecolor{cPLOT1}{HTML}{4D908E}%

\usepackage{calc}
\newcommand{\rotatedCentering}[3]{\rotatebox{#1}{\hspace{(#2-\widthof{#3})/2}#3}}
\newcommand{\scircled}[1]{{\normalsize\textcircled{\scriptsize #1}}}

\definecolor{ForestGreen}{rgb}{0.13, 0.55, 0.13}
\usepackage{pifont}%
\newcommand{\cmark}{{\color{ForestGreen} \ding{51}}}%
\newcommand{\xmark}{{\color{red} \ding{55}}}%

\definecolor{iccvblue}{rgb}{0.21,0.49,0.74}
\usepackage[pagebackref,breaklinks,colorlinks,allcolors=iccvblue]{hyperref}
\usepackage[capitalize]{cleveref}
\newtheorem{theorem}{Theorem}
\newtheorem{lemma}[theorem]{Lemma}

\newtheorem{definition}[theorem]{Definition}

\def\confName{ICCV}
\def\confYear{2025}

\title{
Fast Globally Optimal and Geometrically Consistent 3D Shape Matching
}

\newcommand{\authorspace}{\hspace{0.7cm}}
\newcommand{\affiliationspace}{\hspace{0.8cm}}
\author{Paul Roetzer$^{1,2}$\authorspace Florian Bernard$^{1,2}$\\
	$^1$ University of Bonn \affiliationspace $^2$ Lamarr Institute 
}

\begin{document}

\maketitle
\newlength{\teaserheight}
\setlength{\teaserheight}{3.3cm}
\begin{strip}
	\vspace{-0.5cm}
	\centering
	\footnotesize%
	\renewcommand{\arraystretch}{0.9}
	\begin{tabular}{ccc}
		\setlength{\tabcolsep}{0pt}
		\hspace{-0.7cm}
		\begin{tabular}{c}
			\includegraphics[width=5.9cm,trim={1.2cm 0.7cm 0 -1cm}]{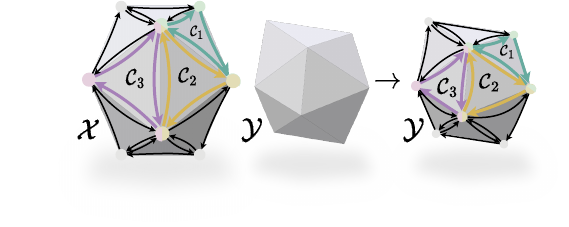}
		\end{tabular}
		&
		\hspace{-1cm}
		\adjustbox{trim=0 0 0 2cm}{%
			\begin{tabular}{ccc}
				\includegraphics[height=\teaserheight]{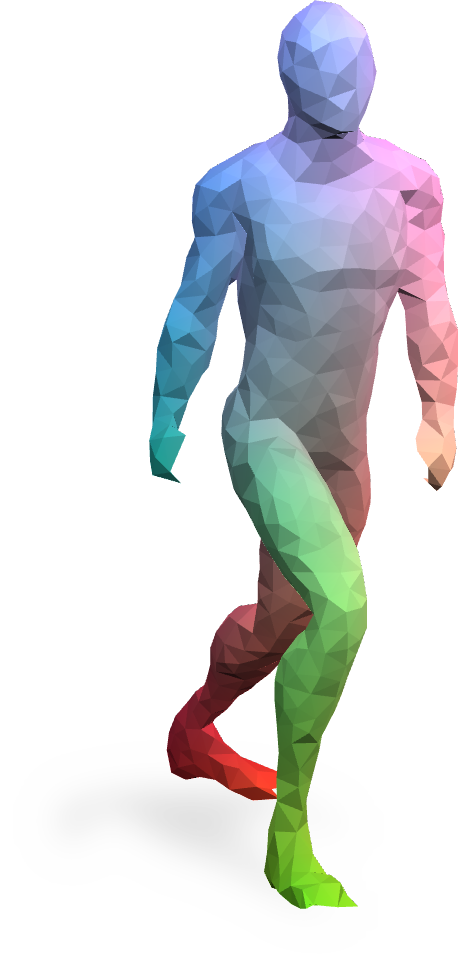}
				&
				\hspace{-0.85cm}
				\includegraphics[height=\teaserheight,trim={0 0 0 0.5cm}]{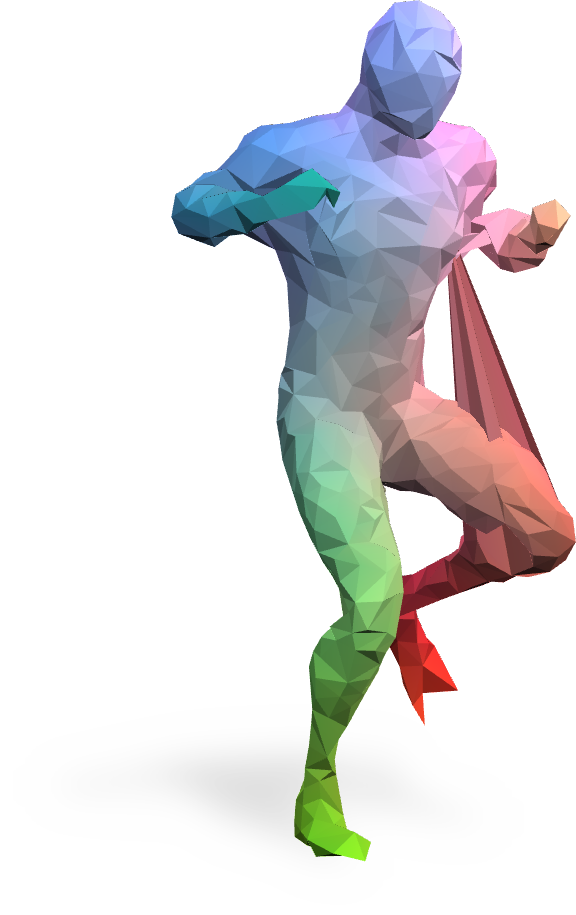}
				&
				\hspace{-0.95cm}
				\includegraphics[height=\teaserheight,trim={0 0 0 0.5cm}]{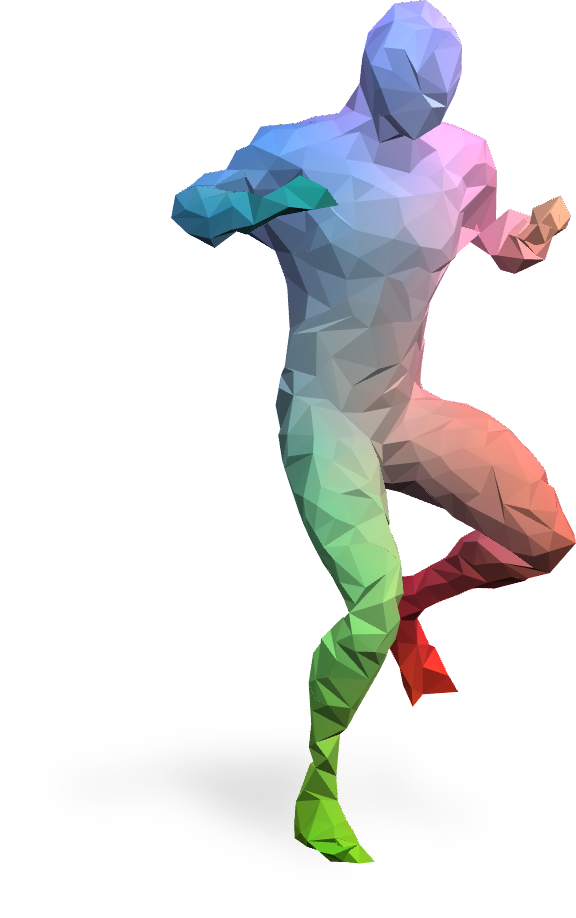}
				\\
				Source& 
				Cao~\etal 
				&Ours 
			\end{tabular}
		}
		&
		\hspace{-1cm}
		\begin{tabular}{c}
			\adjustbox{trim=0 0.85cm 0 0.8cm}{%
				\newcommand{\runtimeLineWidth}{2pt}
\newcommand{\rtCplotWidth}{1.3\columnwidth}
\newcommand{\rtCplotHeight}{0.6\columnwidth}
\pgfplotsset{%
	label style = {font=\large},
	tick label style = {font=\large},
	title style =  {font=\Large},
	legend style={  fill= gray!10,
		fill opacity=0.6, 
		font=\large,
		draw=gray!20, %
		text opacity=1}
}
\begin{tikzpicture}[scale=0.5, transform shape]
	\begin{axis}[%
		width=\rtCplotWidth,
		height=\rtCplotHeight,
		scale only axis,
		grid=major,
	      legend style={
                at={(1.02,1.26)},
		 	anchor=north east,
		 	legend columns=3,
		 	legend cell align={left}},
		ylabel={{\large Runtime [s]  \textcolor{gray!40}{$\leftarrow$}}},
		xlabel={$\#$ of Faces of individual Triangle Meshes},
            ylabel shift = -0.3cm,
            xlabel style={yshift=0.15cm},
		xmin=50,
		xmax=1750,
		xtick={250, 500, 750, 1000, 1250, 1500, 1750},
		xtick scale label code/.code={},
		ymin=0,
		ymax=1000,
            ytick={0, 250, 500, 750, 1000},
		ylabel near ticks,
		]
		
		\addplot [color=windheusercolor, line width=\runtimeLineWidth]
		table[row sep=crcr]{%
50	5.3454752\\
100	78.5283646\\
150	683.8947292\\
200	2232.7588358\\
250	2964.2542276\\
		};
		\addlegendentry{\textcolor{black}{\windheuser}}

		\addplot [color=cPLOT0, tension=0.2, line width=\runtimeLineWidth]
		table[row sep=crcr]{%
0 0\\
50	1.66142253875732\\
100	4.6863883972168\\
150	21.2299663066864\\
200	37.7827697277069\\
250	58.980616235733\\
300	102.505135154724\\
350	222.499098205566\\
400	213.992396688461\\
450	288.231658172607\\
500	430.349188232422\\
550	511.202847003937\\
600	804.884549236298\\
650	810.355409288406\\
700	925.835303497314\\
750	1133.83003811836\\
800	1154.35633182526\\
		};
		\addlegendentry{\textcolor{black}{\smcomb}}
\node[color=discocolor,rotate=50] at (axis cs:950, 397) (b) {\textbf{]}};
\node[color=discocolor] at (axis cs:950, 450) (b) {\large OOM};
    \addplot [color=discocolor, line width=\runtimeLineWidth]
		table[row sep=crcr]{%
50	4.61538190841675\\
100	8.21245727539062\\
150	26.7098184585571\\
200	28.5115680217743\\
250	33.2058145523071\\
300	57.9440919399261\\
350	47.961816072464\\
400	63.5641473293304\\
450	64.881423664093\\
500	145.214042568207\\
550	117.714063596725\\
600	149.570569038391\\
650	158.672130632401\\
700	196.373429059982\\
750	235.609471654892\\
800	232.494430494308\\
850	521.386315727234\\
900	276.886437511444\\
950	397.95380282402\\
		};
		\addlegendentry{\textcolor{black}{\discomatch\ (GPU)}}

\addplot [color=cPLOT5,tension=0.2,line width=\runtimeLineWidth]
		table[row sep=crcr]{%
50	2.15893530845642\\
150	22.4732964038849\\
250	343.254711389542\\
350	87.7516279220581\\
450	323.576714754105\\
550	299.947786092758\\
650	564.656882047653\\
750	673.805225729942\\
800	855.316\\
850	1002.63\\
};
\addlegendentry{\textcolor{black}{\spidermatch\ (geo.)}}%
    \addplot [color=cPLOT5, dashed,line width=\runtimeLineWidth]
		table[row sep=crcr]{%
0 0\\
50	0.0727925300598145\\
150	0.80660080909729\\
250	2.79375910758972\\
350	6.1397168636322\\
450	11.0672917366028\\
550	20.6555211544037\\
650	33.8513643741608\\
750	43.3372287750244\\
850	53.6141972541809\\
950	77.4060039520264\\
1050	80.8316583633423\\
1150	131.65684223175\\
1250	142.600990772247\\
1350	195.615426063538\\
1450	201.480194091797\\
1550	226.529071331024\\
1650	373.353313565254\\
1750	379.037929415703\\
1850	412.055595397949\\
1950	578.553\\
		};
		\addlegendentry{\textcolor{black}{\spidermatch\ }}%
\addplot [color=ourcolor, smooth, tension=0.2, line width=\runtimeLineWidth]
  table[row sep=crcr]{%
50	0.382844209671021\\
150	3.03250408172607\\
250	9.37670803070068\\
350	12.2255246639252\\
450	21.535619020462\\
550	34.291276216507\\
650	48.6560304164886\\
750	66.5410633087158\\
850	88.2784218788147\\
950	111.767062902451\\
1050	139.437685012817\\
1150	169.960187673569\\
1250	204.876774072647\\
1350	253.960168838501\\
1450	303.68256020546\\
1550	351.137157678604\\
1650	423.827531814575\\
1750	498.286844372749\\
1850	586.620145082474\\
};
\addlegendentry{\ours}
\end{axis}
\end{tikzpicture}%
			}
		\end{tabular}
		\\
		Key Idea of Our Approach
		&
		\hspace{-0.5cm}
		Triangulation Transfer via Computed Matchings
		&
		Runtime Comparison on FAUST Dataset
	\end{tabular}
	\vspace{-0.1cm}
	
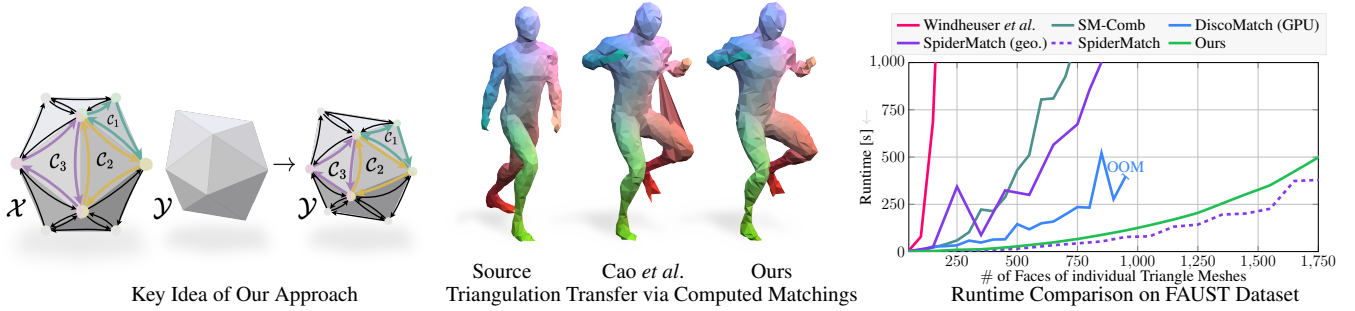
\captionof{figure}{
		\textbf{Left:} Illustration of our approach for geometrically consistent matching of 3D shape $\meshX$ to 3D shape $\meshY$. We represent shape $\meshX$ using $n$ surface cycles $\contour_1,\dots,\contour_n$ and then consistently match these  to shape $\meshY$ while  preserving neighbourhood relations between the cycles.
		\textbf{Middle:} The triangulation- {and colour-}transfer between shapes using computed matchings of Cao~\etal~\cite{cao2023unsupervised}  (not geometrically consistent) and ours (geometrically consistent) illustrates the importance of geometric consistency.
		\textbf{Right:} Our method is the first 3D shape matching approach that yields globally geometrically consistent matchings, that is globally optimal, and scalable in practice, opposed to all competing methods ({OOM stands for `out of memory' and} the dashed line indicates a weak notion of geometric consistency, see explanation of methods in \cref{sec:experiments}).
	}
	\label{fig:teaser}
\end{strip}

\begin{abstract}
Geometric consistency, i.e.~the preservation of neighbourhoods, is 
a natural and strong prior in 3D shape matching.
Geometrically consistent matchings are crucial for many downstream applications, such as texture transfer or statistical shape modelling.
Yet, in practice, geometric consistency is often overlooked, or only achieved under severely limiting assumptions (e.g.~a good initialisation).
In this work, we propose a novel formalism for computing globally optimal and geometrically consistent matchings between 3D shapes which is scalable in practice.
Our key idea is to represent the surface of the source shape as a collection of cyclic graphs, which are then consistently matched to the target shape. Mathematically, we construct a hyper product graph (between source and target shape), and then cast
3D shape matching 
as a minimum-cost circulation flow problem in this hyper graph, which yields global geometrically consistent matchings between both shapes.
We empirically show that our formalism is efficiently solvable and that it leads to high-quality results.
Our code is publicly available.\footnote{\url{https://github.com/paul0noah/geco}}
\end{abstract}
\section{Introduction}
\label{sec:intro}

The availability of correspondences between visual data are a key prerequisite for many visual computing tasks, including 3D reconstruction~\cite{leroy2024grounding}, loss computation for deep learning~\cite{carion2020end}, protein alignment~\cite{krissinel2004secondary}, anomaly detection~\cite{van2022hierarchical}, object recognition~\cite{kim19913}, shape modelling~\cite{egger20203d} and others.
In many practical cases, it is desirable that correspondences preserve neighbourhood relations.
For example, in 3D shape matching this could mean that when bringing a pair of neighbouring points on a  source shape into correspondence with a pair of points on a target shape, the pair of points must remain neighbours on the target.
Such a neighbourhood preservation can serve as 
strong topological prior, for example to resolve ambiguities, achieve robustness, or ensure well-posedness of an otherwise ill-posed problem.
However, despite its crucial importance, neighbourhood preservation is in practice often-times overlooked.
This is because many of the well-known formalisms are not efficiently solvable (e.g.~the quadratic assignment problem~\cite{rendl1994quadratic}, or general integer   
programming~\cite{schrijver2003combinatorial}, which have been used to tackle image keypoint matching~\cite{ma2021image}, graph matching~\cite{yan2016ashortsurvey,haller2022comparative}, or 3D shape matching~\cite{windheuser2011geometrically,maron2016point,bernard2020mina,gao2023sigma}).

In this work we specifically focus on the task of \emph{non-rigid geometrically consistent 3D shape matching}, in which neighbourhood-preserving correspondences between two given 2D manifolds (embedded in 3D space) are sought for. 
So far, there does not exist any 3D shape matching approach that combines the following three desirable properties:
(i) global optimality, 
(ii) neighbourhood preservation, and 
(iii) scalability, see \cref{table:comparison}.
While the recent approach \spidermatch~\cite{roetzer2024spidermatch} fulfils (i)-(iii) to some extent, its scalability relies heavily on the choice of the cycle that is used to represent one of the shapes (cf.~\cref{fig:teaser} right: solid vs.~dashed purple line), thereby trading off neighbourhood preservation with scalability.
Furthermore, proper geometric consistency (\cref{def:geo-cons}) can only be enforced at intersection points of the cycle and thus also the quality of geometric consistency depends on the choice of the cycle.
Inspired by this idea of representing a shape using a cycle, we represent one shape using a \emph{collection of cycles} and show that this leads to an (in practice) efficiently solvable formalism which ensures proper geometric consistency.
\begin{table}[t]
\centering
\small
\setlength{\tabcolsep}{2pt}
\renewcommand{\arraystretch}{0.9}
\begin{tabular}{@{}lccccccccc@{}}
\toprule
                                                &\textbf{Globally}    &\textbf{Geometr.}      &\textbf{}        \\ 
\textbf{Method}                                 &\textbf{Optimal}     &\textbf{Consistent}    & \textbf{Scalable}     \\ \midrule
MINA~\cite{bernard2020mina}                     & (\cmark)              &\xmark                 & \xmark\\
SIGMA~\cite{gao2023sigma}                       & (\cmark)              &\xmark                 & \xmark\\
PMSDP~\cite{maron2016point}                     &  \xmark               &\xmark                 & \cmark\\
\windheuser~\cite{windheuser2011geometrically}  &  \xmark               &\cmark                 & \xmark\\
\spidermatch~\cite{roetzer2024spidermatch}      & \cmark                &(\cmark)               & (\cmark)\\
\ours                                           & \cmark                &\cmark                 & \cmark\\
\bottomrule
\end{tabular}%
\caption{\textbf{Comparison} of axiomatic 3D shape matching methods.
}
\vspace{-0.3cm}
\label{table:comparison}
\end{table}
We summarise our main contributions as follows:
\begin{itemize}[leftmargin=1em,topsep=0em,parsep=0em,itemsep=0em]
    \item For the first time, we present a globally geometrically consistent formalism for 3D shape matching that is efficiently solvable to global optimality in practice.
    \item To achieve this, we introduce a novel 3D shape representation in which we represent a 3D shape using a \emph{collection of surface cycles}, so that 3D shape matching can be cast as finding a minimum-cost flow circulation in a \emph{hyper product graph}.
    \item We experimentally show that our formalism leads to {high quality} and geometrically consistent {matchings between two 3D shapes}.
    \item %
    In addition to 3D shape matching, we also show, in a proof-of-concept manner, that our formalism can generalise to specific instances of other matching problems, such as graph matching when the source graph is planar.
\end{itemize}

\section{Related Work}
\label{sec:related_works}
In the following we discuss works that are most relevant to our approach.
We start by discussing efficiently solvable matching problems, continue with 3D shape matching, and conclude with geometrically consistent 3D shape matching.

\textbf{Efficiently Solvable Matching Problems.} There are various instances of correspondence problems that can be solved efficiently.
Among them is the linear assignment problem (LAP)~\cite{kuhn1955hungarian}, which matches points without considering their neighbourhood relations.
For matching time series data, sequences, or (open) contours, the popular dynamic time warping algorithm can efficiently compute solutions while preserving neighbourhood relations~\cite{sankoff1983time}.
Closed contours can be matched analogously via graph cuts in product graphs~\cite{schmidt2009planar}.
Certain tracking problems can be formulated as efficiently solvable flow problems~\cite{leal2013pedestrian}, and model-based image segmentation can also be solved efficiently by matching a 2D contour to an image~\cite{coughlan2000efficient,felzenszwalb2005representation,schoenemann2009combinatorial}.
The matching of a 2D contour to a 3D shape can be solved using variants of Dijkstra's algorithm for finding minimum cost cycles in product graphs~\cite{lahner2016efficient,roetzer2023conjugate}.
The mentioned approaches show that there are formalisms to efficiently solve a diverse range of matching problems.
Yet, these approaches do not directly generalise to 3D shape matching.
In this work, we propose a novel geometrically consistent 3D shape matching formalism that is (in practice) efficiently solvable to global optimality.

\textbf{3D Shape Matching}
is the task of finding correspondences between two non-rigidly deformed surfaces.
For an in-depth overview on {3D} shape matching we refer the reader to survey papers~\cite{van2011survey, tam2012registration, deng2022survey}.
Many axiomatic shape matching approaches~\cite{ovsjanikov2010one,bronstein2010gromov,rodola2012game} build on the functional maps framework, efficiently solving shape matching in the spectral domain~\cite{ovsjanikov2012functional,maron2016point,gasparetto2017spatial,eisenberger2020smooth,ren2021discrete,maggioli2024rematching}.
In addition, functional maps have been adopted in several deep shape matching variants, either trained in a supervised~\cite{litany2017deep,wiersma2020cnns,li2020shape,groueix20183d,trappolini2021shape} or in a unsupervised manner~\cite{halimi2019unsupervised,sharp2022diffusionnet,donati2022deep,li2022learning,cao2023unsupervised,attaiki2023understanding,dutt2024,bastian2024hybrid}.
There are also alternative approaches based on consensus maximisation~\cite{probst2019unsupervised}, or  using convex relaxations~\cite{chen2015robust,maron2016point} or mixed integer programming~\cite{bernard2020mina, gao2023sigma}.
Yet, many of the existing 3D shape matching methods neglect geometric consistency, as it leads to hard-to-solve formalisms. 

\textbf{Geometrically Consistent 3D Shape Matching.} 
There are several approaches that have recognised the importance of geometric consistency in 3D shape matching.
Some of them incorporate neighbourhood information by 
using formalisms based on the quadratic assignment problem (QAP) -- yet, the QAP is NP-hard~\cite{rendl1994quadratic}, so that respective approaches consider relaxations~\cite{dym2017relax,burghard2017efficient,kushinsky2019sinkhorn} or heuristics~\cite{solomon2016entropic,holzschuh2020simulated,benkner2021q}.
Opposed to such discrete formulations, there are also approaches that tackle 3D shape matching by deforming a continuous shape parametrisation using local optimisation.
However, due to the severe non-convexity of resulting problems, they rely on a good initialisation, e.g.~in the form of sparse sets of landmark correspondences~\cite{schreiner2004inter, sharma2011topologically,schmidt2020inter,takayama2022compatible,schmidt2023surface}, or  in the form of dense correspondences~\cite{ezuz2017deblurring,vestner2017product,ezuz2019elastic}.
Windheuser~\etal~\cite{windheuser2011geometrically,windheuser2011large} have modelled geometrically consistent shape matching in the discrete domain by matching triangles to triangles using an (expensive to solve) integer linear program. Recently, approximative solvers~\cite{roetzer2022scalable,roetzer2024discomatch} and extensions for partial shapes~\cite{ehm2024geometrically,ehm2024partial} have been proposed.
Furthermore, 3D shape matching has recently been formulated as a shortest path problem {with additional constraints}~\cite{roetzer2024spidermatch}.
In this work, the surface mesh of the source 3D shape is represented using a long self-intersecting curve that traces the shape surface{, and is then matched to the other shape while preserving intersections of the curve}.
While their idea of considering alternative 3D shape representations is promising, in the presented framework proper geometric consistency is only enforced at intersection points of the curve~(see \cref{tab:dir-energy}).
In addition, runtimes grow drastically with an increasing number of intersection points, which can be seen in \cref{fig:teaser} right when comparing SpiderMatch~\cite{roetzer2024spidermatch} (using a curve with few intersections) to SpiderMatch~(geo.)~(using a curve with many intersections).
Inspired by the idea of an alternative path-based 3D shape representation, we represent a 3D shape as a collection of cyclic graphs,
and then tackle 3D shape matching by solving \emph{coupled} matching problems of the individual cyclic paths. Based on the couplings, we are able to ensure global geometry consistency. 
Furthermore, we empirically observe that the resulting formalism is efficiently solvable to global optimality.

\section{Background}
\label{sec:background}
We consider the task of finding a matching between a source shape $\meshX$ and a target shape $\meshY$ (\cref{sec:shapesasgraphs}) such that the matching is geometrically consistent (\cref{sec:geo-cons}).
Our main notation is summarised in \cref{table:notation}.
\begin{table}
\renewcommand{\arraystretch}{0.9}
\small\centering
	\begin{tabularx}{\columnwidth}{lp{5.6cm}}
        \toprule
        \textbf{Symbol} & \textbf{Description} \\
        \toprule
        $\meshX=(\verts_\meshX,\faces_\meshX )$ &3D shape\\
        $\graph_\meshX=(\verts_\meshX, \edges_\meshX)$ &3D shape graph of $\meshX$\\
        $\contour = (\contourVerts, \contourEdgs)$ & \surfacecycle~(cycle in $\graph_\meshX$)\\
        $\{\contour_1,\dots,\contour_{n}\}$ & Representation of $\meshX$ with $n$ surface cycles\\
        $n$ & Number of surface cycles\\
        $\meshY=(\verts_\meshY,\faces_\meshY )$ &3D shape $\meshY$\\
        $\graph_\meshY=(\verts_\meshY,\faces_\meshY )$ &3D shape graph of $\meshY$\\
        $\prodGraph_i = (\verts_{\prodGraph_i}, \edges_{\prodGraph_i})$ & Product graph (of $\contour_i$ and mesh $\meshY$)\\
        $\{\prodGraph_1,\dots,\prodGraph_{n}\}$& Product graph collection (of $\{\contour_1,\dots,\contour_{n}\}$\\
        & and mesh $\meshY$)\\
        $\hyperProdGraph = (\verts_\hyperProdGraph, \edges_\hyperProdGraph)$ & Hyper product graph (coupled product\\
        &  graphs $\{\prodGraph,\dots,\prodGraph_{n}\}$)\\
        $m$ & Number of hyper edges $|\edges_\hyperProdGraph|$\\
        $H$ & Vertex edge incidence matrices of $\hyperProdGraph$\\
        \bottomrule
	\end{tabularx}
        \vspace{-0.2cm}
	\caption{Summary of the \textbf{notation} used in this paper.
	}
	\label{table:notation}
\end{table}

\subsection{Shapes and Graphs}\label{sec:shapesasgraphs}

\begin{figure}
    \centering
    \includegraphics[width=0.8\columnwidth]{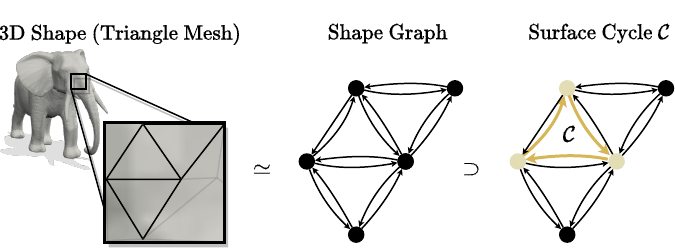}
    \caption{We represent a 3D shape (left) as a (directed) shape graph $\graph_\meshX$ (middle).
    We call a cycle $\contour$ in $\graph_\meshX$ a \emph{\surfacecycle}~(right).
    }
    \label{fig:shape-graph-to-surface-cycle}
\end{figure}

In the following, we define shapes and other relevant concepts, which are also illustrated in \cref{fig:shape-graph-to-surface-cycle}. %
We consider \emph{3D shapes} represented as triangular surface mesh:
\begin{definition}[3D shape]
    A \emph{3D shape} $\meshX$ is defined as a tuple $(\verts_\meshX, \faces_\meshX)$ of vertices $\verts_\meshX$ and consistently oriented (e.g.~clock-wise) triangles $\faces_\meshX \subset \verts_\meshX \times \verts_\meshX \times \verts_\meshX$, such that $\meshX$
    forms an orientable continuous 2D manifold (possibly with boundary) embedded in 3D space. %
\end{definition}
For our formalism it is advantageous to interpret a 3D shape as a graph, which we denote the \emph{shape graph}:
\begin{definition}[Shape graph]
    The \emph{shape graph} (of 3D shape $\meshX$) is a tuple $\graph_\meshX = (\verts_\meshX, \edges_\meshX)$ of vertices $\verts_\meshX$ and directed edges $\edges_\meshX \subset \verts_\meshX \times \verts_\meshX$, such that 
        each triangle in $\faces_\meshX$ is represented by three unique and consistently directed edges. 
\end{definition}

In addition to the shape graph (representing the whole 3D shape $\meshX$), we consider certain subgraphs of $\graph_\meshX$:
\begin{definition}[Surface cycle]
    A \emph{\surfacecycle}~(on 3D shape $\meshX$) is a tuple $\contour = (\contourVerts, \contourEdgs)$ with (non-empty) vertices $\contourVerts \subset \verts_\meshX$ and (non-empty) directed edges $\contourEdgs \subset \contourVerts \times \contourVerts $, such that (i) $\contourEdgs \subset \edges_\meshX$, and (ii) each vertex $v\in\contourVerts$ has exactly one incoming and one outgoing edge in $\contourEdgs$.
\end{definition}

\subsection{Geometrically Consistent 3D Shape Matching}\label{sec:geo-cons}

We define geometrically consistent 3D shape matching in terms of the respective shape graphs:
\begin{definition}[Geometrically Consistent Matching]\label{def:geo-cons}
The mapping $\phi: \verts_\meshX \rightarrow \verts_\meshY$ is a
\emph{geometrically consistent matching} from the source shape $\meshX$ to the target shape $\meshY$ 
if $\phi$ preserves neighbourhoods as follows: 
whenever two vertices $x,\bar{x} \in \verts_\meshX$ are connected (i.e.~$(x,\bar{x})\in\edges_\meshX$ or $(\bar{x},x)\in\edges_\meshX$),
then their corresponding vertices $\phi(x),\phi(\bar{x})\in\verts_\meshY$ must either be (i) connected (i.e.~$(\phi(x),\phi(\bar{x}))\in \edges_\meshY$ or $(\phi(\bar{x}),\phi(x))\in\edges_\meshY$), or (ii) the same vertex (i.e.~$\phi(x)=\phi(\bar{x})$).
\end{definition}
The intuition is that neighbouring elements of $\meshX$ must be matched to neighbouring elements of $\meshY$.
{We note that we require both shapes to have equal genus such that our definition of geometric consistency makes sense.}

\section{Our 3D Shape Matching Approach}
\label{sec:method}
In this section we develop our geometrically consistent 3D shape matching formulation which ensures global geometric consistency.
Our approach is based on representing the source 3D shape using a collection of surface cycles, which we then match to the target shape (\cref{sec:surf-cycles-repr}).
To this end, we formulate an individual subproblem for each surface cycle (\cref{sec:individual-sc-matching}).
Further, we couple individual subproblems (\cref{sec:coupling,sec:injectivity}) so that our matchings are geometrically consistent according to \cref{def:geo-cons}.
Finally, we show that our overall formalism can be interpreted as a minimum-cost circulation flow problem on a hyper product graph (\cref{sec:hyper-product-graph}). 

\subsection{Surface Cycle-based 3D Shape Representation}\label{sec:surf-cycles-repr}
\begin{figure}
    \centering%
    \includegraphics[width=0.8\columnwidth, trim={0 0.7cm 0cm 0cm}]{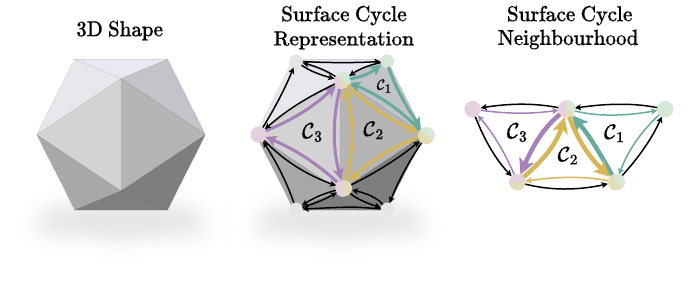}
    \vspace{-0.1cm}
    \caption{\textbf{Surface cycle collection} (middle) for representing the 3D shape of an icosahedron (left). 
    Individual surface cycles are glued together via shared edges (neighbouring \surfacecycles~share opposite edges, cf.~pairs of thickened edges on the right: 
    \textcolor{scPURPLE}{\rule[0mm]{0.6mm}{0.3cm}}\,\textcolor{scYELLOW}{\rule[0mm]{0.6mm}{0.3cm}} 
    and 
    \textcolor{scYELLOW}{\rule[0mm]{0.6mm}{0.3cm}}\,\textcolor{scGREEN}{\rule[0mm]{0.6mm}{0.3cm}}).
    }
    \label{fig:surface-cycle-representation}
\end{figure}

We aim to represent the surface of 3D shape $\meshX$ with a collection of surface cycles which are glued together at \emph{opposite edges}:
\begin{definition}[Opposite edge]
    The opposite edge to an edge $(v, w)$ is defined as $ -(v, w) \coloneqq (w, v)$.
\end{definition}
We note that for every non-boundary edge $e\in\edges_\meshX$ of shape $\meshX$, the opposite edge $-e\in\edges_\meshX$ is also part of the shape graph $\graph_\meshX$ (cf.~\cref{fig:surface-cycle-representation} right) because the pair of opposite edges $e$ and $-e$ belong to two neighbouring triangles.
\begin{definition}[Shape as collection of \surfacecycles]\label{def:surface-cycle-repr}
We represent a 3D shape $\meshX$ with a collection of $n \in \mathbb{N}^+$ 
 surface cycles $\contour_1, \dots, \contour_{n}$ (on shape $\meshX$) that
 partition the surface of $\meshX$ into $n$ polygonal patches, such that: 
\\
(i) $\edges_{\contour_i} \cap \edges_{\contour_j} = \emptyset$ for all $i, j \in \{1,\ldots,n\}, i \neq j$, and  \\
(ii) for every non-boundary edge $e \in \edges_{\contour_i}$ of shape $\meshX$ there exists $j$ such that $-e \in \edges_{\contour_j}$.
\end{definition}
Condition (i) ensures that each edge of $\meshX$ can be part of at most one \surfacecycle{}, and (ii) ensures that neighbouring \surfacecycles{} cover opposite edges, see~\cref{fig:surface-cycle-representation}.
We note that from here on we assume that each surface cycle represents an individual triangle of shape $\meshX$. 
We discuss more general polygonal surface cycles in \cref{sec:supp:other-surface-cycles} in the supplementary.

\subsection{Our Coupled Product Graph Formalism} 
\begin{figure*}[t]
    \centering
    \includegraphics[width=0.88\linewidth,trim={1.8cm 0.1cm 1cm 0.2cm}]{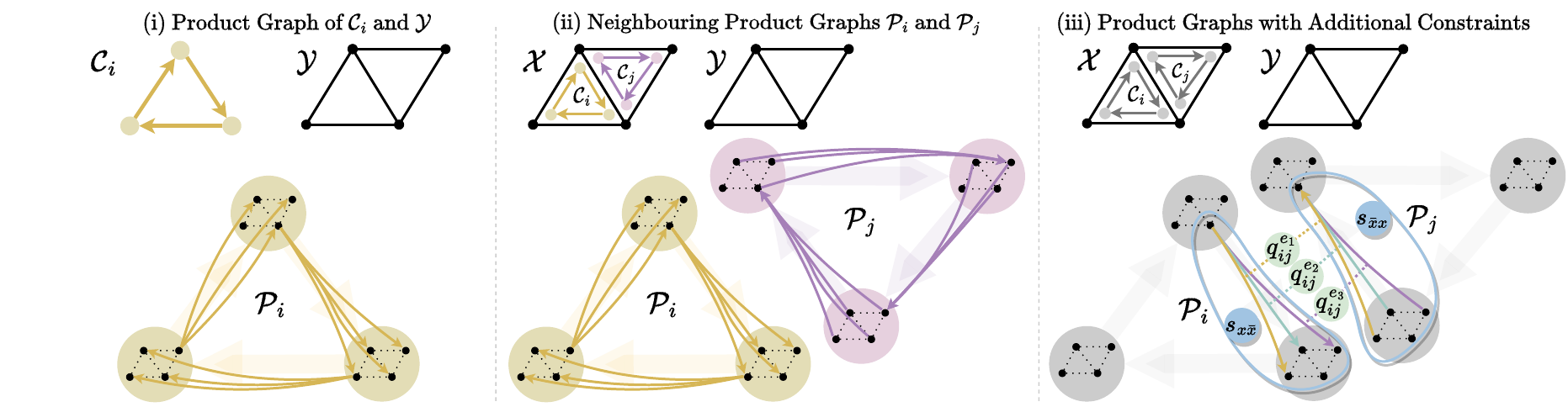}
    \caption{Conceptual summary of our formalism.
    \textbf{Left:} Illustration of product graph $\prodGraph_i$ between the \surfacecycle~$\contour_i$ and shape graph $\meshY$. 
    Each black dot inside the big yellow circle
    represents a product vertex. Overall, the product graph can be thought of having one copy of $\meshY$ for each vertex of $\contour_i$, which are appropriately connected via edges (only a subset of edges are drawn to reduce visual clutter).
    \textbf{Middle:} Two neighbouring product graphs $\prodGraph_i$ and $\prodGraph_j$ that arise from two neighbouring \surfacecycles~$\contour_i$ and $\contour_j$.
    \textbf{Right:} Illustration of constrained neighbouring product graphs. We add additional coupling constraints so that opposite edges of the product graphs (i.e.~edges with same colour, see coloured lines \textcolor{scYELLOW}{\rule[0.6mm]{0.25cm}{0.7mm}},\textcolor{scGREEN}{\rule[0.6mm]{0.25cm}{0.7mm}},\textcolor{scPURPLE}{\rule[0.6mm]{0.25cm}{0.7mm}}) are coupled via coupling constraints $q_{ij}^{e_1}, q_{ij}^{e_2}, q_{ij}^{e_3}\in\verts_\linking$ (see green dots \cir[scLIGHTGREEN,fill=scLIGHTGREEN]{4pt}).
    Furthermore, we bundle edges of the product graphs so that each edge of the surface cycles is matched exactly once.
    To this end, we add injectivity constraints $s_{x\bar{x}}$, $s_{\bar{x}x} \in \verts_\injectivity$ (see blue dots \cir[scLIGHTBLUE,fill=scLIGHTBLUE]{4pt} within bundles of edges).
    }
    \label{fig:hyper-product-graph}
\end{figure*}

On a high-level, we use the product graph formalism introduced in~\cite{lahner2016efficient} to address the matching between individual surface cycles and shape $\meshY$, which results in geometric consistency \emph{along the path within each individual cycle}.
Further, we introduce coupling constraints that have the effect of glueing neighbouring product graphs together.
Combined with additional injectivity constraints, our matching is globally geometrically consistent, i.e.~it \emph{preserves neighbourhoods between surface cycles}, see \cref{fig:hyper-product-graph} for an overview.

\subsubsection{Individual Surface Cycle Matching Subproblems}\label{sec:individual-sc-matching}
Each individual surface cycle $\contour_i$ of $\meshX$ can be matched to shape $\meshY$ by finding a cyclic path in their \emph{product graph} $\prodGraph_i=(\verts_{\prodGraph_i}, \edges_{\prodGraph_i})$.
It has been introduced in \cite{lahner2016efficient} for the problem of matching a 2D shape to a 3D shape and reads:
{
\begin{definition}[Product graph $\prodGraph_i$]\label{def:prodgraph}
    The product graph $\prodGraph_i = (\verts_{\prodGraph_i}, \edges_{\prodGraph_i})$ between the $i$-th \surfacecycle~$\contour_i$ and the 3D shape $\meshY$ is a directed graph defined as
    \begin{equation}\label{eq:prod-graph}
    \begin{aligned}
        \verts_{\prodGraph_i} &= \verts_{\contour_i} \times \verts_\meshY,\\
        \edges_{\prodGraph_i} &= \{\left( v, \bar{v}\right) \in \verts_{\prodGraph_i}\times \verts_{\prodGraph_i}\; |\; 
        v = \twovector{x}{y}, \bar{v} = \twovector{\bar{x}}{\bar{y}}\\
        &\qquad(x, \bar{x}) \in \edges_{\contour_i},(y, \bar{y}) \in \edges_\meshY^+\},
    \end{aligned}
    \end{equation}
    with the \emph{extended edge set} $\edges^+_\meshY := \edges_\meshY \cup \{(y, y) \; | \; y \in \verts_\meshY\}$.
\end{definition}
}

The key idea is that each edge $(v,\bar{v})\in\edges_{\prodGraph_i}$ in the product graph can be interpreted as a (potential) matching between an edge of $\contour_i$ and an edge (or a vertex to account for stretching/compression)
of shape $\meshY$.
As shown in~\cite{lahner2016efficient}, matching surface cycle $\contour_i$ to shape $\meshY$ amounts to solving a cyclic shortest path problem in the  product graph $\prodGraph_i$.

However, for our setting of \emph{3D-to-3D shape matching},
the neighbourhood between pairs of surface cycles cannot be ensured when considering vanilla shortest path algorithms, since they would solve the $n$ individual surface cycle matching subproblems independently.
To tackle this, we consider the linear programming~(LP) formalism of cyclic shortest path problems and add additional constraints to couple
the individual product graphs $\prodGraph_1,\dots,\prodGraph_n$, as explained next.

\subsubsection{Subproblem Coupling}\label{sec:coupling}
We couple the individual surface cycle matching subproblems $\prodGraph_1,\dots,\prodGraph_n$ by glueing them together at opposite edges.
To this end, we introduce \emph{coupling constraints} $\verts_\linking$\, which serve the
purpose of ensuring that matchings of opposite edges are consistent, i.e.~resulting matchings (or rather resulting shortest cyclic paths of neighbouring surface cycles) must go through opposite edges, see \cref{fig:hyper-product-graph}~(iii).
De-facto, this enforces the \emph{glueing} of neighbouring product graphs, and thus ensures geometric consistency of neighbouring surface cycles.
In addition to the coupling, we want to ensure injectivity for each surface cycle edge, i.e.~each such edge is matched exactly once.

\subsubsection{Surface Cycle Matching Injectivity}\label{sec:injectivity}
To enforce that each surface cycle edge is matched exactly once, i.e.~it is matched to exactly one edge (or vertex) of $\meshY$, we introduce \emph{injectivity constraints} $\verts_\injectivity$.
For that, let us denote the set of all edges (in the product graph) that are potential matching candidates of the edge $(x,\bar{x}) \in \edges_{\contour_i}$ as the \emph{edge bundle of $(x,\bar{x})$}.
For each such edge bundle, we introduce one
injectivity constraint (see \cref{fig:hyper-product-graph} (iii)), which has the purpose 
to ensure that only a single edge of each edge bundle is part of the final matching.

\subsection{Constraint Matrices}
As mentioned, we use the LP formalism of the independent cyclic shortest path problems and add additional constraints for coupling an matching injectivity.
To this end, we consider the constraint matrix $H = \begin{bsmallmatrix}P\\L\\S \end{bsmallmatrix}$ consisting of submatrices $P$, $L$ and $S$.
Here, submatrix $P$ represents the collection of incidence matrices of individual product graphs, submatrix $L$ the couplings, and submatrix $S$ the injectivity components.

\textbf{The matrix $P$.}
The vertex edge incidence matrix $P \in \{-1,0,1\}^{c \times m}$ (with $m=|\edges_\hyperProdGraph|$ columns{, i.e.~one column for each product edge,} and $c=\sum_{i=1}^n|\verts_{\contour_i}||\verts_\meshY|$ many rows) 
represents all $n$ product graphs and is defined as $P \coloneqq \text{diag}(P_1,\ldots,P_{n})$,
i.e.~it contains the vertex edge incidence matrices $P_i$ of individual product graphs $\prodGraph_i$ as diagonal blocks.
To illustrate the structure of the matrix representation of one $\prodGraph_i$, we now discuss the simplified case of non-degenerate matchings (i.e.~we do not allow edge to vertex matchings, see \cref{sec:supp:algebraic} in the supplementary for more details).
In this case, the incidence matrix $P_i$ reads
\begin{equation}
    P_i\coloneqq C_i^+\otimes Y^+ - C_i^-\otimes Y^-,
\end{equation}
with $\otimes$ being the Kroenecker product and with $C_i \in \{-1,0,1\}^{|\verts_{\contour_i}|\times|\edges_{\contour_i}|}$ and $Y\in\{-1,0,1\}^{|\verts_{\meshY}|\times|\edges_{\meshY}|}$ being the incidence matrices of surface cycle $\contour_i$ and shape graph $\graph_\meshY$ respectively (we note that $C_i$ has exactly two non-zeros of opposite sign in each row and column since it is the incidence matrix of a cycle).
We use the notation $+$ and $-$ to split the incidence matrix into incoming and outgoing incidence matrix, respectively.
In other words,
$C_i^+$ and $C_i^-$ as well as $Y^+$ and $Y^-$ are the non-negative and non-positive entries of the incidence matrices $C_i$  and $Y$, respectively.
We note that the sign splitting of incidence matrices $C_i$ and $Y$ is necessary to account for proper edge directions in the resulting product graph.
By definition, their edge directions induce edge directions of product edges of respective product graph $\prodGraph_i$.
Each $P_i$ has a block structure with $|\verts_\meshY|\times|\edges_\meshY|$ sized blocks and each block contains either only $+1$ or $-1$ entries.
Further, each block contains exactly one non-zero per column, see
\cref{fig:matrix-structure-of-graphs} in the supplementary.

\textbf{The matrix $L$.}
The coupling of opposite edges of neighbouring product graphs is captured in the matrix
\begin{equation}
   L\coloneqq K\otimes \boldsymbol{I}_{|\edges_\meshY^+|}.
\end{equation}
Here $K\in \{-1,0,1\}^{p\times
|\edges_\meshX|}$ ($p$ is the number of undirected non-boundary edges of $\meshX$) is an incidence matrix which represents the incidence of opposite edges across all $n$ surface cycles.
$K$ has at most one non-zero element per column and exactly two non-zeros per row (with opposite sign) and thus $L$ has identical block structure with non-negative and non-positive blocks respectively.
We note that the definition of $P$ requires
columns of the non-positive blocks of $L$ to be permuted (so that it maps opposite edges of neighbouring product graphs), see \cref{sec:supp:algebraic} in the supplementary.

\textbf{The matrix $S$.} 
The bundling of edges of all $n$ product graphs can be described using the matrix
\begin{equation}
    S \coloneqq \boldsymbol{I}_{|\edges_\meshX|} \otimes 1_{|\edges^+_\meshY|}^T.
\end{equation}
Here, $S$ has block diagonal structure consisting of $|\edges_\meshX|$ many non-negative blocks (i.e. $1_{|\edges^+_\meshY|}^T$ blocks).
{We note that $Sx = 1$ enforces matching injectivity by enumerating product edges such that all $|\edges^+_\meshY|$-many matching candidates of a single surface cycle are grouped.}

\subsection{Resulting Integer Linear Program}\label{sec:poly-algo}
We use the previously described constraint matrix $H$ to cast  geometrically consistent shape matching as a linear program which essentially represents $n$ many coupled cyclic shortest path problems.  

\begin{figure}
    \centering
    \includegraphics[width=0.8\columnwidth,trim={1.5cm 0 0 0}]{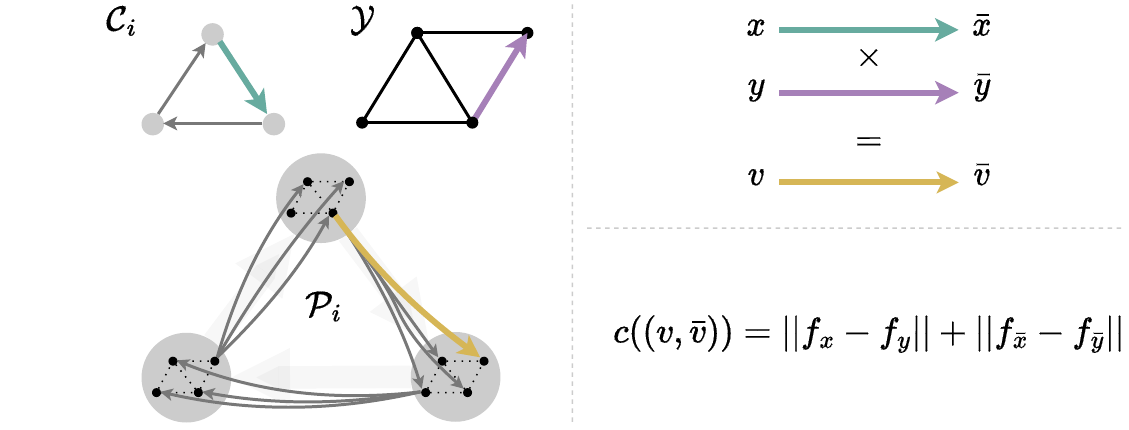}
    \caption{
    Visualisation of \textbf{edge cost $c\left((v,{\bar{v}})\right)$ computation} for the product edge $(v, \bar{v})$. 
    The product edge defines a potential matching between edges $(x,\bar{x})\in\edges_\meshX$ of shape $\meshX$ and $(y,\bar{y})\in\edges_\meshY^+$ of shape $\meshY$. Thus, we define the matching cost $c((v, {\bar{v}}))$ as feature difference between features $\|f_x-f_y\|$ and $\|f_{\bar{x}}-f_{\bar{y}}\|$ %
    at respective source vertices $x$, $y$ and target vertices $\bar{x}$, $\bar{y}$.
    }
    \label{fig:matching-cost}
\end{figure}

\textbf{Matching costs.} 
The matching cost $c(e)$ {of product edge} $e=\left(\twovector{x}{y}, \twovector{\bar{x}}{\bar{y}}\right)$ measures how well the edge $(x,\bar{x})\in\edges_\meshX$ of shape $\meshX$ and the edge $(y,\bar{y})\in\edges_\meshY^+$ of shape $\meshY$ fit together, e.g.~by comparing geometric properties or feature descriptors, see also \cref{fig:matching-cost}.

\textbf{Optimisation problem.} We collect matching costs of all $m$ edges of the $n$ product graphs in the vector $c\in\mathbb{R}_+^m$.
With that, we can use an indicator representation $x$ for each edge ($x_k=1$ means that $k$-th edge is part of the final matching) so that our matching formalism reads
\begin{equation}\tag{GeCo3D}\label{eq:cycle-surfing}
    \underset{x\in \{0, 1\}^{m}}{\min}
    c^T x \quad\text{s.t.}\quad 
    H x = b.
\end{equation}
Here, $H$ is the previously introduced constraint matrix and $b\coloneqq \begin{bmatrix}
    0_{|\verts_\hyperProdGraph|-|\verts_\injectivity|}; 1_{|\verts_\injectivity|}
\end{bmatrix}$
is a column vector with all zeros except for ones in rows belonging to rows of submatrix $S$ of $H$.
We solve \eqref{eq:cycle-surfing} by considering its LP-relaxation, i.e. we replace~$x\in \{0, 1\}^{m}$ with $x\in [0, 1]^{m}$.
We empirically observe in our experiments, that all solutions are integral and consequently globally optimal.
\begin{lemma}
    Matchings between shapes $\meshX$ and $\meshY$ obtained by solving \eqref{eq:cycle-surfing} are globally geometrically consistent according to \cref{def:geo-cons}.
\end{lemma}
\begin{proof}
    We represent each triangle of the source shape $\meshX$ using a surface cycle.
    Furthermore, matchings obtained by solving \eqref{eq:cycle-surfing} preserve neighbourhood relations along the path of a matched surface cycle as well as between the surface cycles. 
    Overall, this ensures that connected vertices on $\meshX$ are only matched to connected vertices (or the same vertex) on shape $\meshY$.
\end{proof}

\subsection{Hyper Product Graph Interpretation}\label{sec:hyper-product-graph}
\begin{figure}[b]
    \centering
     \includegraphics[width=0.8\columnwidth,trim={0.6cm 0.1cm 0 0.2cm}]{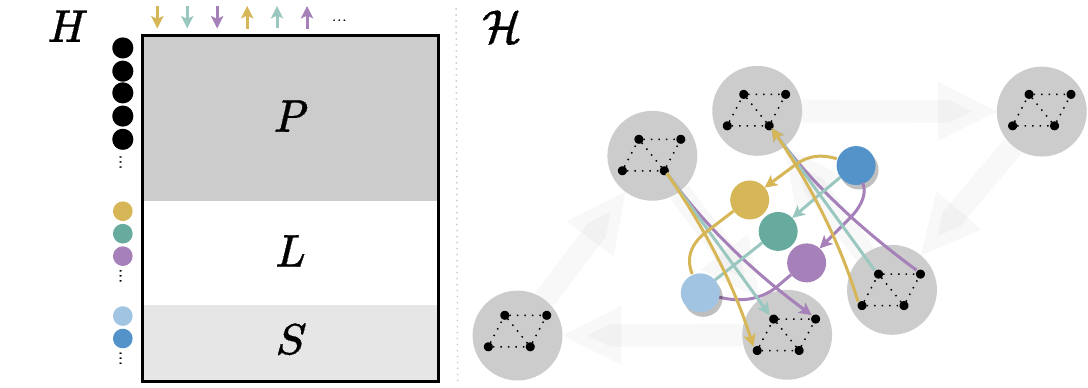}
    \caption{Illustration of interpreting our constraint matrix $H$ as the \textbf{vertex hyper edge incidence matrix} of a hyper product graph $\hyperProdGraph$, {see \cref{sec:detailed-prod-graph} in the supplementary}.
    To obtain $\hyperProdGraph$, rows of $H$ are interpreted as vertices, and columns as directed hyper edges of $\hyperProdGraph$.}
    \label{fig:matrix-structure-schematic}
\end{figure}

Our optimisation problem~\eqref{eq:cycle-surfing} can be interpreted as 
finding a minimum-cost flow circulation in a directed hyper product graph $\hyperProdGraph$, see \cite{beckenbach2019matchings} for definitions of flows in directed hyper graphs.
Here, $\hyperProdGraph$ can be obtained by interpreting $H$ as a vertex hyper edge incidence matrix, i.e.~by interpreting rows of $H$ as vertices and columns of $H$ as directed hyper edges, see \cref{fig:matrix-structure-schematic} for an illustration.
For a formal definition of $\hyperProdGraph$ we refer to \cref{sec:detailed-prod-graph} in the supplementary.

In general, such flow circulation problems on hyper-graphs are NP-hard~\cite{beckenbach2018hypergraph}.
However, there are certain subclasses that are known to be solvable in polynomial time~\cite{truemper1977unimodular,evans1978network,soun1980single,jeroslow1992gainfree,bohmova2018sequence}.
Unfortunately, the specific structure of our hyper graph is not listed among the known polynomial-time solvable subclasses~\cite{truemper1977unimodular,evans1978network,soun1980single,jeroslow1992gainfree,bohmova2018sequence}.
Nevertheless, we empirically observe that all instances considered in our experiments yield an integral (and thus globally optimal) solution when solved using LP-relaxations.

\section{Experiments}\label{sec:experiments}

In this section, we experimentally evaluate our method's performance for 3D shape matching. Further, we show as proof of concept that our method is applicable to planar graph matching.

\textbf{Setup.} We conduct runtime experiments on an Intel Core i9 12900K with 128 GB DDR5 RAM.
For solving our linear programm \eqref{eq:cycle-surfing}, we use off-the-shelf solver Gurobi~\cite{gurobi} (version $10$).
We explicitly resolve coupling constraints during problem construction to obtain smaller constraint matrices, see \cref{sec:supp:reduced-problem-size} in the supplementary.
To better account for shrinking and stretching, 
we integrate a distortion bound, 
see \cref{sec:supp:distortion-bound} in the supplementary.
For quantitive comparison, we decimate shapes to $1000$ triangles using algorithms provided in~\cite{libigl}.
We use the deep features~\cite{cao2023unsupervised} for all methods except for \smooths~\cite{eisenberger2020smooth} and \discreteopt~\cite{ren2021discrete} for which we use the original (axiomatic) features as reported in respective papers.

\textbf{Metrics.} We evaluate matching accuracy using geodesic errors, i.e.~geodesic distance to ground-truth matchings.
We follow the Princeton protocol~\cite{kim2011blended} and normalise errors by the square-root of the shape area (see~\cite[Sec.~8.2]{kim2011blended}).
Furthermore, we evaluate geometric consistency of matchings using Dirichlet energies, i.e.~the deformation energy induced by the matching (see \cite[Sec.~8]{roetzer2024spidermatch}).

\textbf{Methods.} We compare the matching quality of the methods ULRSSM~\cite{cao2023unsupervised}, \smooths~\cite{eisenberger2020smooth}, \discreteopt~\cite{ren2021discrete} and \spidermatch~\cite{roetzer2024spidermatch}.
ULRSSM~\cite{cao2023unsupervised} is an unsupervised deep shape matching method achieving best matchings results on numerous benchmarks. Because of its good performance and its extensive evaluation, we consider it to be representative for deep shape matching methods.
\smooths~\cite{eisenberger2020smooth} is a functional map based shape alignment method which iteratively adds geometric information to compute matchings.
\discreteopt~\cite{ren2021discrete} is a functional map based method which relates functional maps to point-wise maps.
\spidermatch~\cite{roetzer2024spidermatch} is the only other method that is scalable and at the same time considers geometric consistency. 
Furthermore, for runtime evaluations, we only compare to other geometrically consistent methods since these aim to solve a much harder to solve problem (yet most of these methods do not scale well to relevant shape resolutions, see also \cref{fig:teaser} right).
In this sense, we consider Windheuser~\etal~\cite{windheuser2011geometrically}, SM-Comb~\cite{roetzer2022scalable}, DiscoMatch~\cite{roetzer2024discomatch} and \spidermatch~(geo.).
Windheuser~\etal~\cite{windheuser2011geometrically} propose a geometrically consistent 3D shape matching formalism which they solve using LP-relaxations and iterative variable fixations.
The methods SM-Comb~\cite{roetzer2022scalable} and DiscoMatch~\cite{roetzer2024discomatch} aim to solve this formalism approximately on CPU and GPU, respectively.
We use \spidermatch~(geo.) to indicate the \spidermatch~\cite{roetzer2024spidermatch} method when using a different curve to represent the source shape:
we consider a curve that covers \emph{all edges} of the source shape and thus, contains intersections at every vertex of the source shape.
This is necessary to guarantee global geometric consistency since \spidermatch~\cite{roetzer2024spidermatch} can only ensure geometric consistency (according to \cref{def:geo-cons}) at intersection points of the curve.
Thus, \spidermatch~(geo.) yields a stronger notion of geometric consistency compared to \spidermatch~\cite{roetzer2024spidermatch}.

\textbf{Datasets.} We evaluate shape matching on four different datasets: remeshed FAUST~\cite{ren2018continuous,donati2020deep,bogo2014faust} (100 near-isometric deformed human shapes from which we sample 100 test set pairs), SMAL~\cite{zuffi20173d} (49 non-isometric deformed animal shapes of eight species from which we sample 100 test set pairs), DT4D-H~\cite{magnet2022smooth} (9 different classes of humanoid/game character shapes in different poses taken from DeformingThings4D~\cite{li20214dcomplete} from which we sample 100 intra class test set pairs and 100 inter class test set pairs) and BeCoS~\cite{ehm2024beyond} (2543 animal and humanoid shapes from various datasets of which we consider the 141 full-to-full test set pairs).

\subsection{3D Shape Matching}

In the following, we evaluate our method's shape matching performance w.r.t.~runtime and matching quality. 

\textbf{Runtime.} In \cref{fig:teaser} right, we show runtime comparison to formalisms which consider geometric consistency. 
Curves show median runtimes over five instances of FAUST dataset.
Among all methods, ours and \spidermatch~\cite{roetzer2024spidermatch} scale best. They can handle 3D shapes with approximately twice as many vertices as their fastest competitors DiscoMatch~\cite{roetzer2024discomatch}, which tackles the problem proposed by Windheuser~\etal~\cite{windheuser2011geometrically}.
However, we note that when 
considering \spidermatch~(geo.) and consequently curves with more intersections (which is necessary for global geometric consistency)
the scalability of \spidermatch~(geo.) degrades drastically.
This stems from the branch and bound algorithm (having exponential worst-case runtime in general) which is used to solve \spidermatch~\cite{roetzer2024spidermatch}.
In contrast, we empirically observe (in all of our tested instances) that the linear programming relaxation of \eqref{eq:cycle-surfing} is always tight (i.e.~yields an integral optimal solution) which explains the scalability of our approach.
We emphasise that matchings computed from \eqref{eq:cycle-surfing} are provably geometrically consistent according to \cref{def:geo-cons}, while the matchings of \spidermatch~\cite{roetzer2024spidermatch} are not (cf.~also \cref{tab:dir-energy}).

\textbf{Full Shape Matching.} In \cref{tab:geoerrors} and \cref{tab:dir-energy}, we quantitatively compare the shape matching performance of various methods. Our method consistently produces best results across all datasets w.r.t.~mean geodesic errors and smoothest results w.r.t~Dirichlet energies. This showcases the beneficial effect of strong priors induced by geometric consistency. In addition, in \cref{fig:sm-qualitative}, we show qualitative results, which further showcases high-quality matching results computed with our method. We provide more results as well as ablation studies in \cref{sec:supp:more-sm} in the supplementary.

\begin{table}
    \centering
    {\footnotesize
        \setlength{\tabcolsep}{1.8pt}
        \renewcommand{\arraystretch}{0.9}
        \begin{tabular}{@{}l|ccccc@{}}
         \toprule
         \textbf{Method} &  \textbf{FAUST} & \textbf{SMAL} & \textbf{DT4D Intra} & \textbf{DT4D Inter} &\textbf{BeCoS}\\
         \midrule
         ULRSSM~\cite{cao2023unsupervised}      & 0.031  & 0.048& 0.033      &0.041       &0.057\\
         \smooths~\cite{eisenberger2020smooth}  
                     & 0.379  & 0.376& 0.373      &0.420       &0.370\\
         \discreteopt~\cite{ren2021discrete}   
                     & 0.110  & 0.268& 0.075      &0.170       &0.270\\
         \spidermatch~\cite{roetzer2024spidermatch}
                     & 0.029  & \textbf{0.044}& 0.027      &0.041       &0.057\\
         Ours        & \textbf{0.027}  & \textbf{0.044}& \textbf{0.024}      &\textbf{0.039}       &\textbf{0.056}\\
         \bottomrule
        \end{tabular}
    }
    \vspace{-0.1cm}
    \caption{Comparison of \textbf{mean geodesic errors} ($\downarrow$) of various shape matching methods. We can see that our method consistently outperforms other methods on all five datasets.
    }
    \label{tab:geoerrors}
\end{table}

\begin{table}
    \centering
    {\footnotesize
        \setlength{\tabcolsep}{1.8pt}
        \renewcommand{\arraystretch}{0.9}
        \begin{tabular}{@{}l|ccccc@{}}
         \toprule
         \textbf{Method} &  \textbf{FAUST} & \textbf{SMAL} & \textbf{DT4D Intra} & \textbf{DT4D Inter} &\textbf{BeCoS}\\
         \midrule
         ULRSSM~\cite{cao2023unsupervised}      & 2.1   & 2.6   & 3.4      &2.6       &4.1\\
         \smooths~\cite{eisenberger2020smooth}  
                                                & 1.9   & 2.8   & 1.6      &2.3       &5.7\\
         \discreteopt~\cite{ren2021discrete}   
                                                & 8.9   & 13.6  & 7.5      &10.7      &1.1\\
         \spidermatch~\cite{roetzer2024spidermatch}
                                                & 1.7   & 1.8   & 1.7      &2.1       &0.81\\
         Ours                                   & \textbf{0.46}   & \textbf{0.53}   & \textbf{0.48}      &\textbf{0.62}       &\textbf{0.77}\\
         \bottomrule
        \end{tabular}
    }
    \caption{Comparison of \textbf{mean Dirichlet energies} ($\downarrow$) of various shape matching methods to measure smoothness of the matching.
    Our method consistently yields best results, which shows the effect of strong priors induced by global geometric consistency.
    }
    \label{tab:dir-energy}
    \vspace{-0.2cm}
\end{table}

\begin{figure}
    \def\pathOurs{figs/qualitative/ours/}
\def\pathCao{figs/qualitative/caoetal/}
\def\pathRen{figs/qualitative/discrete/}
\def\pathEisenb{figs/qualitative/smooths/}
\def\pathSpiderMatch{figs/qualitative/spidermatch/}
\def\srcEnd{_M.png}
\def\trgtEnd{_N.png}

\def\columnOne{Shuffling165-Shuffling092}
\def\columnTwo{GoalkeeperScoop055-GoalkeeperScoop058}
\def\columnTwoTwo{Running045-Standing2HMagicAttack01081}
\def\columnThree{Falling202-Standing2HMagicAttack01036}
\def\columnFour{Standing2HMagicAttack01058-GoalkeeperScoop058}
\def\columnFive{tr_reg_080-tr_reg_099}
\def\columnSix{tr_reg_085-tr_reg_090}
\def\columnSeven{tr_reg_088-tr_reg_097}
\def\columnEight{tr_reg_097-tr_reg_089}
\def\columnNine{cow_04-dog_08}
\def\columnTen{horse_08-cow_06}
\def\columnEleven{dog_01-dog_08}
\def\heightQ{1.95cm}
\def\widthQ{2cm}
\def\hspaceCols{-0.6cm}
{\scriptsize
\hspace{-0.3cm}
\renewcommand{\arraystretch}{0.1}
\begin{tabular}{ccccccccccccc}
    \setlength{\tabcolsep}{0pt}
    \rotatedCentering{90}{\heightQ}{Source}&
    \hspace{\hspaceCols}
    \includegraphics[height=\heightQ, width=\widthQ]{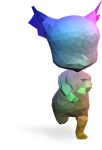}&
    \hspace{\hspaceCols}
    \includegraphics[height=\heightQ, width=\widthQ]{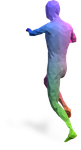}&
    \hspace{\hspaceCols}
    \includegraphics[height=\heightQ, width=\widthQ]{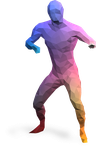}&
    \hspace{\hspaceCols}
    \includegraphics[height=\heightQ, width=\widthQ]{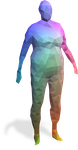}&
    \hspace{\hspaceCols}
    \includegraphics[height=\heightQ, width=\widthQ]{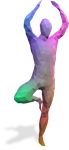}&
    \hspace{\hspaceCols}
    \includegraphics[height=\heightQ, width=\widthQ]{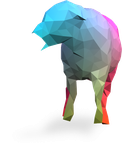}&
    \\
    \rotatedCentering{90}{\heightQ}{\caoetal~\cite{cao2023unsupervised}}&
    \hspace{\hspaceCols}
    \includegraphics[height=\heightQ, width=\widthQ]{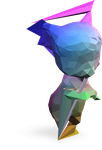}&
    \hspace{\hspaceCols}
    \begin{overpic}[height=\heightQ, width=\widthQ]{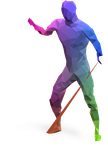}
        \put(8,48){\includegraphics[height=0.7cm]{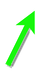}}
    \end{overpic}&
    \hspace{\hspaceCols}
    \includegraphics[height=\heightQ, width=\widthQ]{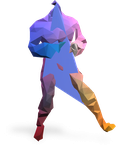}&
    \hspace{\hspaceCols}
    \begin{overpic}[height=\heightQ, width=\widthQ]{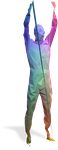}
        \put(37,58){\includegraphics[height=0.7cm]{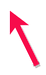}}
    \end{overpic}&
    \hspace{\hspaceCols}
    \begin{overpic}[height=\heightQ, width=\widthQ]{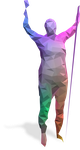}
    \end{overpic}&
    \hspace{\hspaceCols}
    \begin{overpic}[height=\heightQ, width=\widthQ]{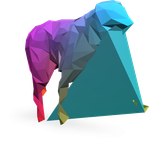}
        \put(73,-1){\includegraphics[height=0.5cm]{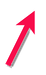}}
    \end{overpic}&
    \\
    \rotatedCentering{90}{\heightQ}{\spidermatch~\cite{roetzer2024spidermatch}}&
    \hspace{\hspaceCols}
    \includegraphics[height=\heightQ, width=\widthQ]{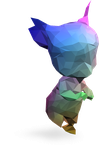}&
    \hspace{\hspaceCols}
    \begin{overpic}[height=\heightQ, width=\widthQ]{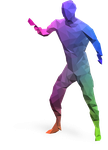}
        \put(8,48){\includegraphics[height=0.7cm]{figs/qualitative/red_arrow_right.png}}
    \end{overpic}&
    \hspace{\hspaceCols}
    \includegraphics[height=\heightQ, width=\widthQ]{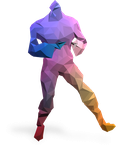}&
    \hspace{\hspaceCols}
    \begin{overpic}[height=\heightQ, width=\widthQ]{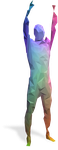}
        \put(37,58){\includegraphics[height=0.7cm]{figs/qualitative/red_arrow_left.png}}
    \end{overpic}&
    \hspace{\hspaceCols}
    \begin{overpic}[height=\heightQ, width=\widthQ]{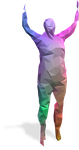}
    \end{overpic}&
    \hspace{\hspaceCols}
    \begin{overpic}[height=\heightQ, width=\widthQ]{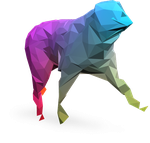}
        \put(73,-1){\includegraphics[height=0.5cm]{figs/qualitative/red_arrow_right.png}}
    \end{overpic}&
    \\
    \rotatedCentering{90}{\heightQ}{\ours}&
    \hspace{\hspaceCols}
    \includegraphics[height=\heightQ, width=\widthQ]{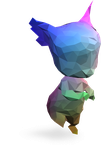}&
    \hspace{\hspaceCols}
    \begin{overpic}[height=\heightQ, width=\widthQ]{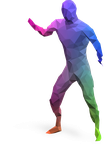}
        \put(8,48){\includegraphics[height=0.7cm]{figs/qualitative/green_arrow_right.png}}
    \end{overpic}&
    \hspace{\hspaceCols}
    \includegraphics[height=\heightQ, width=\widthQ]{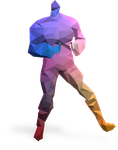}&
    \hspace{\hspaceCols}
    \begin{overpic}[height=\heightQ, width=\widthQ]{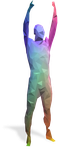}
        \put(37,58){\includegraphics[height=0.7cm]{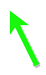}}
    \end{overpic}&
    \hspace{\hspaceCols}
    \includegraphics[height=\heightQ, width=\widthQ]{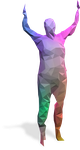}&
    \hspace{\hspaceCols}
    \begin{overpic}[height=\heightQ, width=\widthQ]{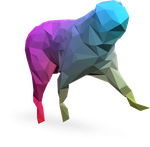}
        \put(73,-1){\includegraphics[height=0.5cm]{figs/qualitative/green_arrow_right.png}}
    \end{overpic}&
    \\
\end{tabular}
}
    \vspace{-0.1cm}
    \caption{\textbf{Qualitative shape matching results} computed with ULRSSM~\cite{cao2023unsupervised}, \spidermatch~\cite{roetzer2024spidermatch} and ours. We visualise matchings using colour and triangulation transfer from source to target shape. At legs and arms of shapes we can see stronger geometric consistency of ours compared to \spidermatch~\cite{roetzer2024spidermatch} (see red and green arrows and distorted triangles).
    }
    \vspace{-0.1cm}
    \label{fig:sm-qualitative}
\end{figure}

\textbf{Partial-to-Full Shape Matching.} As proof of concept we show qualitative results of matchings computed with our method in the partial-to-full setting, see \cref{fig:partial-sm}.

\newcommand{\maxpartialheight}{1.9cm}
\begin{figure}[b]
    \centering
    \setlength{\tabcolsep}{0pt}
    \renewcommand{\arraystretch}{1}
    \begin{tabular}{cccccccc}
        \hspace{-0.3cm}
        \includegraphics[width=0.2\columnwidth,height=\maxpartialheight]{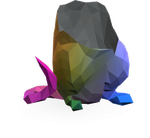}
         &
         \hspace{-0.6cm}
         \includegraphics[width=0.2\columnwidth,height=\maxpartialheight]{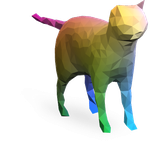}
         &
         \hspace{0.2cm}
         \rotatebox{90}{\textcolor{gray!50}{\rule{\maxpartialheight}{0.2pt}}}
         &
         \includegraphics[width=0.2\columnwidth,height=\maxpartialheight]{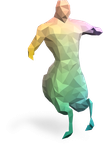}
         &
         \hspace{-0.5cm}
         \includegraphics[width=0.2\columnwidth,height=\maxpartialheight]{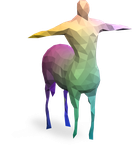}
         &
         \hspace{0.2cm}
         \rotatebox{90}{\textcolor{gray!50}{\rule{\maxpartialheight}{0.2pt}}}
         &
         \includegraphics[width=0.2\columnwidth,height=1.5cm]{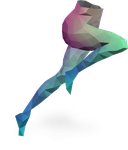}
         &
         \hspace{-0.4cm}
         \includegraphics[width=0.2\columnwidth,height=\maxpartialheight]{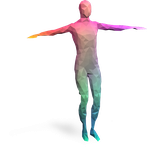}\\
    \end{tabular}
    \vspace{-0.3cm}
    \caption{Qualitative results computed with ours for \textbf{partial-to-full} shape matching of test set shapes~\cite{ehm2024partial} from SHREC'16~\cite{cosmo2016shrec}.}
    \label{fig:partial-sm}
\end{figure}

\subsection{Application to Planar Graph Matching}
As a proof of concept we show that our formalism directly applies to planar graph matching. 
To this end, we use keypoints on images of the WILLOW~\cite{cho2013learning} dataset and obtain graphs by computing Delaunay triangulations of keypoints.
Furthermore, we represent one of the resulting graphs using surface cycles to match it to the other graph, see \cref{fig:graph-sm} and in \cref{sec:supp:planar-graph-matching-results} in the supplementary.

\begin{figure}
    \centering
    \begin{tabular}{cc}
        \setlength{\tabcolsep}{0pt}
        \hspace{-0.39cm}
         \includegraphics[width=0.48\columnwidth,height=0.25\columnwidth,keepaspectratio=false]{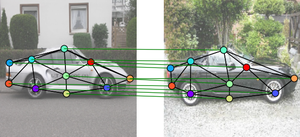}&
         \hspace{-0.2cm}
         \includegraphics[width=0.48\columnwidth,height=0.25\columnwidth,keepaspectratio=false]{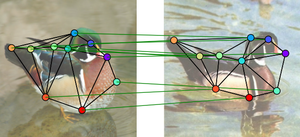}
    \end{tabular}
    \caption{Examples of \textbf{planar graph matching} results using our method on ducks and cars from WILLOW~\cite{cho2013learning} dataset.}
    \vspace{-0.2cm}
    \label{fig:graph-sm}
\end{figure}

\section{Discussion \& Limitations}\label{sec:discussion}

Our algorithm enforces geometric consistency and thus introduces a strong prior to resolve ambiguities in 3D shape matching problems.
Yet, in theory our formalism contains undesirable matchings in the solution space:
On the one hand, multiple surface cycles could be matched to the same vertex on the target shape. While this ensures our optimsation problem is always feasible (since it contains the extreme case of matching all surface cycles to a single vertex), this is an undesirable solution.
In the future, this could be resolved by generalising our approach towards a symmetric formalism.
On the other hand, our formalism allows for inside-out flips, as each individual matching element (i.e.~an edge) cannot disambiguate extrinsic orientations.

We have empirically shown that results can be computed efficiently using off-the-shelf LP-solvers~\cite{gurobi} and that
in all considered instances the linear the programming relaxations are tight, i.e.~we find a globally optimal solution for all instances.
Yet, we observe that our constraint matrix is not totally unimodular.
We leave an in-depth analysis of our formalism, along with answering the question whether there exists a polynomial time algorithm for geometrically consistent 3D shape matching, for future works.

\section{Conclusion}\label{sec:conclusion}

We have presented a novel formalism for non-rigid 3D shape matching which globally enforces geometric consistency. Our key idea is to consider a shape representation that allows to cast 3D shape matching as an integer linear program, which is efficiently solvable in practice.
In addition, we illustrate that the resulting problem can be interpreted as a minimum-cost flow circulation problem in a hyper graph.
Overall, we consider our work to be important for the 3D shape analysis community.
Geometric consistency is crucial in most practical 3D shape matching settings, {which is often times ignored (especially by learning-based methods).}
Our work allows to obtain such matchings efficiently {and with that poses a step towards integrating geometric consistency into learning-based methods}.
Furthermore, we hope to inspire follow-up works on neighbourhood preserving matching formalisms within the broader field of visual computing and beyond.

\subsection*{Acknowledgements}
We thank Johan Thunberg, Tobias Wei{\ss}berg and Zorah L\"ahner for insightful discussions about our formalism.
This work is supported by the ERC starting grant no.~101160648 (Harmony).

{\small
\bibliographystyle{ieee_fullname}
\bibliography{main}
}
\clearpage

\clearpage
\setcounter{page}{1}

{
\newpage
   \twocolumn[
    \centering
    \Large
    \textbf{Fast Globally Optimal and Geometrically Consistent 3D Shape Matching}\\
    \vspace{0.5em}Supplementary Material \\
    \vspace{1.0em}
   ] %
}\renewcommand{\thesubsection}{\Alph{subsection}}
\setcounter{figure}{0}
\renewcommand\thefigure{A.\arabic{figure}}

\subsection{More Details on Constraint Matrix $H$}\label{sec:supp:algebraic}
In the following, we provide more details of submatrices $P$ and $L$ of our constraint matrix $H = \begin{bsmallmatrix}P\\L\\S \end{bsmallmatrix}$.

\begin{figure*}
    \centering
    \includegraphics[width=0.9\linewidth, trim={1.8cm 0 1cm 0}]{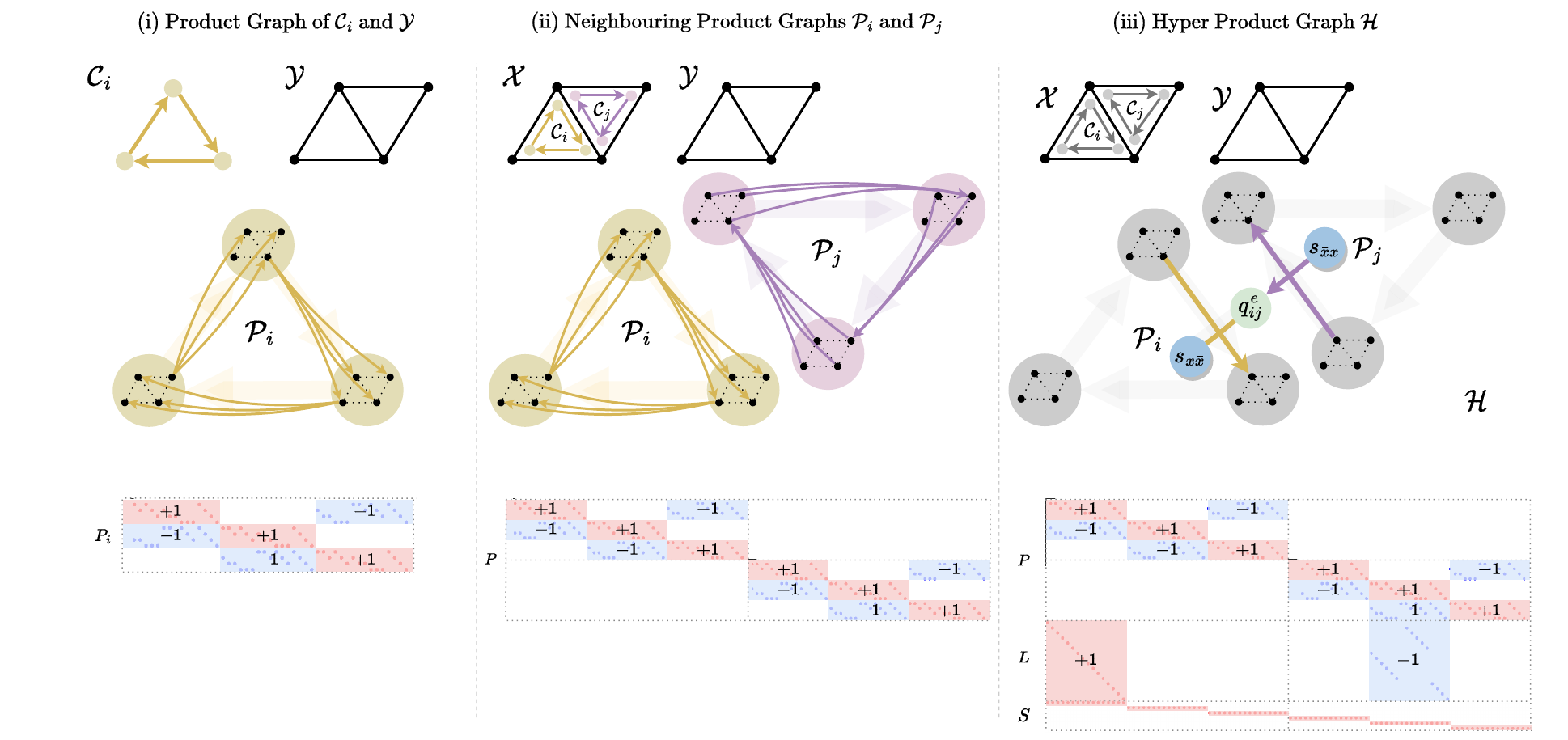}
    \captionof{figure}{Visualisation of the involved \textbf{graph structures} (top) and respective \textbf{vertex-edge incidence matrices} (bottom): (i) for an individual subproblem product graph $\prodGraph_i$, (ii) for two (uncoupled) subproblem product graphs $\prodGraph_i,\prodGraph_j$, and (iii) for two coupled subproblems leading to the hyper product graph $\hyperProdGraph$. 
    We note that in (iii) we illustrate two \emph{directed hyper edges} (which have multiple source and target vertices, see thick coloured lines \textcolor{scYELLOW}{\rule[0.6mm]{0.25cm}{0.7mm}},\textcolor{scPURPLE}{\rule[0.6mm]{0.25cm}{0.7mm}}) of our hyper product graph $\hyperProdGraph$.
    These hyper edges of $\hyperProdGraph$ contain source and target vertices of product edges (see middle) in their respective sets of source and target vertices.
    Furthermore, each hyper edge additionally contains a vertex $q_{ij}^{e}\in\verts_\linking$
    in its set of source and target vertices
    (so that opposite product edges are coupled, see green dot \cir[scLIGHTGREEN,fill=scLIGHTGREEN]{4pt}).
     Finally, each hyper edge contains an injectivity vertex $s_{x\bar{x}}$, $s_{\bar{x}x} \in \verts_\injectivity$ in its set of source vertices (so that each edge of the surface cycles is matched exactly once, see blue dots 
     \cir[scLIGHTBLUE,fill=scLIGHTBLUE]{4pt}).
    }
    \label{fig:matrix-structure-of-graphs}
\end{figure*}

\textbf{The matrix $P$.}
As mentioned in the main paper, $\prodGraph_i$ can be represented algebraically via matrix $P_i$. 
Yet, to account for degenerate edges (i.e.~the extended edge set $\edges_\meshY^+$), we have to consider a slightly altered definition of $P_i$ which reads 
\begin{equation}
    P_i \coloneqq C_i^+ \otimes \Tilde{Y}^+ - C_i^- \otimes \Tilde{Y}^-.
\end{equation}
Here, $C_i^+$ and $C_i^-$ are the non-negative and non-positive entries of the vertex edge incidence matrix $C_i \in \{-1,0,1\}^{|\verts_{\contour_i}|\times|\edges_{\contour_i}|}$.
In contrast to the elaborations in the main paper, which, as mentioned, neglects degenerate edges, $\Tilde{Y}^+$ and $\Tilde{Y}^-$ are the non-negative and non-positive entries of the vertex edge incidence matrix $Y \in \{-1,0,1\}^{|\verts_{\meshY}|\times|\edges_{\meshY}|}$ concatenated column-wise with $\boldsymbol{I}_{|\verts_\meshY|}$ and $-\boldsymbol{I}_{|\verts_\meshY|}$, respectively. 
Thus $\Tilde{Y}^+$ and $\Tilde{Y}^-$ are defined as
\begin{equation}
    \begin{aligned}
        \Tilde{Y}^+ &\coloneqq \begin{bmatrix}Y^+ &\phantom{-}\boldsymbol{I}_{|\verts_\meshY|}\end{bmatrix},\\
        \Tilde{Y}^- &\coloneqq \begin{bmatrix}Y^- &-\boldsymbol{I}_{|\verts_\meshY|}\end{bmatrix}.
    \end{aligned}
\end{equation}
With the interpretation that non-zero entries in matrices $\Tilde{Y}^+$ and $\Tilde{Y}^-$ resemble incoming and outgoing edges at vertices of $\meshY$, one can see that appending identities to respective matrices can be interpreted as adding self-edges, i.e.~accounting for $\edges_\meshY^+$, see also \cref{fig:matrix-structure-of-graphs}~(i).

\textbf{The matrix $L$.}
As mentioned in the main paper, the definition of $P$ requires
columns of the non-positive blocks of $L$ to be permuted.
We can incorporate this permutation by considering the following alternative definition of $L$ which reads
\begin{equation}
    L\coloneqq K^+\otimes \boldsymbol{I}_{|\edges_\meshY^+|} - K^-\otimes \Tilde{\boldsymbol{I}}_{|\edges_\meshY^+|}.
\end{equation}
Here $K^+$ and $K^-$ are the non-negative and non-positive entries of the matrix $K\in \{-1,0,1\}^{p\times
|\edges_\meshX|}$ ($p$ is the number of undirected non-boundary edges of $\meshX$ and $K$ represents the incidence of opposite edges across all $n$ surface cycles). Furthermore, $\Tilde{\boldsymbol{I}}_{|\edges_\meshY^+|}$ is a (column) permuted identity matrix with all non-positive entries.
The column permutation of $\Tilde{\boldsymbol{I}}_{|\edges_\meshY^+|}$ is such that opposite edges in $\edges_\meshY^+$ can be mapped to each other via $\Tilde{\boldsymbol{I}}_{|\edges_\meshY^+|}$, i.e.~such that $\Tilde{Y}^+=\Tilde{Y}^-\Tilde{\boldsymbol{I}}_{|\edges_\meshY^+|}$ and such that $\Tilde{Y}^-=\Tilde{Y}^+\Tilde{\boldsymbol{I}}_{|\edges_\meshY^+|}$.

\subsection{Detailed Hyper Product Graph Formalism}\label{sec:detailed-prod-graph}
In this section we provide detailed definitions of the hyper product graph $\hyperProdGraph$ which arises from our constraint matrix $H$.
We provide an overview of the major concepts in \cref{fig:matrix-structure-of-graphs}, where we also visualise the matrix structures of respective submatrices $P$, $L$, and $S$ of $H$.

\subsubsection{Individual Surface Cycle Matching Subproblems}\label{sec:supp:surf-cycle-matching}
Each surface cycle $\contour_i$ can be matched to shape $\meshY$ by finding a cyclic path in the \emph{product graph}, which has been introduced in \cite{lahner2016efficient} for matching a 2D to a 3D shape{, see \cref{def:prodgraph}}.

In a nutshell, each vertex $v=\twovector{x}{y}\in\verts_{\prodGraph_i}$ in the product graph resembles a (potential) matching between vertices $x\in\verts_\meshX$ and $y\in\verts_\meshY$ of both shapes.
Thus, each edge $(v,\bar{v})$ in the product graph can be interpreted as a (potential) matching between edges of $\contour_i$ and edges  of shape $\meshY$ (or vertices of $\meshY$ via the extended edge set $\edges^+_\meshY$ to account for stretching and compression). 
With that, matching surface cycle $\contour_i$ to shape $\meshY$ amounts to solving a cyclic shortest path problem in the respective product graph $\prodGraph_i$~\cite{lahner2016efficient}.

However, a major downside for our setting of \emph{3D-to-3D shape matching}
is that the neighbourhood between pairs of surface cycles cannot be ensured when considering vanilla shortest path algorithms (that solve the $n$ individual surface cycle matching subproblems independently).
To tackle this, we couple the individual product graphs $\prodGraph_1,\dots,\prodGraph_n$ appropriately, which we explain next.

\subsubsection{Subproblem Coupling}\label{sec:supp:coupling}
We couple the individual surface cycle matching subproblems of $\prodGraph_i$ by glueing them together via opposite edges.
To this end, we introduce \emph{coupling vertices}:

\begin{definition}[Coupling vertices] For every pair of opposite product edges $e\in\edges_{\prodGraph_i}$ and $-e\in\edges_{\prodGraph_j}$
we add a coupling vertex $q^e_{ij}$. The set of coupling vertices is $\verts_\linking \coloneqq \{q^e_{ij} \;|\; e\in\edges_{\prodGraph_i},-e\in\edges_{\prodGraph_j}\}$.
\end{definition}
The purpose of coupling vertices is that matchings of opposite edges are consistent, i.e.~so that matchings of surface cycles cover opposite edges and thus are neighbouring, see \cref{fig:matrix-structure-of-graphs} (iii). 
Consequently, this de-facto enforces the \emph{glueing}, which in turn results in global geometric consistency.

In addition to the coupling, we want to ensure matching injectivity for each edge of surface cycles.
Hence, we need to ensure  that each such edge is matched exactly once, which we tackle next.

\subsubsection{Surface Cycle Matching Injectivity}\label{sec:supp:injectivty}
We want to enforce that each surface cycle edge is matched exactly once, i.e.~it is matched to exactly one edge (or vertex) of $\meshY$.
In other words, among all the potential matching candidates  of a single surface cycle edge, exactly one is part of the final matching.
For that, let us denote the set of all edges (in the product graph) that are potential matching candidates of the edge $(x,\bar{x}) \in \edges_{\contour_i}$ as the \emph{edge bundle of $(x,\bar{x})$}.
For each such edge bundle, we introduce one
injectivity vertex (see \cref{fig:matrix-structure-of-graphs}
(iii) as well as \cref{fig:hyper-product-graph}), which has the purpose 
to ensure that only a single edge of each edge bundle is part of the final matching:

\begin{definition}[Injectivity vertices] 
    For every directed edge $(x,\bar{x})\in \edges_\meshX$ of shape $\meshX$ we introduce one injectivity vertex. The set of injectivity vertices is %
    $\verts_\injectivity \coloneqq\{s_{x\bar{x}} \;|\; (x,\bar{x})\in \edges_\meshX\}$.
\end{definition}

\subsubsection{Resulting Hyper Product Graph}\label{sec:supp:hyper-produ-graph}
Finally, we present our directed hyper product graph $\hyperProdGraph$.
We note that in contrast to ordinary edges (which have exactly one source and one target vertex), a \emph{directed hyper edge} denoted as $\hyperedge{v_1, v_2,\dots}{v_3, v_4,\dots}$ has a \emph{set} of source vertices $\{v_1, v_2,\dots\}$ and a \emph{set} of target vertices $\{v_3, v_4,\dots\}$~\cite{gallo1993directed}.

Our hyper graph has two different types of hyper edges, \emph{coupled} and \emph{uncoupled} ones.
The \emph{coupled hyper edges} refer to hyper edges that belong to \emph{non-boundary} edges of $\meshX$ and serve the purpose of enforcing neighbourhood preservation between surface cycles:
\begin{definition}[Coupled hyper edges]
   The set of \emph{coupled hyper edges} is defined as 
    \begin{equation}
    \begin{aligned}
        \Tilde{\edges}_\hyperProdGraph &=\Big\{
        \hyperedge{v,s_{x\bar{x}}}{\bar{v},q_{ij}^e},
        \hyperedge{\bar{v},s_{\bar{x}x},q_{ij}^e}{v}
         \Big|\; 
        s_{x\bar{x}}, s_{\bar{x}x} \in \verts_\injectivity,\\
        & \quad q_{ij}^e\in\verts_\linking,\; e\coloneqq(v, \bar{v})=\left(\!\twovector{x}{y}, \twovector{\bar{x}}{ \bar{y}}\!\right)\in\edges_{\prodGraph_i},
        -e\in\edges_{\prodGraph_j}\Big\}.
    \end{aligned}
\end{equation}
\end{definition}
From above definition we can see that one coupling vertex always connects exactly two hyper edges.
The \emph{uncoupled hyper edges} refer to hyper edges that belong to boundary edges of $\meshX$ and with which we can account for partiality of the source shape $\meshX$:
\begin{definition}[Uncoupled hyper edges]
    The set of \emph{uncoupled hyper edges} is defined as
    \begin{equation}
        \begin{aligned}
            \hat{\edges}_\hyperProdGraph &= 
            \Big\{ 
            \hyperedge{v,s_{x\bar{x}}}{\bar{v}}\;|\;
            v = \twovector{x}{y},\; \bar{v} =  \twovector{\bar{x}}{ \bar{y}},\;s_{x\bar{x}}\in\verts_\injectivity,\\
            &\quad(x,\bar{x}) \in \edges_\meshX\text{ is boundary edge},\;
            (y,\bar{y}) \in \edges_\meshY^+
            \Big\}.
        \end{aligned}
    \end{equation}
\end{definition}

This gives rise to our hyper product graph, which we also visualise in \cref{fig:matrix-structure-of-graphs} (iii):

\begin{definition}[Hyper product graph] 
    Our hyper product graph $\hyperProdGraph=\left(\verts_\hyperProdGraph,\edges_\hyperProdGraph\right)$ for matching the
    source shape $\meshX$ (represented with $n$ \surfacecycles) to the target shape $\meshY$
    comprises the vertex set $\verts_\hyperProdGraph$ and the set of (directed) hyper edges $\edges_\hyperProdGraph$, and is defined as
    \begin{equation}\label{eq:hyper-prod-graph}\tag{HPG}
    \begin{aligned}
        \verts_\hyperProdGraph &= \verts_{\prodGraph_1}\;\cup\; \dots\;\cup\; \verts_{\prodGraph_{n}} \;\cup\;\verts_\injectivity  \;\cup\;\verts_\linking ,\\
        \edges_\hyperProdGraph &= \Tilde{\edges}_\hyperProdGraph \cup \hat{\edges}_\hyperProdGraph \quad \text{with} \quad m := |\edges_\hyperProdGraph|.
    \end{aligned}
    \end{equation}
\end{definition}

\subsection{Practical Considerations}
In this section, we discuss the implementation of the distortion bound as well as the problem size reduction of our linear program \eqref{eq:cycle-surfing}.
Furthermore, we discuss different choices of surface cycles.

\subsubsection{Distortion Bound}\label{sec:supp:distortion-bound}

\begin{figure}
    \centering
    \includegraphics[width=0.85\columnwidth]{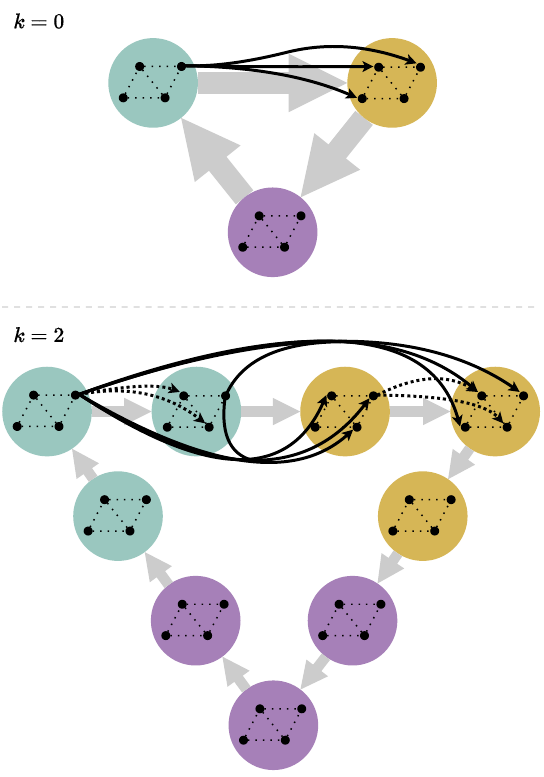}
    \caption{Visualisation of a product graph with distortion bound $k=0$ (\textbf{top}) and with distortion bound $k=2$ (\textbf{bottom}).
    The {distortion bound} $k\in2\mathbb{Z}^+_0$ is achieved by duplicating product vertices $k$ times (duplicates are visible when comparing top and bottom graphs).
    We connect duplicated product vertices such that: (i) all duplicates of the same group of product vertices (e.g.~all vertices within greenish circles \cir[scGREEN,fill=scGREEN]{4pt}) are connected (see dashed black arrows resembling potential matchings of edges of $\meshY$ to vertices of $\meshX$) and (ii) $k-1$ duplicates as well as the original vertices of the same group are connected to $k-1$ duplicates as well as the original vertex of the subsequent group (see solid black arrows at bottom figure which connect vertices in greenish circles \cir[scGREEN,fill=scGREEN]{4pt} with vertices in yellow circles \cir[scYELLOW,fill=scYELLOW]{4pt} and which resemble potential matchings of edges of $\meshX$ and edges or vertices of $\meshY$).
    We note that we only allow for distortion-bounds of integer multiples of $2$ (so that neighbouring product graphs can still be coupled). 
    }
    \label{fig:distortion-bound}
\end{figure}

Our hyper product graph $\hyperProdGraph$ already allows to map edges of $\meshX$ to vertices and edges of $\meshY$.
Thus, our formalism already accounts for shrinking and stretching.
Yet, for more flexibility, we want to additionally allow for matchings of edges of $\meshY$ to vertices of $\meshX$.
To this end, we integrate a distortion bound by
creating $k$ duplicates of vertices of every product graph $\prodGraph_i$.
We connect the duplicates such that resulting product edges resemble edge to edge, edge to vertex or vertex to edge matchings between shapes $\meshX$ and $\meshY$, see also the product graph definition in \cite{lahner2016efficient} and \cref{fig:distortion-bound} for a visualisation of additional product edges.
This effectively allows for at most $k$ consecutive edges of $\meshY$ to be matched to a vertex of shape $\meshX$ (in addition to the already allowed matchings between edges of $\meshX$ and edges or vertices of $\meshY$).
In all experiments we set $k=2$.
We note that similar concepts have been used for image segmentation, see~\cite[Section 7.1.2]{schoenemann2009combinatorial}.

\subsubsection{Reduced Problem Size}\label{sec:supp:reduced-problem-size}
\begin{figure*} 
    \centering
    \begin{tabular}{c}
         \includegraphics[width=0.35\linewidth,trim={1.3cm 0.5cm 0 0}]{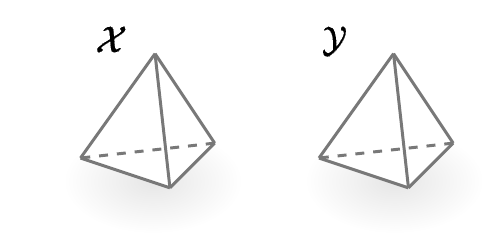}  \\
         \includegraphics[width=0.9\linewidth,trim={1cm 0 0 -0.5cm}]{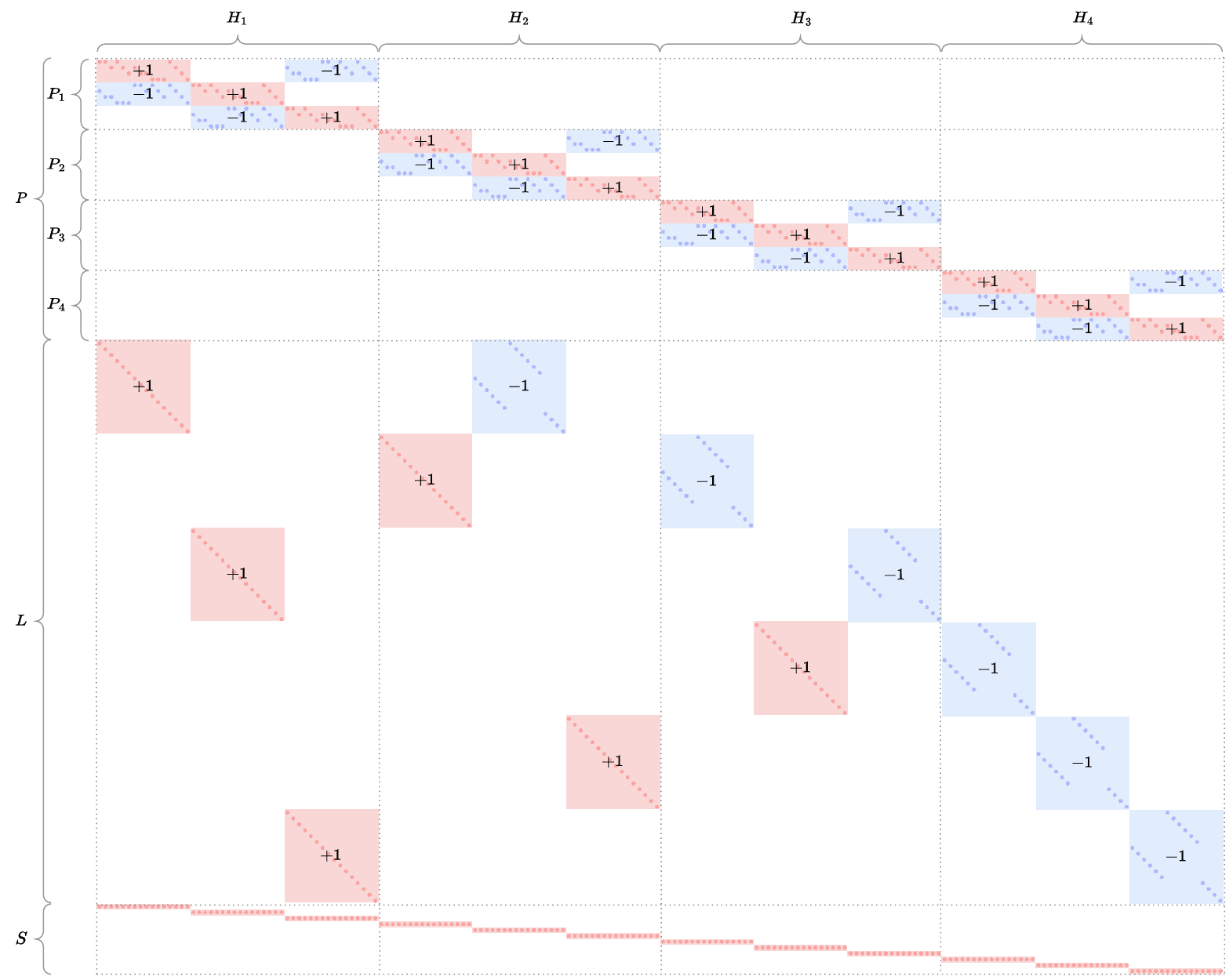} 
    \end{tabular}
    \caption{For the matching of two tetrahedron shapes (\textbf{top}) we show the resulting $H$ matrix (\textbf{bottom}). In light red and light blue we indicate the blocks of $H$, where within each block there is exactly one non-zero element per column.
    }
    \label{fig:matrix-structure-overlay}
\end{figure*}

\begin{figure*}
\centering
    \begin{tabular}{c}
         \includegraphics[width=0.35\linewidth]{figs/tet-match.drawio.pdf}  \\
         \includegraphics[width=0.8\linewidth]{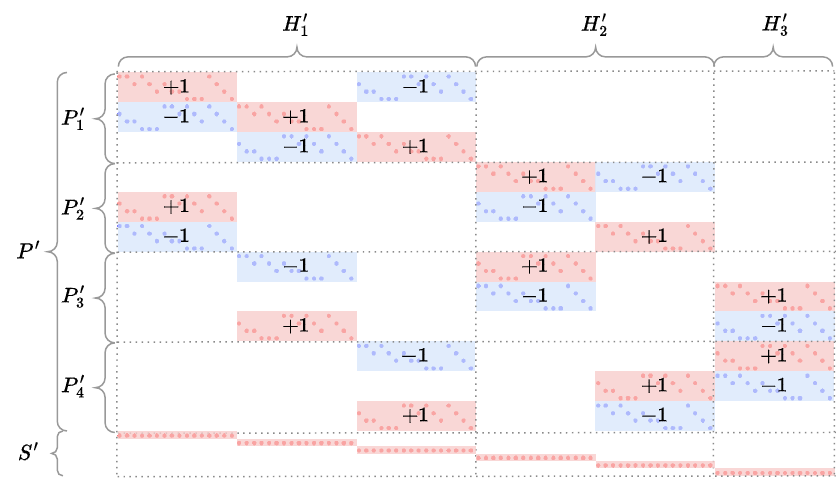}
    \end{tabular}
    \caption{We show the \textbf{reduced matrix structure} after resolving coupling constraints for the same matching problem of matching two tetrahedron as shown in \cref{fig:matrix-structure-overlay}.}
    \label{fig:reduced-matrix-strucutre}
\end{figure*}

For improved solver runtimes we reduce the problem size of \eqref{eq:cycle-surfing} by approximately $50\%$.
To this end, we resolve coupling constraints during construction of matrix $H$ (these effectively ensure that two variables hold the same value during optimisation and thus we only need to consider approximately $50\%$ of the variables). 
In \cref{fig:matrix-structure-overlay} we show an example of full-sized matrix $H$ for the problem of matching two tetrahedron and furthermore, in \cref{fig:reduced-matrix-strucutre}, we visualise the resulting smaller matrix for the same problem of matching two tetrahedron.
We emphasise that problem size reduction, as described above, does \emph{not} prune the problem but rather describes \eqref{eq:cycle-surfing} with a smaller but equivalent optimisation problem.
In other words, a solution to the reduced problem yields a uniquely defined solution to the larger problem.

\subsubsection{Other Choices of Surface Cycles}\label{sec:supp:other-surface-cycles}
As mentioned in the main paper, we assume that each surface cycle represents a single triangle of $\meshX$.
We note that a single surface cycle can represent the boundary of polygonal patches of shape $\meshX$ that contain multiple triangles.
Yet, in this case, the union of vertices of all surface cycles might not contain all vertices of shape $\meshX$, so that one would have to adjust the notion of geometric consistency in \cref{def:geo-cons} accordingly.

\subsection{Additional Experiments}\label{sec:supp:more-exp}
In this section, we provide additional results for shape matching and planar graph matching.

\subsubsection{Shape Matching}\label{sec:supp:more-sm}

\begin{figure}[t]
    \centering
    \newcommand{\pckLineWidth}{2.0pt}
\newcommand{\plotWidth}{1.5\columnwidth}
\newcommand{\plotHeight}{0.8\columnwidth}

\pgfplotsset{%
    label style = {font=\large},%
    tick label style = {font=\large},
    title style =  {font=\Large},
    legend style={  fill= gray!10,
                    fill opacity=0.6, 
                    font=\large,
                    draw=gray!20, %
                    text opacity=1}
}
\begin{tikzpicture}[scale=0.5, transform shape]
	\begin{axis}[
		width=\plotWidth,
		height=\plotHeight,
		grid=major,
		title=Discretisation: FAUST,
		legend style={
			at={(0.97,0.03)},
			anchor=south east,
			legend columns=1},
		legend cell align={left},
	ylabel={\large Mean Geo.~Err.},
        xlabel={\large \# of Faces of \textbf{one} Triangle Mesh (other is fixed to 1000)},
        xmin=0,
        xmax=1000,
        ylabel near ticks,
        xtick={200, 400, 600, 800, 1000},
        ymin=0,
        ymax=0.4,
        ytick={0, 0.10, 0.20, 0.30, 0.40, 1},
	]
    \addplot [color=ourcolor, smooth, tension=0.2, line width=\pckLineWidth, forget plot]
    table[row sep=crcr]{%
50   0.382\\
100  0.0242\\
150  0.0271\\
200  0.0369\\
250  0.0343\\
300  0.0259\\
350  0.0248\\
400  0.0251\\
450  0.0248\\
500  0.0247\\
550  0.0245\\
600  0.0245\\
650  0.0245\\
700  0.0241\\
750  0.0239\\
800  0.0240\\
850  0.0237\\
900  0.0236\\
950  0.0235\\
1000 0.0235\\
    };

	\end{axis}
\end{tikzpicture}
    \caption{{Evaluation of the performance of our method under varying \textbf{discretisation}. We fix the resolution of one shape to 1000 triangles and vary the resolution of the other shape.}}
    \label{fig:discretisation}
\end{figure}
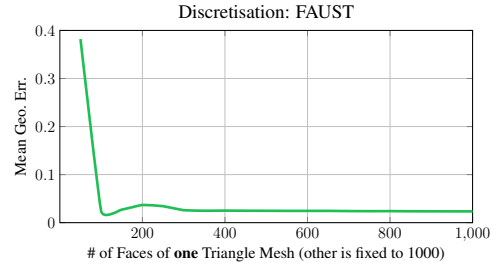

\textbf{Ablation Studies.} In \cref{tab:supp:ablation}, we conduct ablation studies of our method. 
To this end, we use $10$ pairs from FAUST dataset, decimate shapes to $500$ triangles and evaluate mean geodesic errors in different settings.
We consider our formalism~\eqref{eq:cycle-surfing} with and without coupling constraints (i.e.~without coupling constraints means that we only consider constraints $Px = 0$ and $Sx = 1$ and drop the constraints $Lx=0$).
Furthermore, we use different methods to compute features: wave kernel signatures (WKS)~\cite{aubry2011wave}, features based upon image foundation models which we call Diff3F~\cite{dutt2024} (we note that we use an empty text prompt for feature extraction with Diff3F) and features computed with deep feature extractor ULRSSM~\cite{cao2023unsupervised}.
For all types of features, our coupling constraints help to improve results. This shows the importance of priors induced by global geometric consistency.
Overall, we obtain best results with features computed with ULRSSM~\cite{cao2023unsupervised} (which is the feature extractor that we use for our method for the rest of our shape matching experiments).
{Additionally, we evaluate our methods performance to shapes with different discretisation.
We fix one shape to $1000$ triangles, vary the resolution of the other shape and plot mean geodesic errors in \cref{fig:discretisation} of five instances from FAUST. 
We can see that our method performs reasonably under varying shape discretisation.}

\begin{table}[t]
    \centering
    {\footnotesize
        \setlength{\tabcolsep}{3pt}
        \renewcommand{\arraystretch}{0.9}
        \begin{tabular}{@{}l|c@{}}
         \toprule
         \textbf{Method} &  \textbf{Mean Geodesic Error} \\
         \midrule
         Ours (no coupling) $+$ WKS~\cite{aubry2011wave}          & {0.2335}  \\
         Ours $+$ WKS~\cite{aubry2011wave}                        & {0.2253}  \\
         Ours (no coupling) $+$ Diff3F~\cite{dutt2024}            & {0.1105}  \\
         Ours $+$  Diff3F~\cite{dutt2024}                         & {0.0511}  \\
         Ours (no coupling) $+$ ULRSSM~\cite{cao2023unsupervised} & {0.0318}  \\
         Ours $+$ ULRSSM~\cite{cao2023unsupervised}               & \textbf{0.0316}  \\
         \bottomrule
        \end{tabular}
    }
    \caption{\textbf{Ablation studies} conducted on ten pairs of FAUST dataset. We consider different methods to compute features and furthermore consider our method with and without coupling constraints. Using coupling constraints improves results for all types of features. Using features computed with ULRSSM~\cite{cao2023unsupervised} yields best results overall.}
    \label{tab:supp:ablation}
\end{table}

{
\textbf{Memory Footprint of Constraints.} In \cref{tab:memory-foot}, we compare the memory footprint of our method to other geometrically consistent approaches.
Our has the lowest memory requirements.
}
\begin{table}
    \centering
    {\footnotesize
    \begin{tabular}{c|ccc}
    \toprule
     \textbf{\# Faces}&  \textbf{Windheuser~et~al.} & \textbf{SpiderMatch} & \textbf{Ours}\\
     \midrule
     500  & 0.2 GB &  0.2 GB & \textbf{0.1} GB\\
     1000 & 0.8 GB &  1.0 GB & \textbf{0.3} GB\\
     1500 & 1.8 GB &  2.2 GB & \textbf{0.7} GB\\
     2000 & 3.1 GB &  4.0 GB & \textbf{1.2} GB\\
     \bottomrule
    \end{tabular}
    }
    \caption{{Comparison of the \textbf{memory footprint} of the constraint matrices of Windheuser~et~al.~\cite{windheuser2011geometrically}, SpiderMatch~\cite{roetzer2024spidermatch} and our approach.}}
    \label{tab:memory-foot}
\end{table}

\begin{figure}[t]
    \centering
    \newcommand{\runtimeLineWidth}{2pt}
\newcommand{\rtCplotWidth}{1.3\columnwidth}
\newcommand{\rtCplotHeight}{0.7\columnwidth}
\pgfplotsset{%
	label style = {font=\large},
	tick label style = {font=\large},
	title style =  {font=\Large},
	legend style={  fill= gray!10,
		fill opacity=0.6, 
		font=\large,
		draw=gray!20, %
		text opacity=1}
}
\begin{tikzpicture}[scale=0.5, transform shape]
	\begin{axis}[%
		width=\rtCplotWidth,
		height=\rtCplotHeight,
		title=Runtime Comparison: \faust,
		title style={yshift=0cm},
		scale only axis,
		grid=major,
	      legend style={
		 	anchor=north east,
                at={(0.98,0.98)},
		 	legend columns=1,
		 	legend cell align={left}},
		ylabel={{\large Runtime [s]  \textcolor{gray!40}{$\leftarrow$}}},
		xlabel={$\#$ of Faces of individual Triangle Meshes},
		xmin=50,
		xmax=1750,
		xtick={250, 500, 750, 1000, 1250, 1500, 1750},
		xtick scale label code/.code={},
		ymin=0,
		ymax=3600,
            ytick={0, 1000, 2000, 3000},
		ylabel near ticks,
		]
\addplot [color=cRED, smooth, tension=0.2, line width=\runtimeLineWidth]
  table[row sep=crcr]{%
50	0.378964185714722\\
150	4.72082161903381\\
250	18.0825681686401\\
350	44.8990142345428\\
450	106.598613500595\\
550	218.056921243668\\
650	429.495904922485\\
750	848.802618980408\\
850	1682.68661427498\\
950	2832.45302224159\\
1050	4799.50301659107\\
};
\addlegendentry{\ours~(Mosek)}

\addplot [color=ourcolor, smooth, tension=0.2, line width=\runtimeLineWidth]
  table[row sep=crcr]{%
50	0.382844209671021\\
150	3.03250408172607\\
250	9.37670803070068\\
350	12.2255246639252\\
450	21.535619020462\\
550	34.291276216507\\
650	48.6560304164886\\
750	66.5410633087158\\
850	88.2784218788147\\
950	111.767062902451\\
1050	139.437685012817\\
1150	169.960187673569\\
1250	204.876774072647\\
1350	253.960168838501\\
1450	303.68256020546\\
1550	351.137157678604\\
1650	423.827531814575\\
1750	498.286844372749\\
1850	586.620145082474\\
};
\addlegendentry{\ours~(Gurobi)}
\end{axis}
\end{tikzpicture}%
    \caption{\textbf{Runtime comparison} of off-the-shelf linear programming solvers Gurobi~\cite{gurobi} and Mosek~\cite{mosek} when solving linear program~\eqref{eq:cycle-surfing} with varying resolution of input shapes. Curves are median runtimes over five pairs from FAUST dataset.}
    \label{fig:supp:runtime}
\end{figure}
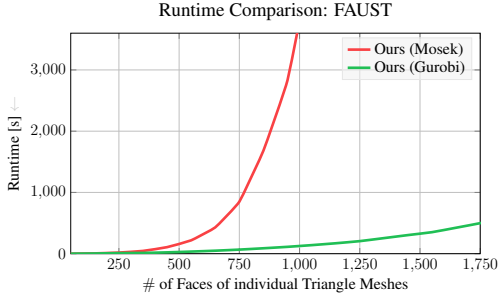

\textbf{Runtime.}
In \cref{fig:supp:runtime}, we compare runtimes of off-the-shelf linear programming solvers Gurobi~\cite{gurobi} and Mosek~\cite{mosek}.
Plotted curves are median runtimes of five pairs of FAUST dataset.
We can see that Mosek~\cite{mosek} scales poor compared to Gurobi~\cite{gurobi}.

\begin{figure*}
    \def\pathOurs{figs/qualitative/ours/}
\def\pathCao{figs/qualitative/caoetal/}
\def\pathRen{figs/qualitative/discrete/}
\def\pathEisenb{figs/qualitative/smooths/}
\def\pathSpiderMatch{figs/qualitative/spidermatch/}
\def\srcEnd{_M.png}
\def\trgtEnd{_N.png}

\def\columnOne{Shuffling165-Shuffling092}
\def\columnTwo{GoalkeeperScoop055-GoalkeeperScoop058}
\def\columnTwoTwo{Running045-Standing2HMagicAttack01081}
\def\columnThree{Falling202-Standing2HMagicAttack01036}
\def\columnFour{Standing2HMagicAttack01058-GoalkeeperScoop058}
\def\columnFive{tr_reg_080-tr_reg_099}
\def\columnSix{tr_reg_085-tr_reg_090}
\def\columnSeven{tr_reg_088-tr_reg_097}
\def\columnEight{tr_reg_097-tr_reg_089}
\def\columnNine{cow_04-dog_08}
\def\columnTen{horse_08-cow_06}
\def\columnEleven{dog_01-dog_08}
\def\heightQ{2.05cm}
\def\widthQ{2cm}
\def\hspaceCols{-0.6cm}
{\scriptsize
\hspace{-0.2cm}
\begin{tabular}{ccccccccccccc}
    \setlength{\tabcolsep}{0pt}
    &\hspace{\hspaceCols} \scircled{1} &\hspace{\hspaceCols}  \scircled{2} &\hspace{\hspaceCols}  \scircled{3} &\hspace{\hspaceCols}  \scircled{4} &\hspace{\hspaceCols}  \scircled{5} &\hspace{\hspaceCols}  \scircled{6} &\hspace{\hspaceCols}  \scircled{7} &\hspace{\hspaceCols}  \scircled{8} &\hspace{\hspaceCols}  \scircled{9} &\hspace{\hspaceCols}  \scircled{10} &\hspace{\hspaceCols}\scircled{11}&\hspace{\hspaceCols}\scircled{12}\\
    \rotatedCentering{90}{\heightQ}{Source}&
    \hspace{\hspaceCols}
    \includegraphics[height=\heightQ, width=\widthQ]{\pathOurs\columnOne\srcEnd}&
    \hspace{\hspaceCols}
    \includegraphics[height=\heightQ, width=\widthQ]{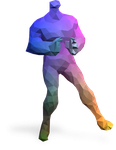}&
    \hspace{\hspaceCols}
    \includegraphics[height=\heightQ, width=\widthQ]{\pathOurs\columnTwoTwo\srcEnd}&
    \hspace{\hspaceCols}
    \includegraphics[height=\heightQ, width=\widthQ]{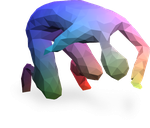}&
    \hspace{\hspaceCols}
    \includegraphics[height=\heightQ, width=\widthQ]{\pathOurs\columnFour\srcEnd}&
    \hspace{\hspaceCols}
    \includegraphics[height=\heightQ, width=\widthQ]{\pathOurs\columnFive\srcEnd}&
    \hspace{\hspaceCols}
    \includegraphics[height=\heightQ, width=\widthQ]{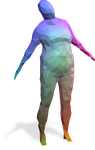}&
    \hspace{\hspaceCols}
    \includegraphics[height=\heightQ, width=\widthQ]{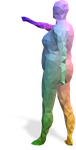}&
    \hspace{\hspaceCols}
    \includegraphics[height=\heightQ, width=\widthQ]{\pathOurs\columnEight\srcEnd}&
    \hspace{\hspaceCols}
    \includegraphics[height=\heightQ, width=\widthQ]{\pathOurs\columnNine\srcEnd}&
    \hspace{\hspaceCols}
    \includegraphics[height=\heightQ, width=\widthQ]{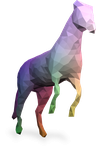}&
    \hspace{\hspaceCols}
    \includegraphics[height=\heightQ, width=\widthQ]{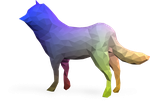}\\
    \rotatedCentering{90}{\heightQ}{\caoetal~\cite{cao2023unsupervised}}&
    \hspace{\hspaceCols}
    \begin{overpic}[height=\heightQ, width=\widthQ]{\pathCao\columnOne\trgtEnd}

    \end{overpic}&
    \hspace{\hspaceCols}
    \begin{overpic}[height=\heightQ, width=\widthQ]{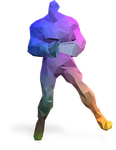}

    \end{overpic}&
    \hspace{\hspaceCols}
    \begin{overpic}[height=\heightQ, width=\widthQ]{\pathCao\columnTwoTwo\trgtEnd}
        \put(8,48){\includegraphics[height=0.7cm]{figs/qualitative/green_arrow_right.png}}
    \end{overpic}&
    \hspace{\hspaceCols}
    \begin{overpic}[height=\heightQ, width=\widthQ]{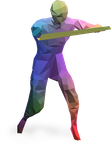}
        \put(47,46){\includegraphics[height=0.7cm]{figs/qualitative/red_arrow_left.png}}
    \end{overpic}&
    \hspace{\hspaceCols}
    \begin{overpic}[height=\heightQ, width=\widthQ]{\pathCao\columnFour\trgtEnd}

    \end{overpic}&
    \hspace{\hspaceCols}
    \begin{overpic}[height=\heightQ, width=\widthQ]{\pathCao\columnFive\trgtEnd}
        \put(37,58){\includegraphics[height=0.7cm]{figs/qualitative/red_arrow_left.png}}
    \end{overpic}&
    \hspace{\hspaceCols}
    \begin{overpic}[height=\heightQ, width=\widthQ]{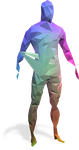}

    \end{overpic}&
    \hspace{\hspaceCols}
    \begin{overpic}[height=\heightQ, width=\widthQ]{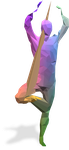}

    \end{overpic}&
    \hspace{\hspaceCols}
    \begin{overpic}[height=\heightQ, width=\widthQ]{\pathCao\columnEight\trgtEnd}
    \end{overpic}&
    \hspace{\hspaceCols}
    \begin{overpic}[height=\heightQ, width=\widthQ]{\pathCao\columnNine\trgtEnd}
        \put(72,-7){\includegraphics[height=0.7cm]{figs/qualitative/red_arrow_right.png}}
    \end{overpic}&
    \hspace{\hspaceCols}
    \begin{overpic}[height=\heightQ, width=\widthQ]{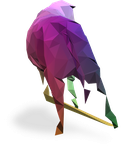}

    \end{overpic}&
    \hspace{\hspaceCols}
    \begin{overpic}[height=\heightQ, width=\widthQ]{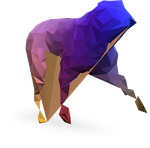}
         \put(72,-7){\includegraphics[height=0.7cm]{figs/qualitative/red_arrow_right.png}}
    \end{overpic}\\
    \rotatedCentering{90}{\heightQ}{\discreteopt~\cite{ren2021discrete}}&
    \hspace{\hspaceCols}
    \begin{overpic}[height=\heightQ, width=\widthQ]{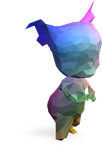}          
    
    \end{overpic}&
    \hspace{\hspaceCols}
    \begin{overpic}[height=\heightQ, width=\widthQ]{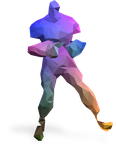}          
    
    \end{overpic}&
    \hspace{\hspaceCols}
    \begin{overpic}[height=\heightQ, width=\widthQ]{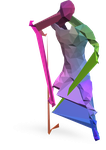}          
        \put(8,48){\includegraphics[height=0.7cm]{figs/qualitative/red_arrow_right.png}}
    \end{overpic}&
    \hspace{\hspaceCols}
    \begin{overpic}[height=\heightQ, width=\widthQ]{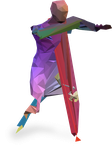}
        \put(47,46){\includegraphics[height=0.7cm]{figs/qualitative/red_arrow_left.png}}
    \end{overpic}&
    \hspace{\hspaceCols}
    \begin{overpic}[height=\heightQ, width=\widthQ]{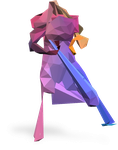}          
    
    \end{overpic}&
    \hspace{\hspaceCols}
    \begin{overpic}[height=\heightQ, width=\widthQ]{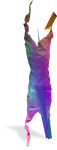}          
        \put(37,58){\includegraphics[height=0.7cm]{figs/qualitative/red_arrow_left.png}}
    \end{overpic}&
    \hspace{\hspaceCols}
    \begin{overpic}[height=\heightQ, width=\widthQ]{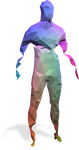}          
    
    \end{overpic}&
    \hspace{\hspaceCols}
    \begin{overpic}[height=\heightQ, width=\widthQ]{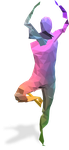}          
    
    \end{overpic}&
    \hspace{\hspaceCols}
    \begin{overpic}[height=\heightQ, width=\widthQ]{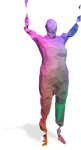}          
    
    \end{overpic}&
    \hspace{\hspaceCols}
    \begin{overpic}[height=\heightQ, width=\widthQ]{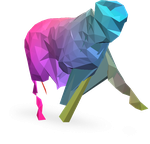}
        \put(72,-7){\includegraphics[height=0.7cm]{figs/qualitative/green_arrow_right.png}}
    \end{overpic}&
    \hspace{\hspaceCols}
    \begin{overpic}[height=\heightQ, width=\widthQ]{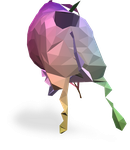}          
    
    \end{overpic}&
    \hspace{\hspaceCols}
    \begin{overpic}[height=\heightQ, width=\widthQ]{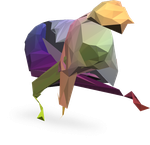}
         \put(72,-7){\includegraphics[height=0.7cm]{figs/qualitative/red_arrow_right.png}}
    \end{overpic}\\
    \rotatedCentering{90}{\heightQ}{\smooths~\cite{eisenberger2020smooth}}&
    \hspace{\hspaceCols}
    \begin{overpic}[height=\heightQ, width=\widthQ]{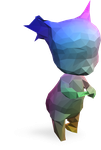}          
    
    \end{overpic}&
    \hspace{\hspaceCols}
    \begin{overpic}[height=\heightQ, width=\widthQ]{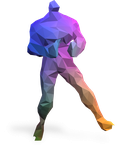}
    \end{overpic}&
    \hspace{\hspaceCols}
    \begin{overpic}[height=\heightQ, width=\widthQ]{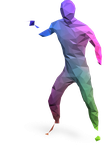}          
        \put(8,48){\includegraphics[height=0.7cm]{figs/qualitative/red_arrow_right.png}}
    \end{overpic}&
    \hspace{\hspaceCols}
    \begin{overpic}[height=\heightQ, width=\widthQ]{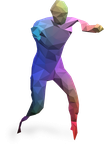}          
        \put(47,46){\includegraphics[height=0.7cm]{figs/qualitative/red_arrow_left.png}}
    \end{overpic}&
    \hspace{\hspaceCols}
    \begin{overpic}[height=\heightQ, width=\widthQ]{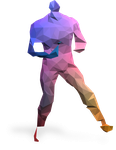}          
    
    \end{overpic}&
    \hspace{\hspaceCols}
    \begin{overpic}[height=\heightQ, width=\widthQ]{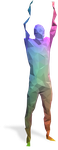}          
        \put(37,58){\includegraphics[height=0.7cm]{figs/qualitative/red_arrow_left.png}}
    \end{overpic}&
    \hspace{\hspaceCols}
    \begin{overpic}[height=\heightQ, width=\widthQ]{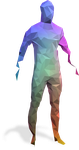}          
    
    \end{overpic}&
    \hspace{\hspaceCols}
    \begin{overpic}[height=\heightQ, width=\widthQ]{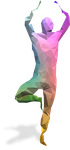}
    \end{overpic}&
    \hspace{\hspaceCols}
    \begin{overpic}[height=\heightQ, width=\widthQ]{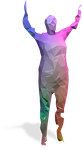}          
    
    \end{overpic}&
    \hspace{\hspaceCols}
    \begin{overpic}[height=\heightQ, width=\widthQ]{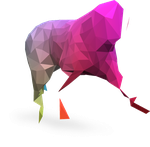}
        \put(72,-7){\includegraphics[height=0.7cm]{figs/qualitative/red_arrow_right.png}}
    \end{overpic}&
    \hspace{\hspaceCols}
    \begin{overpic}[height=\heightQ, width=\widthQ]{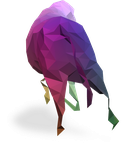}          
    
    \end{overpic}&
    \hspace{\hspaceCols}
    \begin{overpic}[height=\heightQ, width=\widthQ]{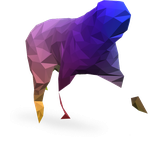}
         \put(72,-7){\includegraphics[height=0.7cm]{figs/qualitative/red_arrow_right.png}}
    \end{overpic}\\
    \rotatedCentering{90}{\heightQ}{\spidermatch~\cite{roetzer2024spidermatch}}&
    \hspace{\hspaceCols}
    \begin{overpic}[height=\heightQ, width=\widthQ]{\pathSpiderMatch\columnOne\trgtEnd}          
    
    \end{overpic}&
    \hspace{\hspaceCols}
    \begin{overpic}[height=\heightQ, width=\widthQ]{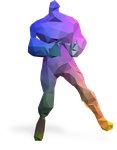}
    \end{overpic}&
    \hspace{\hspaceCols}
    \begin{overpic}[height=\heightQ, width=\widthQ]{\pathSpiderMatch\columnTwoTwo\trgtEnd}          
        \put(8,48){\includegraphics[height=0.7cm]{figs/qualitative/red_arrow_right.png}}
    \end{overpic}&
    \hspace{\hspaceCols}
    \begin{overpic}[height=\heightQ, width=\widthQ]{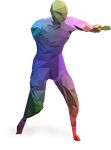}          
        \put(47,46){\includegraphics[height=0.7cm]{figs/qualitative/green_arrow_left.png}}
    \end{overpic}&
    \hspace{\hspaceCols}
    \begin{overpic}[height=\heightQ, width=\widthQ]{\pathSpiderMatch\columnFour\trgtEnd}          
    
    \end{overpic}&
    \hspace{\hspaceCols}
    \begin{overpic}[height=\heightQ, width=\widthQ]{\pathSpiderMatch\columnFive\trgtEnd}          
        \put(37,58){\includegraphics[height=0.7cm]{figs/qualitative/red_arrow_left.png}}
    \end{overpic}&
    \hspace{\hspaceCols}
    \begin{overpic}[height=\heightQ, width=\widthQ]{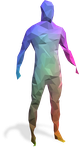}          
    
    \end{overpic}&
    \hspace{\hspaceCols}
    \begin{overpic}[height=\heightQ, width=\widthQ]{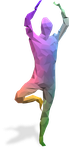}
    \end{overpic}&
    \hspace{\hspaceCols}
    \begin{overpic}[height=\heightQ, width=\widthQ]{\pathSpiderMatch\columnEight\trgtEnd}          
    
    \end{overpic}&
    \hspace{\hspaceCols}
    \begin{overpic}[height=\heightQ, width=\widthQ]{\pathSpiderMatch\columnNine\trgtEnd}
        \put(72,-7){\includegraphics[height=0.7cm]{figs/qualitative/red_arrow_right.png}}
    \end{overpic}&
    \hspace{\hspaceCols}
    \begin{overpic}[height=\heightQ, width=\widthQ]{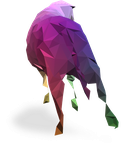}          
    
    \end{overpic}&
    \hspace{\hspaceCols}
    \begin{overpic}[height=\heightQ, width=\widthQ]{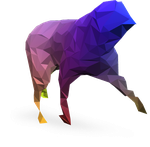}
         \put(72,-7){\includegraphics[height=0.7cm]{figs/qualitative/green_arrow_right.png}}
    \end{overpic}\\
    \rotatedCentering{90}{\heightQ}{\ours}&
    \hspace{\hspaceCols}
    \begin{overpic}[height=\heightQ, width=\widthQ]{\pathOurs\columnOne\trgtEnd}          
    
    \end{overpic}&
    \hspace{\hspaceCols}
    \begin{overpic}[height=\heightQ, width=\widthQ]{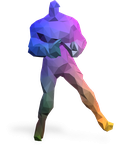}          
    
    \end{overpic}&
    \hspace{\hspaceCols}
    \begin{overpic}[height=\heightQ, width=\widthQ]{\pathOurs\columnTwoTwo\trgtEnd}          
        \put(8,48){\includegraphics[height=0.7cm]{figs/qualitative/green_arrow_right.png}}
    \end{overpic}&
    \hspace{\hspaceCols}
    \begin{overpic}[height=\heightQ, width=\widthQ]{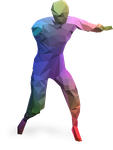}          
         \put(47,46){\includegraphics[height=0.7cm]{figs/qualitative/green_arrow_left.png}}
    \end{overpic}&
    \hspace{\hspaceCols}
    \begin{overpic}[height=\heightQ, width=\widthQ]{\pathOurs\columnFour\trgtEnd}          
    
    \end{overpic}&
    \hspace{\hspaceCols}
    \begin{overpic}[height=\heightQ, width=\widthQ]{\pathOurs\columnFive\trgtEnd}          
        \put(37,58){\includegraphics[height=0.7cm]{figs/qualitative/green_arrow_left.png}}
    \end{overpic}&
    \hspace{\hspaceCols}
    \begin{overpic}[height=\heightQ, width=\widthQ]{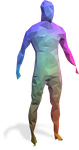}          
    
    \end{overpic}&
    \hspace{\hspaceCols}
    \begin{overpic}[height=\heightQ, width=\widthQ]{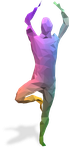}          
    
    \end{overpic}&
    \hspace{\hspaceCols}
    \begin{overpic}[height=\heightQ, width=\widthQ]{\pathOurs\columnEight\trgtEnd}          
    
    \end{overpic}&
    \hspace{\hspaceCols}
    \begin{overpic}[height=\heightQ, width=\widthQ]{\pathOurs\columnNine\trgtEnd}          
        \put(72,-7){\includegraphics[height=0.7cm]{figs/qualitative/green_arrow_right.png}}
    \end{overpic}&
    \hspace{\hspaceCols}
    \begin{overpic}[height=\heightQ, width=\widthQ]{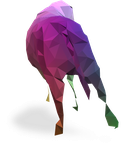}          
    
    \end{overpic}&
    \hspace{\hspaceCols}
    \begin{overpic}[height=\heightQ, width=\widthQ]{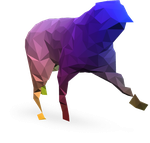}
        \put(72,-7){\includegraphics[height=0.7cm]{figs/qualitative/red_arrow_right.png}}
    \end{overpic}\\
\end{tabular}
}
    \caption{Additional \textbf{qualitative shape matching results} on datasets DT4D-H (columns \scircled{1}-\scircled{5}), FAUST (columns \scircled{6}-\scircled{9}), and SMAL (columns \scircled{10}-\scircled{12}). 
    Columns \scircled{1}, \scircled{3}, \scircled{5}, \scircled{6}, \scircled{9}, \scircled{10} are also shown in the main paper and are repeated here for comparison with methods \discreteopt~\cite{ren2021discrete} and \smooths~\cite{eisenberger2020smooth}.
    All matchings are visualised by transfering colour and triangulation from source to target shape. 
    In column \scircled{12}, we can see a failure mode of ours very likely caused by the fact that our solution space still contains matchings which allow for inside out flips (one side of the leg is matched to the other side of the leg).}
    \label{fig:supp:qualitative}
\end{figure*}

\begin{figure*}[t]
    \centering
    \begin{tabular}{ccccc}
        \hspace{-0.87cm}
         \newcommand{\pckLineWidth}{1.5pt}
\newcommand{\plotWidth}{0.9\columnwidth}
\newcommand{\plotHeight}{0.75\columnwidth}
\newcommand{\pckTitle}{$\%$ of  Corr. Points: \faust}

\pgfplotsset{%
    label style = {font=\large},
    tick label style = {font=\large},
    title style =  {font=\Large},
    legend style={  fill= gray!10,
                    fill opacity=0.6, 
                    font=\large,
                    draw=gray!20, %
                    text opacity=1}
}
\begin{tikzpicture}[scale=0.5, transform shape]
	\begin{axis}[
		width=\plotWidth,
		height=\plotHeight,
		grid=major,
		title=\pckTitle,
		legend style={
			at={(0.97,0.03)},
			anchor=south east,
			legend columns=1},
		legend cell align={left},
        title style={yshift=-0.25cm},
	ylabel={{\large$\%$ Correct Matchings \textcolor{gray!40}{$\rightarrow$}}},
        xlabel={Geodesic Error Threshold},
        xmin=0,
        xmax=0.25,
        ylabel near ticks,
        xtick={0, 0.05, 0.1, 0.15, 0.2, 0.25},
        xticklabels={$0$, $0.05$, $0.1$, $0.15$, $0.2$, $0.25$},
        ymin=0,
        ymax=1,
        ytick={0, 0.20, 0.40, 0.60, 0.80, 1},
        yticklabels={$0$, $20$, $40$, $60$, $80$, $100$},
	]
    \addplot [color=ulrssmcolour, smooth, line width=\pckLineWidth]
    table[row sep=crcr]{%
0	0.113194063981325\\
0.00862068965517241	0.115028227055096\\
0.0172413793103448	0.202686929801577\\
0.0258620689655172	0.479002405850266\\
0.0344827586206897	0.650865867892618\\
0.0431034482758621	0.777518401181487\\
0.0517241379310345	0.88208951668612\\
0.0603448275862069	0.937781377289726\\
0.0689655172413793	0.96824754055406\\
0.0775862068965517	0.984588266120388\\
0.0862068965517241	0.992472785307639\\
0.0948275862068965	0.995164479169148\\
0.103448275862069	0.996331673852457\\
0.112068965517241	0.996998642242919\\
0.120689655172414	0.997570329434744\\
0.129310344827586	0.997856173030657\\
0.137931034482759	0.998046735427932\\
0.146551724137931	0.998189657225888\\
0.155172413793103	0.998284938424525\\
0.163793103448276	0.998332579023844\\
0.172413793103448	0.998332579023844\\
0.181034482758621	0.998380219623163\\
0.189655172413793	0.998404039922822\\
0.198275862068966	0.998523141421119\\
0.206896551724138	0.998570782020438\\
0.21551724137931	0.998570782020438\\
0.224137931034483	0.998570782020438\\
0.232758620689655	0.998570782020438\\
0.241379310344828	0.998570782020438\\
0.25	0.998594602320097\\
    };
    \addlegendentry{\textcolor{black}{\caoetal: 0.031}}
    \addplot [color=cPLOT1, smooth, line width=\pckLineWidth]
    table[row sep=crcr]{%
0	0.022473240503931\\
0.00862068965517241	0.0227337311736289\\
0.0172413793103448	0.0380079568059108\\
0.0258620689655172	0.0985365160556976\\
0.0344827586206897	0.14002557544757\\
0.0431034482758621	0.174765558397272\\
0.0517241379310345	0.214052287581699\\
0.0603448275862069	0.242019513119257\\
0.0689655172413793	0.262598276025386\\
0.0775862068965517	0.28116415648385\\
0.0862068965517241	0.296343658236241\\
0.0948275862068965	0.310339111490007\\
0.103448275862069	0.322866344605475\\
0.112068965517241	0.333783271762811\\
0.120689655172414	0.343847684001137\\
0.129310344827586	0.354433077578858\\
0.137931034482759	0.363644974898172\\
0.146551724137931	0.373259448707019\\
0.155172413793103	0.382589750876196\\
0.163793103448276	0.391754286255565\\
0.172413793103448	0.400374159325566\\
0.181034482758621	0.410272804774084\\
0.189655172413793	0.419295254333618\\
0.198275862068966	0.427749360613811\\
0.206896551724138	0.435421994884911\\
0.21551724137931	0.443260395945818\\
0.224137931034483	0.451359287676423\\
0.232758620689655	0.459694989106754\\
0.241379310344828	0.466728237188595\\
0.25	0.474329828549777\\
    };
    \addlegendentry{\textcolor{black}{\discreteopt: {0.379}}}
    \addplot [color=cPLOT3, smooth, line width=\pckLineWidth]
    table[row sep=crcr]{%
0	0.0469736309282771\\
0.00862068965517241	0.0479740835139706\\
0.0172413793103448	0.0886829756318334\\
0.0258620689655172	0.229889712012577\\
0.0344827586206897	0.333412734332198\\
0.0431034482758621	0.428646292370358\\
0.0517241379310345	0.53552797694195\\
0.0603448275862069	0.614539910912079\\
0.0689655172413793	0.675281675043472\\
0.0775862068965517	0.726161835115886\\
0.0862068965517241	0.762392510897787\\
0.0948275862068965	0.7875943879374\\
0.103448275862069	0.806650627664896\\
0.112068965517241	0.820918987160859\\
0.120689655172414	0.832424191896334\\
0.129310344827586	0.841166241871323\\
0.137931034482759	0.847240418284462\\
0.146551724137931	0.852457063909864\\
0.155172413793103	0.856673256949572\\
0.163793103448276	0.860008098901884\\
0.172413793103448	0.86320001905624\\
0.181034482758621	0.865724970820133\\
0.189655172413793	0.868345203782664\\
0.198275862068966	0.870441390152688\\
0.206896551724138	0.872823420118625\\
0.21551724137931	0.874252638098187\\
0.224137931034483	0.876277363569234\\
0.232758620689655	0.877682761249136\\
0.241379310344828	0.879374002524952\\
0.25	0.881494009194636\\
    };
    \addlegendentry{\textcolor{black}{\smooths: 0.110}}
    \addplot [color=cPLOT5, smooth, line width=\pckLineWidth]
    table[row sep=crcr]{%
0	0.121189820529864\\
0.00862068965517241	0.123112714841895\\
0.0172413793103448	0.217571930490932\\
0.0258620689655172	0.503038647801728\\
0.0344827586206897	0.679422656917672\\
0.0431034482758621	0.80569271674105\\
0.0517241379310345	0.904163897065806\\
0.0603448275862069	0.952900959073212\\
0.0689655172413793	0.976759092203969\\
0.0775862068965517	0.988510112999715\\
0.0862068965517241	0.993993922704397\\
0.0948275862068965	0.996296647991644\\
0.103448275862069	0.99715126768588\\
0.112068965517241	0.997649795840851\\
0.120689655172414	0.99810084512392\\
0.129310344827586	0.998528154971038\\
0.137931034482759	0.998646852150793\\
0.146551724137931	0.99881302820245\\
0.155172413793103	0.998979204254107\\
0.163793103448276	0.999026683126009\\
0.172413793103448	0.999026683126009\\
0.181034482758621	0.999074161997911\\
0.189655172413793	0.999097901433862\\
0.198275862068966	0.999097901433862\\
0.206896551724138	0.999097901433862\\
0.21551724137931	0.999097901433862\\
0.224137931034483	0.999097901433862\\
0.232758620689655	0.999121640869813\\
0.241379310344828	0.999121640869813\\
0.25	0.999145380305764\\
    };
    \addlegendentry{\textcolor{black}{\spidermatch: {0.029}}}
    \addplot [color=ourcolor, smooth, line width=\pckLineWidth]
    table[row sep=crcr]{%
0	0.125469947175558\\
0.00862068965517241	0.127444915052586\\
0.0172413793103448	0.2265264360158\\
0.0258620689655172	0.521010802836339\\
0.0344827586206897	0.703136153809546\\
0.0431034482758621	0.826345595583686\\
0.0517241379310345	0.918669395136344\\
0.0603448275862069	0.961928330081378\\
0.0689655172413793	0.981892161994956\\
0.0775862068965517	0.991076952362823\\
0.0862068965517241	0.995645552753058\\
0.0948275862068965	0.997073240375006\\
0.103448275862069	0.997739494598582\\
0.112068965517241	0.99833436444106\\
0.120689655172414	0.998738875933946\\
0.129310344827586	0.998953029077238\\
0.137931034482759	0.999119592633132\\
0.146551724137931	0.999238566601628\\
0.155172413793103	0.999333745776424\\
0.163793103448276	0.999357540570123\\
0.172413793103448	0.999381335363822\\
0.181034482758621	0.999428924951221\\
0.189655172413793	0.99945271974492\\
0.198275862068966	0.999476514538619\\
0.206896551724138	0.999476514538619\\
0.21551724137931	0.999476514538619\\
0.224137931034483	0.999500309332318\\
0.232758620689655	0.999524104126017\\
0.241379310344828	0.999524104126017\\
0.25	0.999524104126017\\
    };
    \addlegendentry{\textcolor{black}{\ours: \textbf{0.027}}}
	\end{axis}
\end{tikzpicture}
         & 
         \hspace{-1.225cm}
         \newcommand{\pckLineWidth}{1.5pt}
\newcommand{\plotWidth}{0.9\columnwidth}
\newcommand{\plotHeight}{0.75\columnwidth}
\newcommand{\pckTitle}{$\%$ of  Corr. Points: SMAL}

\pgfplotsset{%
    label style = {font=\large},
    tick label style = {font=\large},
    title style =  {font=\Large},
    legend style={  fill= gray!10,
                    fill opacity=0.6, 
                    font=\large,
                    draw=gray!20, %
                    text opacity=1}
}
\begin{tikzpicture}[scale=0.5, transform shape]
	\begin{axis}[
		width=\plotWidth,
		height=\plotHeight,
		grid=major,
		title=\pckTitle,
		legend style={
			at={(0.97,0.03)},
			anchor=south east,
			legend columns=1},
		legend cell align={left},
       title style={yshift=-0.25cm},
        xlabel={Geodesic Error Threshold},
	xmin=0,
        xmax=0.25,
        ylabel near ticks,
        xtick={0, 0.05, 0.1, 0.15, 0.2, 0.25},
        xticklabels={$0$, $0.05$, $0.1$, $0.15$, $0.2$, $0.25$},
        ymin=0,
        ymax=1,
        ytick={0, 0.20, 0.40, 0.60, 0.80, 1},
        yticklabels={$0$, $20$, $40$, $60$, $80$, $100$},
	]
    \addplot [color=ulrssmcolour, smooth, line width=\pckLineWidth]
    table[row sep=crcr]{%
0	0.0901596493344593\\
0.00862068965517241	0.0931756946957816\\
0.0172413793103448	0.152451039530301\\
0.0258620689655172	0.2651304942293\\
0.0344827586206897	0.444082519001086\\
0.0431034482758621	0.582780391683758\\
0.0517241379310345	0.677323360276672\\
0.0603448275862069	0.758957654723127\\
0.0689655172413793	0.828889693167652\\
0.0775862068965517	0.8810069570113\\
0.0862068965517241	0.915309446254072\\
0.0948275862068965	0.939397595206499\\
0.103448275862069	0.956126593477299\\
0.112068965517241	0.96702457071621\\
0.120689655172414	0.974906502593799\\
0.129310344827586	0.979330035790405\\
0.137931034482759	0.982708006595086\\
0.146551724137931	0.98455784775003\\
0.155172413793103	0.986005549523465\\
0.163793103448276	0.987292395544296\\
0.172413793103448	0.988177102183617\\
0.181034482758621	0.988900953070334\\
0.189655172413793	0.989343306389995\\
0.198275862068966	0.989785659709655\\
0.206896551724138	0.989906301524108\\
0.21551724137931	0.990026943338561\\
0.224137931034483	0.990268226967467\\
0.232758620689655	0.990469296658222\\
0.241379310344828	0.990710580287128\\
0.25	0.99083122210158\\
    };
    \addlegendentry{\textcolor{black}{\caoetal: {0.048}}}
    \addplot [color=cPLOT1, smooth, line width=\pckLineWidth]
    table[row sep=crcr]{%
0	0.0103323094107791\\
0.00862068965517241	0.0106115610164758\\
0.0172413793103448	0.015638089919017\\
0.0258620689655172	0.0280049467427295\\
0.0344827586206897	0.0527386603901544\\
0.0431034482758621	0.0722463797023976\\
0.0517241379310345	0.0897594446882355\\
0.0603448275862069	0.108908126221726\\
0.0689655172413793	0.129173814178003\\
0.0775862068965517	0.146327841385088\\
0.0862068965517241	0.165037698966769\\
0.0948275862068965	0.180037499501336\\
0.103448275862069	0.196114413372163\\
0.112068965517241	0.210715283041449\\
0.120689655172414	0.225356045797263\\
0.129310344827586	0.239238839909044\\
0.137931034482759	0.25364024414569\\
0.146551724137931	0.268400686161088\\
0.155172413793103	0.282961662743846\\
0.163793103448276	0.296724777596043\\
0.172413793103448	0.309330992938924\\
0.181034482758621	0.322934535445007\\
0.189655172413793	0.334942354489967\\
0.198275862068966	0.348466110822994\\
0.206896551724138	0.360832967646707\\
0.21551724137931	0.374316830893206\\
0.224137931034483	0.386404436111222\\
0.232758620689655	0.399050544540631\\
0.241379310344828	0.411297722104759\\
0.25	0.423983723620697\\
    };
    \addlegendentry{\textcolor{black}{\discreteopt: 0.376}}
    \addplot [color=cPLOT3, smooth, line width=\pckLineWidth]
    table[row sep=crcr]{%
0	0.0325391254077714\\
0.00862068965517241	0.0332411116158071\\
0.0172413793103448	0.0511211132675393\\
0.0258620689655172	0.0959656439691126\\
0.0344827586206897	0.171986620968741\\
0.0431034482758621	0.238634017425775\\
0.0517241379310345	0.288268571664533\\
0.0603448275862069	0.340587190816369\\
0.0689655172413793	0.391006317875872\\
0.0775862068965517	0.431184704959326\\
0.0862068965517241	0.464136763430648\\
0.0948275862068965	0.490316719659743\\
0.103448275862069	0.513606144443986\\
0.112068965517241	0.533055291737209\\
0.120689655172414	0.547755708799604\\
0.129310344827586	0.561134740058637\\
0.137931034482759	0.572325226080852\\
0.146551724137931	0.582524672750547\\
0.155172413793103	0.591278853697816\\
0.163793103448276	0.598752942148078\\
0.172413793103448	0.605277284552174\\
0.181034482758621	0.610686707684684\\
0.189655172413793	0.61646777057439\\
0.198275862068966	0.621175207498864\\
0.206896551724138	0.626584630631375\\
0.21551724137931	0.630837841185944\\
0.224137931034483	0.634430358838832\\
0.232758620689655	0.639179089069662\\
0.241379310344828	0.643143246479746\\
0.25	0.647024817277119\\
    };
    \addlegendentry{\textcolor{black}{\smooths: 0.268}}
    \addplot [color=cPLOT5, smooth, line width=\pckLineWidth]
    table[row sep=crcr]{%
0	0.0875075437537719\\
0.00862068965517241	0.0900422450211225\\
0.0172413793103448	0.146811506739087\\
0.0258620689655172	0.257734862200764\\
0.0344827586206897	0.433916716958358\\
0.0431034482758621	0.571233152283243\\
0.0517241379310345	0.666304566485617\\
0.0603448275862069	0.753490243411788\\
0.0689655172413793	0.823979078656206\\
0.0775862068965517	0.87889760611547\\
0.0862068965517241	0.915912291289479\\
0.0948275862068965	0.941742104204385\\
0.103448275862069	0.96012874673104\\
0.112068965517241	0.972319452826393\\
0.120689655172414	0.980768457050895\\
0.129310344827586	0.985676926171796\\
0.137931034482759	0.989700261516797\\
0.146551724137931	0.991631462482398\\
0.155172413793103	0.993482196741098\\
0.163793103448276	0.995051297525649\\
0.172413793103448	0.996057131361899\\
0.181034482758621	0.996861798430899\\
0.189655172413793	0.997465298732649\\
0.198275862068966	0.99802856568095\\
0.206896551724138	0.9981492657413\\
0.21551724137931	0.99818949909475\\
0.224137931034483	0.9983101991551\\
0.232758620689655	0.99843089921545\\
0.241379310344828	0.99851136592235\\
0.25	0.9986320659827\\
};
    \addlegendentry{\textcolor{black}{\spidermatch: \textbf{0.044}}}
    \addplot [color=ourcolor, smooth, line width=\pckLineWidth]
    table[row sep=crcr]{%
0	0.0902003554693812\\
0.00862068965517241	0.093229924058814\\
0.0172413793103448	0.152165131685248\\
0.0258620689655172	0.26712716109226\\
0.0344827586206897	0.445346582646631\\
0.0431034482758621	0.584464372273388\\
0.0517241379310345	0.681491355630958\\
0.0603448275862069	0.765067054451446\\
0.0689655172413793	0.835110680239134\\
0.0775862068965517	0.886896105994506\\
0.0862068965517241	0.921110033931168\\
0.0948275862068965	0.945912102116659\\
0.103448275862069	0.961908224268864\\
0.112068965517241	0.972774276942963\\
0.120689655172414	0.980529972531911\\
0.129310344827586	0.985498465018581\\
0.137931034482759	0.988770399095169\\
0.146551724137931	0.990628534496688\\
0.155172413793103	0.992163515915334\\
0.163793103448276	0.993334949103248\\
0.172413793103448	0.994021651316852\\
0.181034482758621	0.99458717078688\\
0.189655172413793	0.994950719017612\\
0.198275862068966	0.995395055744062\\
0.206896551724138	0.99547584423978\\
0.21551724137931	0.995758603974794\\
0.224137931034483	0.995758603974794\\
0.232758620689655	0.995798998222653\\
0.241379310344828	0.995879786718371\\
0.25	0.99596057521409\\
};
    \addlegendentry{\textcolor{black}{\ours: \textbf{0.044}}}
	\end{axis}
\end{tikzpicture}
         & 
         \hspace{-1.225cm}
         \newcommand{\pckLineWidth}{1.5pt}
\newcommand{\plotWidth}{0.9\columnwidth}
\newcommand{\plotHeight}{0.75\columnwidth}
\newcommand{\pckTitle}{$\%$ of  Corr. Points: DT4D Intra}

\pgfplotsset{%
    label style = {font=\large},
    tick label style = {font=\large},
    title style =  {font=\Large},
    legend style={  fill= gray!10,
                    fill opacity=0.6, 
                    font=\large,
                    draw=gray!20, %
                    text opacity=1}
}
\begin{tikzpicture}[scale=0.5, transform shape]
	\begin{axis}[
		width=\plotWidth,
		height=\plotHeight,
		grid=major,
		title=\pckTitle,
		legend style={
			at={(0.97,0.03)},
			anchor=south east,
			legend columns=1},
		legend cell align={left},
        title style={yshift=-0.25cm},
        xlabel={Geodesic Error Threshold},
        xmin=0,
        xmax=0.25,
        ylabel near ticks,
        xtick={0, 0.05, 0.1, 0.15, 0.2, 0.25},
        xticklabels={$0$, $0.05$, $0.1$, $0.15$, $0.2$, $0.25$},
        ymin=0,
        ymax=1,
        ytick={0, 0.20, 0.40, 0.60, 0.80, 1},
        yticklabels={$0$, $20$, $40$, $60$, $80$, $100$},
	]
    \addplot [color=ulrssmcolour, smooth, line width=\pckLineWidth]
    table[row sep=crcr]{%
0	0.0992552242763142\\
0.00862068965517241	0.109281978355709\\
0.0172413793103448	0.281616814095302\\
0.0258620689655172	0.531104630142929\\
0.0344827586206897	0.704307165754778\\
0.0431034482758621	0.830653908264841\\
0.0517241379310345	0.902407867145508\\
0.0603448275862069	0.938079973004893\\
0.0689655172413793	0.957771939550242\\
0.0775862068965517	0.968594085179204\\
0.0862068965517241	0.974643881510762\\
0.0948275862068965	0.978669044807057\\
0.103448275862069	0.980886500036154\\
0.112068965517241	0.982332666489913\\
0.120689655172414	0.983393188556003\\
0.129310344827586	0.984309093976717\\
0.137931034482759	0.98491166333245\\
0.146551724137931	0.985562438236641\\
0.155172413793103	0.98628552146352\\
0.163793103448276	0.986719371399648\\
0.172413793103448	0.987321940755381\\
0.181034482758621	0.987852201788426\\
0.189655172413793	0.988334257273012\\
0.198275862068966	0.988671696112222\\
0.206896551724138	0.989057340499892\\
0.21551724137931	0.989635807081395\\
0.224137931034483	0.989997348694835\\
0.232758620689655	0.990238376437128\\
0.241379310344828	0.99052760972788\\
0.25	0.990624020824797\\
    };
    \addlegendentry{\textcolor{black}{\caoetal: 0.033}}
    \addplot [color=cPLOT1, smooth, line width=\pckLineWidth]
    table[row sep=crcr]{%
0	0.0251243004347513\\
0.00862068965517241	0.0272380083107151\\
0.0172413793103448	0.0681670789998319\\
0.0258620689655172	0.133692023154709\\
0.0344827586206897	0.184372973362477\\
0.0431034482758621	0.227992217711911\\
0.0517241379310345	0.261114980904571\\
0.0603448275862069	0.284269689909447\\
0.0689655172413793	0.30310090553167\\
0.0775862068965517	0.318977734009079\\
0.0862068965517241	0.332716835202844\\
0.0948275862068965	0.345519179497034\\
0.103448275862069	0.35673624288425\\
0.112068965517241	0.366344005956813\\
0.120689655172414	0.375471380875748\\
0.129310344827586	0.384670814017726\\
0.137931034482759	0.393365839598395\\
0.146551724137931	0.401748612879206\\
0.155172413793103	0.410587754905964\\
0.163793103448276	0.418177887733289\\
0.172413793103448	0.426944971537002\\
0.181034482758621	0.43494343429491\\
0.189655172413793	0.442773761199049\\
0.198275862068966	0.450556049287825\\
0.206896551724138	0.459227055460812\\
0.21551724137931	0.466601013619004\\
0.224137931034483	0.47404703000024\\
0.232758620689655	0.481877356904379\\
0.241379310344828	0.489779742031561\\
0.25	0.497081641966709\\
    };
    \addlegendentry{\textcolor{black}{\discreteopt: 0.373}}
    \addplot [color=cPLOT3, smooth, line width=\pckLineWidth]
    table[row sep=crcr]{%
0	0.0507882177119992\\
0.00862068965517241	0.0557055392180495\\
0.0172413793103448	0.146989345803404\\
0.0258620689655172	0.30012534348937\\
0.0344827586206897	0.434050040977679\\
0.0431034482758621	0.563370775683363\\
0.0517241379310345	0.662753700043388\\
0.0603448275862069	0.735019042568577\\
0.0689655172413793	0.788362339102348\\
0.0775862068965517	0.826736730463289\\
0.0862068965517241	0.852649086438799\\
0.0948275862068965	0.87176396856771\\
0.103448275862069	0.88603384274213\\
0.112068965517241	0.89642288964952\\
0.120689655172414	0.904473798389818\\
0.129310344827586	0.910258882514583\\
0.137931034482759	0.915152099503447\\
0.146551724137931	0.918574940943933\\
0.155172413793103	0.921708528178181\\
0.163793103448276	0.92462517475775\\
0.172413793103448	0.926625849684231\\
0.181034482758621	0.928602420093526\\
0.189655172413793	0.930121004676276\\
0.198275862068966	0.931711902810587\\
0.206896551724138	0.932893024152726\\
0.21551724137931	0.933857204840187\\
0.224137931034483	0.935110639733886\\
0.232758620689655	0.936002506869787\\
0.241379310344828	0.937063105625994\\
0.25	0.938196017933761\\
    };
    \addlegendentry{\textcolor{black}{\smooths: 0.075}}
    \addplot [color=cPLOT5, smooth, line width=\pckLineWidth]
    table[row sep=crcr]{%
0	0.101424515135473\\
0.00862068965517241	0.111988064873189\\
0.0172413793103448	0.289378699648684\\
0.0258620689655172	0.546705808749218\\
0.0344827586206897	0.725853024688387\\
0.0431034482758621	0.855599403243659\\
0.0517241379310345	0.923961692092979\\
0.0603448275862069	0.956470474998797\\
0.0689655172413793	0.973386592232542\\
0.0775862068965517	0.981206987824246\\
0.0862068965517241	0.985851099667934\\
0.0948275862068965	0.989195822705616\\
0.103448275862069	0.991650223783628\\
0.112068965517241	0.993093989123634\\
0.120689655172414	0.994200875884306\\
0.129310344827586	0.994850570287309\\
0.137931034482759	0.995259637133645\\
0.146551724137931	0.995716829491313\\
0.155172413793103	0.996270272871649\\
0.163793103448276	0.996703402473651\\
0.172413793103448	0.99708840656432\\
0.181034482758621	0.99723278309832\\
0.189655172413793	0.997521536166322\\
0.198275862068966	0.997689975455989\\
0.206896551724138	0.998050916790991\\
0.21551724137931	0.998147167813658\\
0.224137931034483	0.998315607103325\\
0.232758620689655	0.998435920881659\\
0.241379310344828	0.998532171904327\\
0.25	0.998628422926994\\
    };
    \addlegendentry{\textcolor{black}{\spidermatch: {0.027}}}
    \addplot [color=ourcolor, smooth, line width=\pckLineWidth]
    table[row sep=crcr]{%
0	0.113096929156228\\
0.00862068965517241	0.124719130142824\\
0.0172413793103448	0.322451509594773\\
0.0258620689655172	0.600816136780392\\
0.0344827586206897	0.7822257806245\\
0.0431034482758621	0.901702006766704\\
0.0517241379310345	0.957230300369328\\
0.0603448275862069	0.979648234716806\\
0.0689655172413793	0.989488364885457\\
0.0775862068965517	0.993594875900721\\
0.0862068965517241	0.995816007644826\\
0.0948275862068965	0.996978227743485\\
0.103448275862069	0.997649732689377\\
0.112068965517241	0.998062966502234\\
0.120689655172414	0.998424546088484\\
0.129310344827586	0.998631162994912\\
0.137931034482759	0.998863607014644\\
0.146551724137931	0.999018569694465\\
0.155172413793103	0.999147705260983\\
0.163793103448276	0.999276840827501\\
0.172413793103448	0.999302667940804\\
0.181034482758621	0.999328495054108\\
0.189655172413793	0.999354322167411\\
0.198275862068966	0.999354322167411\\
0.206896551724138	0.999380149280715\\
0.21551724137931	0.999405976394018\\
0.224137931034483	0.999431803507322\\
0.232758620689655	0.999431803507322\\
0.241379310344828	0.999457630620626\\
0.25	0.999457630620626\\
    };
    \addlegendentry{\textcolor{black}{\ours: \textbf{0.024}}}
	\end{axis}
\end{tikzpicture}
         & 
         \hspace{-1.225cm}
         \newcommand{\pckLineWidth}{1.5pt}
\newcommand{\plotWidth}{0.9\columnwidth}
\newcommand{\plotHeight}{0.75\columnwidth}
\newcommand{\pckTitle}{$\%$ of  Corr. Points: DT4D Inter}

\pgfplotsset{%
    label style = {font=\large},%
    tick label style = {font=\large},
    title style =  {font=\Large},
    legend style={  fill= gray!10,
                    fill opacity=0.6, 
                    font=\large,
                    draw=gray!20, %
                    text opacity=1}
}
\begin{tikzpicture}[scale=0.5, transform shape]
	\begin{axis}[
		width=\plotWidth,
		height=\plotHeight,
		grid=major,
		title=\pckTitle,
		legend style={
			at={(0.97,0.03)},
			anchor=south east,
			legend columns=1},
		legend cell align={left},
        title style={yshift=-0.25cm},
        xlabel={Geodesic Error Threshold},
        xmin=0,
        xmax=0.25,
        ylabel near ticks,
        xtick={0, 0.05, 0.1, 0.15, 0.2, 0.25},
        xticklabels={$0$, $0.05$, $0.1$, $0.15$, $0.2$, $0.25$},
        ymin=0,
        ymax=1,
        ytick={0, 0.20, 0.40, 0.60, 0.80, 1},
        yticklabels={$0$, $20$, $40$, $60$, $80$, $100$},
	]
    \addplot [color=ulrssmcolour, smooth, line width=\pckLineWidth]
    table[row sep=crcr]{%
0	0.0504459610731954\\
0.00862068965517241	0.0569457985772578\\
0.0172413793103448	0.169248546564114\\
0.0258620689655172	0.347091322716932\\
0.0344827586206897	0.497598671144332\\
0.0431034482758621	0.647925468529953\\
0.0517241379310345	0.765211425269924\\
0.0603448275862069	0.841656736359369\\
0.0689655172413793	0.894594301809121\\
0.0775862068965517	0.928140685371755\\
0.0862068965517241	0.950023471635431\\
0.0948275862068965	0.963456469143827\\
0.103448275862069	0.971834037482396\\
0.112068965517241	0.977611670819341\\
0.120689655172414	0.982414328530676\\
0.129310344827586	0.984833712490521\\
0.137931034482759	0.986928104575163\\
0.146551724137931	0.988950276243094\\
0.155172413793103	0.99057523561911\\
0.163793103448276	0.991839092911566\\
0.172413793103448	0.993175170620734\\
0.181034482758621	0.994077925829632\\
0.189655172413793	0.995377893330444\\
0.198275862068966	0.996136207705918\\
0.206896551724138	0.996533419997833\\
0.21551724137931	0.996641750622901\\
0.224137931034483	0.996894522081392\\
0.232758620689655	0.996966742498104\\
0.241379310344828	0.99700285270646\\
0.25	0.997111183331528\\
    };
    \addlegendentry{\textcolor{black}{\caoetal: {0.041}}}
    \addplot [color=cPLOT1, smooth, line width=\pckLineWidth]
    table[row sep=crcr]{%
0	0.00837472356161404\\
0.00862068965517241	0.00942609578363485\\
0.0172413793103448	0.0288221005691912\\
0.0258620689655172	0.0623572490301998\\
0.0344827586206897	0.0956386179893413\\
0.0431034482758621	0.135772033498894\\
0.0517241379310345	0.175760432150237\\
0.0603448275862069	0.20860675053475\\
0.0689655172413793	0.239459087118878\\
0.0775862068965517	0.265235833665664\\
0.0862068965517241	0.286553311822499\\
0.0948275862068965	0.305006707029692\\
0.103448275862069	0.321466120436501\\
0.112068965517241	0.334771417177247\\
0.120689655172414	0.346916579052315\\
0.129310344827586	0.358046622919914\\
0.137931034482759	0.368596599354675\\
0.146551724137931	0.378058949352862\\
0.155172413793103	0.387883841496574\\
0.163793103448276	0.396838632491027\\
0.172413793103448	0.405104593408984\\
0.181034482758621	0.413443062756045\\
0.189655172413793	0.422180328463184\\
0.198275862068966	0.43102635681398\\
0.206896551724138	0.439546097233803\\
0.21551724137931	0.447594532864446\\
0.224137931034483	0.455787985353297\\
0.232758620689655	0.464090200485806\\
0.241379310344828	0.472464924047421\\
0.25	0.480187071747091\\
    };
    \addlegendentry{\textcolor{black}{\discreteopt: 0.42}}
    \addplot [color=cPLOT3, smooth, line width=\pckLineWidth]
    table[row sep=crcr]{%
0	0.0507882177119992\\
0.00862068965517241	0.0557055392180495\\
0.0172413793103448	0.146989345803404\\
0.0258620689655172	0.30012534348937\\
0.0344827586206897	0.434050040977679\\
0.0431034482758621	0.563370775683363\\
0.0517241379310345	0.662753700043388\\
0.0603448275862069	0.735019042568577\\
0.0689655172413793	0.788362339102348\\
0.0775862068965517	0.826736730463289\\
0.0862068965517241	0.852649086438799\\
0.0948275862068965	0.87176396856771\\
0.103448275862069	0.88603384274213\\
0.112068965517241	0.89642288964952\\
0.120689655172414	0.904473798389818\\
0.129310344827586	0.910258882514583\\
0.137931034482759	0.915152099503447\\
0.146551724137931	0.918574940943933\\
0.155172413793103	0.921708528178181\\
0.163793103448276	0.92462517475775\\
0.172413793103448	0.926625849684231\\
0.181034482758621	0.928602420093526\\
0.189655172413793	0.930121004676276\\
0.198275862068966	0.931711902810587\\
0.206896551724138	0.932893024152726\\
0.21551724137931	0.933857204840187\\
0.224137931034483	0.935110639733886\\
0.232758620689655	0.936002506869787\\
0.241379310344828	0.937063105625994\\
0.25	0.938196017933761\\
    };
    \addlegendentry{\textcolor{black}{\smooths: 0.17}}
    \addplot [color=cPLOT5, smooth, line width=\pckLineWidth]
    table[row sep=crcr]{%
0	0.0473184982147365\\
0.00862068965517241	0.0531251127060266\\
0.0172413793103448	0.161322898258016\\
0.0258620689655172	0.330724564503913\\
0.0344827586206897	0.478450607710896\\
0.0431034482758621	0.628881595556678\\
0.0517241379310345	0.747935225592383\\
0.0603448275862069	0.827136004616439\\
0.0689655172413793	0.884552962816028\\
0.0775862068965517	0.921881198831464\\
0.0862068965517241	0.944206008583691\\
0.0948275862068965	0.958235654776932\\
0.103448275862069	0.967576730262921\\
0.112068965517241	0.973275146968659\\
0.120689655172414	0.978360442889602\\
0.129310344827586	0.981606376456162\\
0.137931034482759	0.983626068453132\\
0.146551724137931	0.985970353806759\\
0.155172413793103	0.987917913946695\\
0.163793103448276	0.989360551087388\\
0.172413793103448	0.990586792656977\\
0.181034482758621	0.991921232012118\\
0.189655172413793	0.993327803224294\\
0.198275862068966	0.993760594366502\\
0.206896551724138	0.994517978865366\\
0.21551724137931	0.994914704079056\\
0.224137931034483	0.995347495221264\\
0.232758620689655	0.995527824863851\\
0.241379310344828	0.995960616006059\\
0.25	0.996321275291232\\
    };
    \addlegendentry{\textcolor{black}{\spidermatch: {0.041}}}
    \addplot [color=ourcolor, smooth, line width=\pckLineWidth]
    table[row sep=crcr]{%
0	0.0509785580659088\\
0.00862068965517241	0.0573823571736941\\
0.0172413793103448	0.16988055835372\\
0.0258620689655172	0.346164915815225\\
0.0344827586206897	0.497517628435746\\
0.0431034482758621	0.64836667146352\\
0.0517241379310345	0.76428263059433\\
0.0603448275862069	0.842279464671176\\
0.0689655172413793	0.895092819110663\\
0.0775862068965517	0.929558209814362\\
0.0862068965517241	0.950712332709742\\
0.0948275862068965	0.963699812922723\\
0.103448275862069	0.972046337602533\\
0.112068965517241	0.977622679522233\\
0.120689655172414	0.982623399050223\\
0.129310344827586	0.985321629011369\\
0.137931034482759	0.987300330982875\\
0.146551724137931	0.989315009353864\\
0.155172413793103	0.991185782126925\\
0.163793103448276	0.992552885307238\\
0.172413793103448	0.993812059289106\\
0.181034482758621	0.995071233270974\\
0.189655172413793	0.996222478054396\\
0.198275862068966	0.996977982443517\\
0.206896551724138	0.997337746438337\\
0.21551724137931	0.997625557634192\\
0.224137931034483	0.99776946323212\\
0.232758620689655	0.997949345229529\\
0.241379310344828	0.998129227226939\\
0.25	0.998309109224349\\
    };
    \addlegendentry{\textcolor{black}{\ours: \textbf{0.039}}}
	\end{axis}
\end{tikzpicture}
         & 
         \hspace{-1.225cm}
         \newcommand{\pckLineWidth}{1.5pt}
\newcommand{\plotWidth}{0.9\columnwidth}
\newcommand{\plotHeight}{0.75\columnwidth}
\newcommand{\pckTitle}{$\%$ of  Corr. Points: BeCoS}

\pgfplotsset{%
    label style = {font=\large},%
    tick label style = {font=\large},
    title style =  {font=\Large},
    legend style={  fill= gray!10,
                    fill opacity=0.6, 
                    font=\large,
                    draw=gray!20, %
                    text opacity=1}
}
\begin{tikzpicture}[scale=0.5, transform shape]
	\begin{axis}[
		width=\plotWidth,
		height=\plotHeight,
		grid=major,
		title=\pckTitle,
		legend style={
			at={(0.97,0.03)},
			anchor=south east,
			legend columns=1},
		legend cell align={left},
        title style={yshift=-0.25cm},
        xlabel={Geodesic Error Threshold},
        xmin=0,
        xmax=0.25,
        ylabel near ticks,
        xtick={0, 0.05, 0.1, 0.15, 0.2, 0.25},
        xticklabels={$0$, $0.05$, $0.1$, $0.15$, $0.2$, $0.25$},
        ymin=0,
        ymax=1,
        ytick={0, 0.20, 0.40, 0.60, 0.80, 1},
        yticklabels={$0$, $20$, $40$, $60$, $80$, $100$},
	]
    \addplot [color=ulrssmcolour, smooth, line width=\pckLineWidth]
    table[row sep=crcr]{%
0	0.0363659491470326\\
0.0101010101010101	0.0627252554661993\\
0.0202020202020202	0.181377468093226\\
0.0303030303030303	0.312325979110426\\
0.0404040404040404	0.451853631648128\\
0.0505050505050505	0.581713849589417\\
0.0606060606060606	0.686854267663562\\
0.0707070707070707	0.771613924497901\\
0.0808080808080808	0.832869277627804\\
0.0909090909090909	0.879892018712987\\
0.101010101010101	0.912781083487626\\
0.111111111111111	0.935861376902746\\
0.121212121212121	0.951606292312693\\
0.131313131313131	0.962616426158608\\
0.141414141414141	0.969909403134849\\
0.151515151515152	0.975011660282956\\
0.161616161616162	0.978333074215934\\
0.171717171717172	0.980509660367758\\
0.181818181818182	0.982262236230266\\
0.191919191919192	0.983491866069283\\
0.202020202020202	0.984071346798016\\
0.212121212121212	0.9844529560584\\
0.222222222222222	0.984862832671406\\
0.232323232323232	0.985258575608101\\
0.242424242424242	0.985541249134312\\
0.252525252525253	0.985852190013144\\
    };
    \addlegendentry{\textcolor{black}{\caoetal: {0.057}}}
    \addplot [color=cPLOT1, smooth, line width=\pckLineWidth]
    table[row sep=crcr]{%
0	0.00528792164778462\\
0.0101010101010101	0.00935336678559752\\
0.0202020202020202	0.0272072097684402\\
0.0303030303030303	0.0483446815164395\\
0.0404040404040404	0.0722398328334447\\
0.0505050505050505	0.0972010973858903\\
0.0606060606060606	0.12155112368335\\
0.0707070707070707	0.145289911725824\\
0.0808080808080808	0.166470028003241\\
0.0909090909090909	0.188076589574834\\
0.101010101010101	0.207451420773572\\
0.111111111111111	0.225276834070136\\
0.121212121212121	0.243088032523561\\
0.131313131313131	0.260785512231873\\
0.141414141414141	0.276379195155581\\
0.151515151515152	0.291958663236151\\
0.161616161616162	0.30826308831682\\
0.171717171717172	0.32277644316195\\
0.181818181818182	0.337247153477661\\
0.191919191919192	0.35140513724431\\
0.202020202020202	0.365776343658048\\
0.212121212121212	0.380275483660038\\
0.222222222222222	0.393978592446232\\
0.232323232323232	0.407312115310808\\
0.242424242424242	0.421626462351988\\
0.252525252525253	0.434860481314589\\
    };
    \addlegendentry{\textcolor{black}{\discreteopt: 0.37}}
    \addplot [color=cPLOT3, smooth, line width=\pckLineWidth]
    table[row sep=crcr]{%
0	0.00898876404494382\\
0.0101010101010101	0.0163134689233395\\
0.0202020202020202	0.0480728203669464\\
0.0303030303030303	0.0871568766889489\\
0.0404040404040404	0.134490115204096\\
0.0505050505050505	0.185407481154886\\
0.0606060606060606	0.236509742568625\\
0.0707070707070707	0.284170103825914\\
0.0808080808080808	0.326184042099275\\
0.0909090909090909	0.364912530223297\\
0.101010101010101	0.399729768169535\\
0.111111111111111	0.43060731048215\\
0.121212121212121	0.457914948087043\\
0.131313131313131	0.481041103683687\\
0.141414141414141	0.501735172806144\\
0.151515151515152	0.519726923623951\\
0.161616161616162	0.535997724363533\\
0.171717171717172	0.55073247048784\\
0.181818181818182	0.56286445740293\\
0.191919191919192	0.57445598065709\\
0.202020202020202	0.586033281183331\\
0.212121212121212	0.596686104394823\\
0.222222222222222	0.605902432086474\\
0.232323232323232	0.615289432513156\\
0.242424242424242	0.623936851088039\\
0.252525252525253	0.63369364244062\\
    };
    \addlegendentry{\textcolor{black}{\smooths: 0.27}}
    \addplot [color=cPLOT5, smooth, line width=\pckLineWidth]
    table[row sep=crcr]{%
0	0.0349077136154216\\
0.0101010101010101	0.0599508182820317\\
0.0202020202020202	0.174566833432262\\
0.0303030303030303	0.301492410752141\\
0.0404040404040404	0.440925407727748\\
0.0505050505050505	0.571412419797055\\
0.0606060606060606	0.678425054410809\\
0.0707070707070707	0.764351734079539\\
0.0808080808080808	0.827326945363068\\
0.0909090909090909	0.874473557760253\\
0.101010101010101	0.908957291048362\\
0.111111111111111	0.932756720088188\\
0.121212121212121	0.9497159331807\\
0.131313131313131	0.961347126826649\\
0.141414141414141	0.969798468017751\\
0.151515151515152	0.974702507137002\\
0.161616161616162	0.978348737951892\\
0.171717171717172	0.980723027784844\\
0.181818181818182	0.982334153028633\\
0.191919191919192	0.983634359365725\\
0.202020202020202	0.984510585375505\\
0.212121212121212	0.985104157833743\\
0.222222222222222	0.985570536193787\\
0.232323232323232	0.985966251165946\\
0.242424242424242	0.986305435427796\\
0.252525252525253	0.986842477175726\\
    };
    \addlegendentry{\textcolor{black}{\spidermatch: {0.057}}}
    \addplot [color=ourcolor, smooth, line width=\pckLineWidth]
    table[row sep=crcr]{%
0	0.034139724169116\\
0.0101010101010101	0.0597586479764865\\
0.0202020202020202	0.174075853493104\\
0.0303030303030303	0.301817205516618\\
0.0404040404040404	0.441075062174994\\
0.0505050505050505	0.572561044539905\\
0.0606060606060606	0.679219421207325\\
0.0707070707070707	0.764003504408772\\
0.0808080808080808	0.826856771422112\\
0.0909090909090909	0.874152159167985\\
0.101010101010101	0.90880058783631\\
0.111111111111111	0.932610784535383\\
0.121212121212121	0.949313248926068\\
0.131313131313131	0.961225412615872\\
0.141414141414141	0.969548383450147\\
0.151515151515152	0.974734343205969\\
0.161616161616162	0.978464842866832\\
0.171717171717172	0.980838797196473\\
0.181818181818182	0.982633393624237\\
0.191919191919192	0.984159507121863\\
0.202020202020202	0.985120393398146\\
0.212121212121212	0.985657359258422\\
0.222222222222222	0.986180194438164\\
0.232323232323232	0.986703029617906\\
0.242424242424242	0.987254126158716\\
0.252525252525253	0.987890006782727\\
    };
    \addlegendentry{\textcolor{black}{\ours: \textbf{0.056}}}
	\end{axis}
\end{tikzpicture}
    \end{tabular}
    \caption{\textbf{Quantitative shape matching results} on datasets FAUST, SMAL, DT4D Intra and Inter and BeCoS. 
    We show percentage of correct points ($\uparrow$) w.r.t.~geodesic error thresholds (i.e.~this quantifies if a matched point is within a geodesic threshold radius around ground truth point). 
    Numbers in legends are mean geodesic errors ($\downarrow$).
    Across all five datasets our method performs the best, very likely due to enforced geometric consistency.
    }
    \label{fig:pck}
\end{figure*}
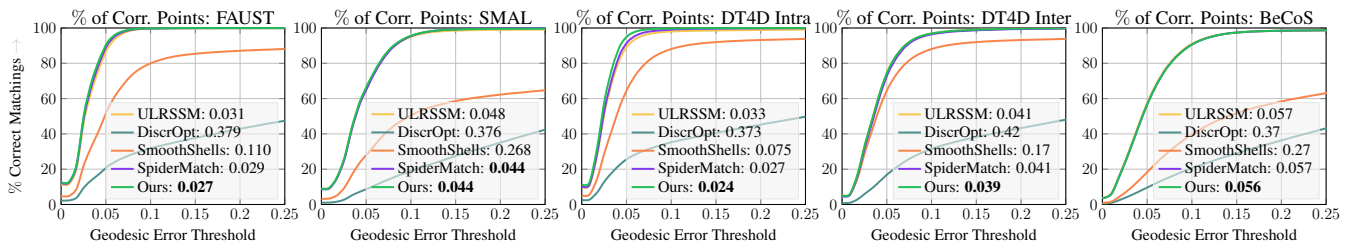

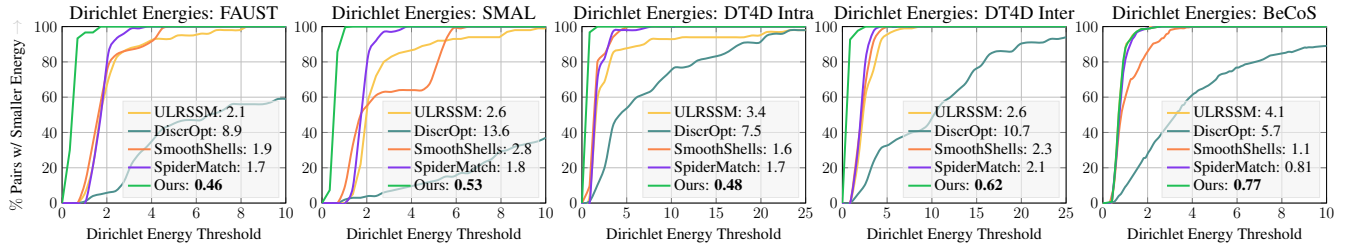
\begin{figure*}[t]
    \centering
    \begin{tabular}{ccccc}
        \hspace{-0.87cm}
         \newcommand{\pckLineWidth}{1.5pt}
\newcommand{\plotWidth}{0.9\columnwidth}
\newcommand{\plotHeight}{0.75\columnwidth}
\newcommand{\pckTitle}{Dirichlet Energies: FAUST}

\pgfplotsset{%
    label style = {font=\large},
    tick label style = {font=\large},
    title style =  {font=\Large},
    legend style={  fill= gray!10,
                    fill opacity=0.6, 
                    font=\large,
                    draw=gray!20, %
                    text opacity=1}
}
\begin{tikzpicture}[scale=0.5, transform shape]
	\begin{axis}[
		width=\plotWidth,
		height=\plotHeight,
		grid=major,
		title=\pckTitle,
		legend style={
			at={(0.97,0.03)},
			anchor=south east,
			legend columns=1},
		legend cell align={left},
        title style={yshift=-0.25cm},
	ylabel={{\large$\%$ Pairs w/ Smaller Energy \textcolor{gray!40}{$\rightarrow$}}},
        xlabel={Dirichlet Energy Threshold},
        xmin=0,
        xmax=10,
        ylabel near ticks,
        xtick={0, 2, 4, 6, 8, 10},
        xticklabels={$0$, $2$, $4$, $6$, $8$, $10$},
        ymin=0,
        ymax=1,
        ytick={0, 0.20, 0.40, 0.60, 0.80, 1},
        yticklabels={$0$, $20$, $40$, $60$, $80$, $100$},
	]
    \addplot [color=ulrssmcolour, smooth, line width=\pckLineWidth]
    table[row sep=crcr]{%
0	0\\
0.344827586206897	0\\
0.689655172413793	0\\
1.03448275862069	0.06\\
1.37931034482759	0.21\\
1.72413793103448	0.51\\
2.06896551724138	0.7\\
2.41379310344828	0.82\\
2.75862068965517	0.86\\
3.10344827586207	0.88\\
3.44827586206897	0.89\\
3.79310344827586	0.91\\
4.13793103448276	0.93\\
4.48275862068965	0.93\\
4.82758620689655	0.93\\
5.17241379310345	0.95\\
5.51724137931035	0.95\\
5.86206896551724	0.95\\
6.20689655172414	0.96\\
6.55172413793103	0.96\\
6.89655172413793	0.98\\
7.24137931034483	0.98\\
7.58620689655172	0.98\\
7.93103448275862	0.98\\
8.27586206896552	1\\
8.62068965517241	1\\
8.96551724137931	1\\
9.31034482758621	1\\
9.6551724137931	1\\
10	1\\
    };
    \addlegendentry{\textcolor{black}{\caoetal: 2.1}}
    \addplot [color=cPLOT1, smooth, line width=\pckLineWidth]
    table[row sep=crcr]{%
0	0\\
0.344827586206897	0\\
0.689655172413793	0\\
1.03448275862069	0.01\\
1.37931034482759	0.04\\
1.72413793103448	0.05\\
2.06896551724138	0.06\\
2.41379310344828	0.07\\
2.75862068965517	0.12\\
3.10344827586207	0.22\\
3.44827586206897	0.28\\
3.79310344827586	0.32\\
4.13793103448276	0.37\\
4.48275862068965	0.42\\
4.82758620689655	0.44\\
5.17241379310345	0.46\\
5.51724137931035	0.47\\
5.86206896551724	0.47\\
6.20689655172414	0.47\\
6.55172413793103	0.5\\
6.89655172413793	0.51\\
7.24137931034483	0.53\\
7.58620689655172	0.56\\
7.93103448275862	0.56\\
8.27586206896552	0.56\\
8.62068965517241	0.56\\
8.96551724137931	0.56\\
9.31034482758621	0.57\\
9.6551724137931	0.59\\
10	0.59\\
    };
    \addlegendentry{\textcolor{black}{\discreteopt: {8.9}}}
    \addplot [color=cPLOT3, smooth, line width=\pckLineWidth]
    table[row sep=crcr]{%
0	0\\
0.344827586206897	0\\
0.689655172413793	0\\
1.03448275862069	0.12\\
1.37931034482759	0.34\\
1.72413793103448	0.57\\
2.06896551724138	0.78\\
2.41379310344828	0.84\\
2.75862068965517	0.86\\
3.10344827586207	0.87\\
3.44827586206897	0.88\\
3.79310344827586	0.9\\
4.13793103448276	0.93\\
4.48275862068965	0.99\\
4.82758620689655	1\\
5.17241379310345	1\\
5.51724137931035	1\\
5.86206896551724	1\\
6.20689655172414	1\\
6.55172413793103	1\\
6.89655172413793	1\\
7.24137931034483	1\\
7.58620689655172	1\\
7.93103448275862	1\\
8.27586206896552	1\\
8.62068965517241	1\\
8.96551724137931	1\\
9.31034482758621	1\\
9.6551724137931	1\\
10	1\\
    };
    \addlegendentry{\textcolor{black}{\smooths: 1.9}}
    \addplot [color=cPLOT5, smooth, line width=\pckLineWidth]
    table[row sep=crcr]{%
0	0\\
0.344827586206897	0\\
0.689655172413793	0\\
1.03448275862069	0\\
1.37931034482759	0.21\\
1.72413793103448	0.49\\
2.06896551724138	0.85\\
2.41379310344828	0.93\\
2.75862068965517	0.96\\
3.10344827586207	0.99\\
3.44827586206897	0.99\\
3.79310344827586	1\\
4.13793103448276	1\\
4.48275862068965	1\\
4.82758620689655	1\\
9.6551724137931	1\\
10	1\\
    };
    \addlegendentry{\textcolor{black}{\spidermatch: {1.7}}}

    \addplot [color=ourcolor, line width=\pckLineWidth]
    table[row sep=crcr]{%
0	0\\
0.344827586206897	0.3\\
0.689655172413793	0.933333333333333\\
1.03448275862069	0.966666666666667\\
1.37931034482759	0.966666666666667\\
1.72413793103448	1\\
2.06896551724138	1\\
2.41379310344828	1\\
2.75862068965517	1\\
10	1\\
    };
    \addlegendentry{\textcolor{black}{\ours: \textbf{0.46}}}
	\end{axis}
\end{tikzpicture}
         & 
         \hspace{-1.15cm}
         \newcommand{\pckLineWidth}{1.5pt}
\newcommand{\plotWidth}{0.9\columnwidth}
\newcommand{\plotHeight}{0.75\columnwidth}
\newcommand{\pckTitle}{Dirichlet Energies: SMAL}

\pgfplotsset{%
    label style = {font=\large},
    tick label style = {font=\large},
    title style =  {font=\Large},
    legend style={  fill= gray!10,
                    fill opacity=0.6, 
                    font=\large,
                    draw=gray!20, %
                    text opacity=1}
}
\begin{tikzpicture}[scale=0.5, transform shape]
	\begin{axis}[
		width=\plotWidth,
		height=\plotHeight,
		grid=major,
		title=\pckTitle,
		legend style={
			at={(0.97,0.03)},
			anchor=south east,
			legend columns=1},
		legend cell align={left},
        title style={yshift=-0.25cm},
        xlabel={Dirichlet Energy Threshold},
        xmin=0,
        xmax=10,
        ylabel near ticks,
        xtick={0, 2, 4, 6, 8, 10},
        xticklabels={$0$, $2$, $4$, $6$, $8$, $10$},
        ymin=0,
        ymax=1,
        ytick={0, 0.20, 0.40, 0.60, 0.80, 1},
        yticklabels={$0$, $20$, $40$, $60$, $80$, $100$},
	]
\addplot [color=cPLOT2, ulrssmcolour, line width=\pckLineWidth]
    table[row sep=crcr]{%
0	0\\
0.344827586206897	0\\
0.689655172413793	0\\
1.03448275862069	0\\
1.37931034482759	0.05\\
1.72413793103448	0.22\\
2.06896551724138	0.59\\
2.41379310344828	0.72\\
2.75862068965517	0.8\\
3.10344827586207	0.83\\
3.44827586206897	0.85\\
3.79310344827586	0.86\\
4.13793103448276	0.87\\
4.48275862068965	0.89\\
4.82758620689655	0.9\\
5.17241379310345	0.92\\
5.51724137931035	0.92\\
5.86206896551724	0.93\\
6.20689655172414	0.93\\
6.55172413793103	0.94\\
6.89655172413793	0.94\\
7.24137931034483	0.94\\
7.58620689655172	0.94\\
7.93103448275862	0.94\\
8.27586206896552	0.97\\
8.62068965517241	0.98\\
8.96551724137931	0.98\\
9.31034482758621	0.98\\
9.6551724137931	0.99\\
10	0.99\\
    };
    \addlegendentry{\textcolor{black}{\caoetal: 2.6}}
    \addplot [color=cPLOT1, smooth, line width=\pckLineWidth]
    table[row sep=crcr]{%
0	0\\
0.344827586206897	0\\
0.689655172413793	0\\
1.03448275862069	0.02\\
1.37931034482759	0.03\\
1.72413793103448	0.03\\
2.06896551724138	0.04\\
2.41379310344828	0.04\\
2.75862068965517	0.06\\
3.10344827586207	0.07\\
3.44827586206897	0.08\\
3.79310344827586	0.09\\
4.13793103448276	0.1\\
4.48275862068965	0.11\\
4.82758620689655	0.12\\
5.17241379310345	0.13\\
5.51724137931035	0.15\\
5.86206896551724	0.15\\
6.20689655172414	0.17\\
6.55172413793103	0.17\\
6.89655172413793	0.2\\
7.24137931034483	0.23\\
7.58620689655172	0.25\\
7.93103448275862	0.29\\
8.27586206896552	0.3\\
8.62068965517241	0.32\\
8.96551724137931	0.33\\
9.31034482758621	0.34\\
9.6551724137931	0.35\\
10	0.37\\
    };
    \addlegendentry{\textcolor{black}{\discreteopt: {13.6}}}
    \addplot [color=cPLOT3, smooth, line width=\pckLineWidth]
    table[row sep=crcr]{%
0	0\\
0.344827586206897	0\\
0.689655172413793	0\\
1.03448275862069	0.1\\
1.37931034482759	0.33\\
1.72413793103448	0.5\\
2.06896551724138	0.56\\
2.41379310344828	0.61\\
2.75862068965517	0.63\\
3.10344827586207	0.63\\
3.44827586206897	0.64\\
3.79310344827586	0.64\\
4.13793103448276	0.64\\
4.48275862068965	0.64\\
4.82758620689655	0.68\\
5.17241379310345	0.8\\
5.51724137931035	0.92\\
5.86206896551724	0.99\\
6.20689655172414	0.99\\
6.55172413793103	1\\
6.89655172413793	1\\
7.24137931034483	1\\
7.58620689655172	1\\
7.93103448275862	1\\
8.27586206896552	1\\
8.62068965517241	1\\
8.96551724137931	1\\
9.31034482758621	1\\
9.6551724137931	1\\
10	1\\
    };
    \addlegendentry{\textcolor{black}{\smooths: 2.8}}
    \addplot [color=cPLOT5, smooth, line width=\pckLineWidth]
    table[row sep=crcr]{%
0	0\\
0.344827586206897	0\\
0.689655172413793	0\\
1.03448275862069	0\\
1.37931034482759	0.09\\
1.72413793103448	0.47\\
2.06896551724138	0.84\\
2.41379310344828	0.92\\
2.75862068965517	0.97\\
3.10344827586207	0.97\\
3.44827586206897	0.98\\
3.79310344827586	1\\
4.13793103448276	1\\
4.48275862068965	1\\
4.82758620689655	1\\
5.17241379310345	1\\
5.51724137931035	1\\
5.86206896551724	1\\
6.20689655172414	1\\
6.55172413793103	1\\
6.89655172413793	1\\
7.24137931034483	1\\
7.58620689655172	1\\
7.93103448275862	1\\
8.27586206896552	1\\
8.62068965517241	1\\
8.96551724137931	1\\
9.31034482758621	1\\
9.6551724137931	1\\
10	1\\
    };
    \addlegendentry{\textcolor{black}{\spidermatch: {1.8}}}

    \addplot [color=ourcolor, line width=\pckLineWidth]
    table[row sep=crcr]{%
0	0\\
0.344827586206897	0.0733333333333333\\
0.689655172413793	0.856666666666667\\
1.03448275862069	0.996666666666667\\
1.37931034482759	1\\
1.72413793103448	1\\
2.06896551724138	1\\
2.41379310344828	1\\
2.75862068965517	1\\
3.10344827586207	1\\
3.44827586206897	1\\
9.6551724137931	1\\
10	1\\
    };
    \addlegendentry{\textcolor{black}{\ours: \textbf{0.53}}}
    
	\end{axis}
\end{tikzpicture}
         & 
         \hspace{-1.15cm}
         \newcommand{\pckLineWidth}{1.5pt}
\newcommand{\plotWidth}{0.9\columnwidth}
\newcommand{\plotHeight}{0.75\columnwidth}
\newcommand{\pckTitle}{Dirichlet Energies: DT4D Intra}

\pgfplotsset{%
    label style = {font=\large},
    tick label style = {font=\large},
    title style =  {font=\Large},
    legend style={  fill= gray!10,
                    fill opacity=0.6, 
                    font=\large,
                    draw=gray!20, %
                    text opacity=1}
}
\begin{tikzpicture}[scale=0.5, transform shape]
	\begin{axis}[
		width=\plotWidth,
		height=\plotHeight,
		grid=major,
		title=\pckTitle,
		legend style={
			at={(0.97,0.03)},
			anchor=south east,
			legend columns=1},
		legend cell align={left},
        title style={yshift=-0.25cm},
        xlabel={Dirichlet Energy Threshold},
        xmin=0,
        xmax=25,
        ylabel near ticks,
        xtick={0, 5, 10, 15, 20 , 25},
        ymin=0,
        ymax=1,
        ytick={0, 0.20, 0.40, 0.60, 0.80, 1},
        yticklabels={$0$, $20$, $40$, $60$, $80$, $100$},
	]
    \addplot [color=ulrssmcolour, smooth, line width=\pckLineWidth]
    table[row sep=crcr]{%
0	0\\
0.862068965517241	0.0404040404040404\\
1.72413793103448	0.585858585858586\\
2.58620689655172	0.696969696969697\\
3.44827586206897	0.848484848484849\\
4.31034482758621	0.868686868686869\\
5.17241379310345	0.878787878787879\\
6.03448275862069	0.888888888888889\\
6.89655172413793	0.898989898989899\\
7.75862068965517	0.929292929292929\\
8.62068965517241	0.929292929292929\\
9.48275862068965	0.929292929292929\\
10.3448275862069	0.929292929292929\\
11.2068965517241	0.939393939393939\\
12.0689655172414	0.939393939393939\\
12.9310344827586	0.939393939393939\\
13.7931034482759	0.939393939393939\\
14.6551724137931	0.939393939393939\\
15.5172413793103	0.939393939393939\\
16.3793103448276	0.939393939393939\\
17.2413793103448	0.939393939393939\\
18.1034482758621	0.939393939393939\\
18.9655172413793	0.94949494949495\\
19.8275862068966	0.94949494949495\\
20.6896551724138	0.96969696969697\\
21.551724137931	0.96969696969697\\
22.4137931034483	0.96969696969697\\
23.2758620689655	0.97979797979798\\
24.1379310344828	0.97979797979798\\
25	0.97979797979798\\
    };
    \addlegendentry{\textcolor{black}{\caoetal: 3.4}}
    \addplot [color=cPLOT1, smooth, line width=\pckLineWidth]
    table[row sep=crcr]{%
0	0\\
0.862068965517241	0.0101010101010101\\
1.72413793103448	0.101010101010101\\
2.58620689655172	0.212121212121212\\
3.44827586206897	0.434343434343434\\
4.31034482758621	0.505050505050505\\
5.17241379310345	0.545454545454545\\
6.03448275862069	0.595959595959596\\
6.89655172413793	0.616161616161616\\
7.75862068965517	0.636363636363636\\
8.62068965517241	0.686868686868687\\
9.48275862068965	0.727272727272727\\
10.3448275862069	0.767676767676768\\
11.2068965517241	0.767676767676768\\
12.0689655172414	0.777777777777778\\
12.9310344827586	0.808080808080808\\
13.7931034482759	0.818181818181818\\
14.6551724137931	0.828282828282828\\
15.5172413793103	0.838383838383838\\
16.3793103448276	0.858585858585859\\
17.2413793103448	0.888888888888889\\
18.1034482758621	0.909090909090909\\
18.9655172413793	0.909090909090909\\
19.8275862068966	0.909090909090909\\
20.6896551724138	0.94949494949495\\
21.551724137931	0.95959595959596\\
22.4137931034483	0.95959595959596\\
23.2758620689655	0.97979797979798\\
24.1379310344828	0.97979797979798\\
25	0.97979797979798\\
    };
    \addlegendentry{\textcolor{black}{\discreteopt: {7.5}}}
    \addplot [color=cPLOT3, smooth, tension=0.1, line width=\pckLineWidth]
    table[row sep=crcr]{%
0	0\\
0.862068965517241	0.101010101010101\\
1.72413793103448	0.797979797979798\\
2.58620689655172	0.848484848484849\\
3.44827586206897	0.939393939393939\\
4.31034482758621	0.97979797979798\\
5.17241379310345	1\\
6.03448275862069	1\\
6.89655172413793	1\\
7.75862068965517	1\\
25	1\\
    };
    \addlegendentry{\textcolor{black}{\smooths: {1.6}}}
    \addplot [color=cPLOT5, smooth, line width=\pckLineWidth]
    table[row sep=crcr]{%
0	0\\
0.862068965517241	0\\
1.72413793103448	0.696969696969697\\
2.58620689655172	0.828282828282828\\
3.44827586206897	0.96969696969697\\
4.31034482758621	0.97979797979798\\
5.17241379310345	0.97979797979798\\
6.03448275862069	0.97979797979798\\
6.89655172413793	0.98989898989899\\
7.75862068965517	1\\
8.62068965517241	1\\
9.48275862068965	1\\
10.3448275862069	1\\
11.2068965517241	1\\
25	1\\
    };
    \addlegendentry{\textcolor{black}{\spidermatch: {1.7}}}
    \addplot [color=ourcolor, line width=\pckLineWidth]
    table[row sep=crcr]{%
0	0\\
0.862068965517241	0.967391304347826\\
1.72413793103448	1\\
2.58620689655172	1\\
3.44827586206897	1\\
25	1\\
    };
    \addlegendentry{\textcolor{black}{\ours: \textbf{0.48}}}        
	\end{axis}
\end{tikzpicture}
         & 
         \hspace{-1.15cm}
         \newcommand{\pckLineWidth}{1.5pt}
\newcommand{\plotWidth}{0.9\columnwidth}
\newcommand{\plotHeight}{0.75\columnwidth}
\newcommand{\pckTitle}{Dirichlet Energies: DT4D Inter}

\pgfplotsset{%
    label style = {font=\large},
    tick label style = {font=\large},
    title style =  {font=\Large},
    legend style={  fill= gray!10,
                    fill opacity=0.6, 
                    font=\large,
                    draw=gray!20, %
                    text opacity=1}
}
\begin{tikzpicture}[scale=0.5, transform shape]
	\begin{axis}[
		width=\plotWidth,
		height=\plotHeight,
		grid=major,
		title=\pckTitle,
		legend style={
			at={(0.97,0.03)},
			anchor=south east,
			legend columns=1},
		legend cell align={left},
        title style={yshift=-0.25cm},
        xlabel={Dirichlet Energy Threshold},
        xmin=0,
        xmax=25,
        ylabel near ticks,
        xtick={0, 5, 10, 15, 20 , 25},
        ymin=0,
        ymax=1,
        ytick={0, 0.20, 0.40, 0.60, 0.80, 1},
        yticklabels={$0$, $20$, $40$, $60$, $80$, $100$},
	]
    \addplot [color=ulrssmcolour, smooth, line width=\pckLineWidth]
    table[row sep=crcr]{%
0	0\\
0.862068965517241	0\\
1.72413793103448	0.23\\
2.58620689655172	0.61\\
3.44827586206897	0.75\\
4.31034482758621	0.92\\
5.17241379310345	0.96\\
6.03448275862069	0.98\\
6.89655172413793	0.99\\
7.75862068965517	0.99\\
8.62068965517241	1\\
9.48275862068965	1\\
10.3448275862069	1\\
25	1\\
    };
    \addlegendentry{\textcolor{black}{\caoetal: 2.6}}
    \addplot [color=cPLOT1, smooth, line width=\pckLineWidth]
    table[row sep=crcr]{%
0	0\\
0.862068965517241	0\\
1.72413793103448	0.05\\
2.58620689655172	0.11\\
3.44827586206897	0.23\\
4.31034482758621	0.31\\
5.17241379310345	0.33\\
6.03448275862069	0.36\\
6.89655172413793	0.38\\
7.75862068965517	0.4\\
8.62068965517241	0.41\\
9.48275862068965	0.44\\
10.3448275862069	0.51\\
11.2068965517241	0.57\\
12.0689655172414	0.6\\
12.9310344827586	0.64\\
13.7931034482759	0.68\\
14.6551724137931	0.75\\
15.5172413793103	0.78\\
16.3793103448276	0.84\\
17.2413793103448	0.86\\
18.1034482758621	0.86\\
18.9655172413793	0.86\\
19.8275862068966	0.9\\
20.6896551724138	0.91\\
21.551724137931	0.91\\
22.4137931034483	0.91\\
23.2758620689655	0.93\\
24.1379310344828	0.93\\
25	0.94\\
    };
    \addlegendentry{\textcolor{black}{\discreteopt: {10.7}}}
    \addplot [color=cPLOT3, smooth, line width=\pckLineWidth]
    table[row sep=crcr]{%
0	0\\
0.862068965517241	0\\
1.72413793103448	0.32\\
2.58620689655172	0.68\\
3.44827586206897	0.91\\
4.31034482758621	0.98\\
5.17241379310345	1\\
6.03448275862069	1\\
6.89655172413793	1\\
7.75862068965517	1\\
8.62068965517241	1\\
25	1\\
    };
    \addlegendentry{\textcolor{black}{\smooths:  {2.3}}}
    \addplot [color=cPLOT5, smooth, line width=\pckLineWidth]
    table[row sep=crcr]{%
0	0\\
0.862068965517241	0\\
1.72413793103448	0.3\\
2.58620689655172	0.81\\
3.44827586206897	0.97\\
4.31034482758621	1\\
5.17241379310345	1\\
6.03448275862069	1\\
6.89655172413793	1\\
25	1\\
    };
    \addlegendentry{\textcolor{black}{\spidermatch: {2.1}}}
    \addplot [color=ourcolor, line width=\pckLineWidth]
    table[row sep=crcr]{%
0	0\\
0.862068965517241	0.926666666666667\\
1.72413793103448	0.976666666666667\\
2.58620689655172	1\\
3.44827586206897	1\\
4.31034482758621	1\\
25	1\\
    };
    \addlegendentry{\textcolor{black}{\ours: \textbf{0.62}}}        
	\end{axis}
\end{tikzpicture}
         & 
         \hspace{-1.15cm}
         \newcommand{\pckLineWidth}{1.5pt}
\newcommand{\plotWidth}{0.9\columnwidth}
\newcommand{\plotHeight}{0.75\columnwidth}
\newcommand{\pckTitle}{Dirichlet Energies: BeCoS}

\pgfplotsset{%
    label style = {font=\large},
    tick label style = {font=\large},
    title style =  {font=\Large},
    legend style={  fill= gray!10,
                    fill opacity=0.6, 
                    font=\large,
                    draw=gray!20, %
                    text opacity=1}
}
\begin{tikzpicture}[scale=0.5, transform shape]
	\begin{axis}[
		width=\plotWidth,
		height=\plotHeight,
		grid=major,
		title=\pckTitle,
		legend style={
			at={(0.97,0.03)},
			anchor=south east,
			legend columns=1},
		legend cell align={left},
        title style={yshift=-0.25cm},
        xlabel={Dirichlet Energy Threshold},
        xmin=0,
        xmax=10,
        ylabel near ticks,
        xtick={0, 2, 4, 6, 8, 10},
        ymin=0,
        ymax=1,
        ytick={0, 0.20, 0.40, 0.60, 0.80, 1},
        yticklabels={$0$, $20$, $40$, $60$, $80$, $100$},
	]
    \addplot [color=ulrssmcolour, smooth, line width=\pckLineWidth]
    table[row sep=crcr]{%
0	0\\
0.101010101010101	0\\
0.202020202020202	0\\
0.303030303030303	0\\
0.404040404040404	0.00704225352112676\\
0.505050505050505	0.124413145539906\\
0.606060606060606	0.291079812206573\\
0.707070707070707	0.504694835680751\\
0.808080808080808	0.65962441314554\\
1.01010101010101	0.880281690140845\\
1.11111111111111	0.908450704225352\\
1.21212121212121	0.924882629107981\\
1.31313131313131	0.948356807511737\\
1.41414141414141	0.960093896713615\\
1.61616161616162	0.97887323943662\\
1.71717171717172	0.985915492957746\\
1.81818181818182	0.985915492957746\\
1.91919191919192	0.985915492957746\\
2.02020202020202	0.988262910798122\\
2.22222222222222	0.992957746478873\\
2.32323232323232	0.995305164319249\\
2.42424242424242	0.997652582159624\\
2.52525252525253	1\\
2.62626262626263	1\\
10	1\\
    };
    \addlegendentry{\textcolor{black}{\caoetal: 4.1}}
    \addplot [color=cPLOT1, smooth, line width=\pckLineWidth]
    table[row sep=crcr]{%
0	0\\
0.101010101010101	0\\
0.202020202020202	0\\
0.303030303030303	0\\
0.404040404040404	0.00469483568075117\\
0.505050505050505	0.00704225352112676\\
0.606060606060606	0.0211267605633803\\
0.707070707070707	0.039906103286385\\
0.808080808080808	0.0516431924882629\\
0.909090909090909	0.0727699530516432\\
1.01010101010101	0.0938967136150235\\
1.11111111111111	0.110328638497653\\
1.71717171717172	0.230046948356808\\
1.81818181818182	0.253521126760563\\
1.91919191919192	0.258215962441315\\
2.02020202020202	0.284037558685446\\
2.12121212121212	0.302816901408451\\
2.22222222222222	0.326291079812207\\
2.32323232323232	0.347417840375587\\
2.42424242424242	0.36150234741784\\
2.52525252525253	0.375586854460094\\
2.62626262626263	0.394366197183099\\
3.23232323232323	0.495305164319249\\
3.33333333333333	0.516431924882629\\
3.43434343434343	0.525821596244131\\
3.53535353535354	0.551643192488263\\
3.63636363636364	0.55868544600939\\
3.73737373737374	0.57981220657277\\
4.04040404040404	0.617370892018779\\
4.14141414141414	0.624413145539906\\
4.24242424242424	0.636150234741784\\
4.34343434343434	0.643192488262911\\
4.44444444444444	0.647887323943662\\
4.54545454545455	0.654929577464789\\
4.64646464646465	0.661971830985915\\
4.74747474747475	0.669014084507042\\
5.05050505050505	0.699530516431925\\
5.15151515151515	0.71830985915493\\
5.25252525252525	0.727699530516432\\
5.35353535353535	0.732394366197183\\
5.45454545454545	0.734741784037559\\
5.55555555555556	0.73943661971831\\
5.65656565656566	0.746478873239437\\
5.75757575757576	0.746478873239437\\
5.85858585858586	0.755868544600939\\
5.95959595959596	0.767605633802817\\
6.06060606060606	0.767605633802817\\
6.16161616161616	0.774647887323944\\
6.26262626262626	0.779342723004695\\
6.36363636363636	0.786384976525822\\
6.46464646464646	0.788732394366197\\
6.96969696969697	0.805164319248826\\
7.07070707070707	0.809859154929577\\
7.17171717171717	0.814553990610329\\
7.27272727272727	0.821596244131455\\
7.37373737373737	0.828638497652582\\
7.47474747474747	0.835680751173709\\
7.57575757575758	0.838028169014085\\
7.67676767676768	0.84037558685446\\
7.77777777777778	0.842723004694836\\
7.87878787878788	0.845070422535211\\
7.97979797979798	0.847417840375587\\
8.08080808080808	0.849765258215962\\
8.18181818181818	0.852112676056338\\
8.28282828282828	0.856807511737089\\
8.38383838383838	0.856807511737089\\
8.48484848484848	0.856807511737089\\
8.98989898989899	0.880281690140845\\
9.09090909090909	0.880281690140845\\
9.19191919191919	0.882629107981221\\
9.29292929292929	0.882629107981221\\
9.39393939393939	0.882629107981221\\
9.49494949494949	0.884976525821596\\
9.8989898989899	0.889671361502347\\
10	0.889671361502347\\
    };
    \addlegendentry{\textcolor{black}{\discreteopt: {5.7}}}
    \addplot [color=cPLOT3, smooth, line width=\pckLineWidth]
    table[row sep=crcr]{%
0	0\\
0.101010101010101	0\\
0.202020202020202	0\\
0.303030303030303	0.00234741784037559\\
0.404040404040404	0.0469483568075117\\
0.505050505050505	0.110328638497653\\
0.606060606060606	0.248826291079812\\
0.707070707070707	0.387323943661972\\
0.808080808080808	0.485915492957746\\
0.909090909090909	0.553990610328638\\
1.21212121212121	0.690140845070423\\
1.31313131313131	0.704225352112676\\
1.41414141414141	0.71830985915493\\
1.51515151515152	0.741784037558685\\
1.61616161616162	0.767605633802817\\
1.71717171717172	0.800469483568075\\
1.81818181818182	0.81924882629108\\
1.91919191919192	0.842723004694836\\
2.22222222222222	0.89906103286385\\
2.32323232323232	0.906103286384977\\
2.42424242424242	0.910798122065728\\
2.52525252525253	0.917840375586854\\
2.62626262626263	0.931924882629108\\
2.72727272727273	0.936619718309859\\
2.82828282828283	0.955399061032864\\
2.92929292929293	0.962441314553991\\
3.03030303030303	0.981220657276995\\
3.43434343434343	0.990610328638498\\
3.53535353535354	0.990610328638498\\
3.63636363636364	0.990610328638498\\
3.73737373737374	0.995305164319249\\
3.83838383838384	0.995305164319249\\
3.93939393939394	0.997652582159624\\
4.04040404040404	0.997652582159624\\
4.14141414141414	0.997652582159624\\
4.24242424242424	0.997652582159624\\
4.34343434343434	0.997652582159624\\
4.44444444444444	1\\
10	1\\
    };
    \addlegendentry{\textcolor{black}{\smooths:  {1.1}}}
    \addplot [color=cPLOT5, line width=\pckLineWidth]
    table[row sep=crcr]{%
0	0\\
0.101010101010101	0\\
0.202020202020202	0\\
0.303030303030303	0\\
0.404040404040404	0.00469483568075117\\
0.606060606060606	0.262910798122066\\
0.707070707070707	0.450704225352113\\
0.909090909090909	0.737089201877934\\
1.01010101010101	0.814553990610329\\
1.11111111111111	0.86150234741784\\
1.31313131313131	0.927230046948357\\
1.41414141414141	0.943661971830986\\
1.51515151515152	0.967136150234742\\
1.61616161616162	0.974178403755869\\
1.81818181818182	0.983568075117371\\
1.91919191919192	0.988262910798122\\
2.12121212121212	0.990610328638498\\
2.22222222222222	0.995305164319249\\
2.42424242424242	0.997652582159624\\
2.52525252525253	0.997652582159624\\
2.62626262626263	1\\
2.72727272727273	1\\
10	1\\
    };
    \addlegendentry{\textcolor{black}{\spidermatch: {0.81}}}
    \addplot [color=ourcolor, line width=\pckLineWidth]
    table[row sep=crcr]{%
0	0\\
0.101010101010101	0\\
0.404040404040404	0.00704225352112676\\
0.505050505050505	0.124413145539906\\
0.606060606060606	0.291079812206573\\
0.707070707070707	0.504694835680751\\
0.909090909090909	0.805164319248826\\
1.01010101010101	0.880281690140845\\
1.11111111111111	0.908450704225352\\
1.21212121212121	0.924882629107981\\
1.41414141414141	0.960093896713615\\
1.51515151515152	0.971830985915493\\
1.61616161616162	0.97887323943662\\
1.71717171717172	0.985915492957746\\
1.91919191919192	0.985915492957746\\
2.02020202020202	0.988262910798122\\
2.12121212121212	0.990610328638498\\
2.32323232323232	0.995305164319249\\
2.42424242424242	0.997652582159624\\
2.52525252525253	1\\
2.62626262626263	1\\
2.72727272727273	1\\
10	1\\
    };
    \addlegendentry{\textcolor{black}{\ours: \textbf{0.77}}}       
	\end{axis}
\end{tikzpicture}
    \end{tabular}
    \caption{We show percentage of matched points ($\uparrow$) which are below a certain \textbf{Dirichlet Energy threshold} on datasets FAUST, SMAL, DT4D and BeCoS. Numbers in legends are mean Dirichlet energies across all pairs ($\downarrow$).
    Our method consistently yields best results on all datasets very likely since it is the only method enforcing \emph{global} geometric consistency.
    }
    \label{fig:dirchilet}
\end{figure*}

\textbf{Full Shape Matching.} We show error plots of geodesic errors in \cref{fig:pck} and Dirichlet energy plots in \cref{fig:dirchilet}. 
Furthermore, we show additional qualitative results, including results for methods by Ren~\etal~\cite{ren2021discrete} and Eisenberger~\etal~\cite{eisenberger2020smooth} in \cref{fig:supp:qualitative}.

\textbf{Partial-to-Full Shape Matching.} In \cref{fig:supp-partial-sm}, we show more qualitative results computed with our method for the partial-to-full shape matching setting {and furthermore in \cref{tab:partial-quant} we report quantitive results on SHREC’16~\cite{cosmo2016shrec}}.

\newcommand{\maxpartialheightsupp}{2cm}
\begin{figure}[b]
    \centering
    \setlength{\tabcolsep}{0pt}
    \renewcommand{\arraystretch}{1}
    \begin{tabular}{cccccccc}
        \hspace{-0.3cm}
        \includegraphics[width=0.2\columnwidth,height=\maxpartialheightsupp]{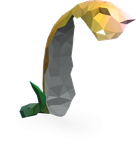}
         &
         \hspace{-0.5cm}
         \includegraphics[width=0.2\columnwidth,height=\maxpartialheightsupp]{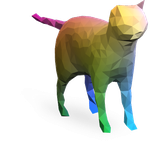}
         &
         \hspace{0.1cm}
         \rotatebox{90}{\textcolor{gray!50}{\rule{\maxpartialheightsupp}{0.2pt}}}
         &
         \includegraphics[width=0.2\columnwidth,height=\maxpartialheightsupp]{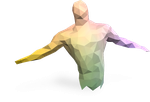}
         &
         \hspace{-0.5cm}
         \includegraphics[width=0.2\columnwidth,height=\maxpartialheightsupp]{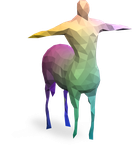}
         &
         \hspace{0.1cm}
         \rotatebox{90}{\textcolor{gray!50}{\rule{\maxpartialheightsupp}{0.2pt}}}
         &
         \includegraphics[width=0.2\columnwidth,height=1.5cm]{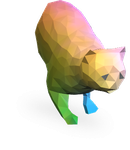}
         &
         \hspace{-0.7cm}
         \includegraphics[width=0.2\columnwidth,height=\maxpartialheightsupp]{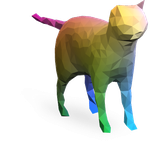}\\
         \hspace{-0.3cm}
         \includegraphics[width=0.2\columnwidth,height=1.7cm]{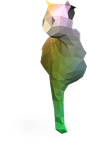}
         &
         \hspace{-0.5cm}
         \includegraphics[width=0.2\columnwidth,height=\maxpartialheightsupp]{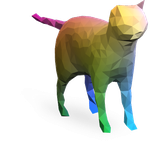}
         &
         \hspace{0.1cm}
         \rotatebox{90}{\textcolor{gray!50}{\rule{\maxpartialheightsupp}{0.2pt}}}
         &
         \includegraphics[width=0.2\columnwidth,height=1.2cm]{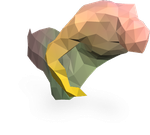}
         &
         \hspace{-0.5cm}
         \includegraphics[width=0.2\columnwidth,height=\maxpartialheightsupp]{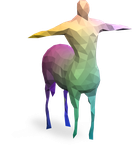}
         &
         \hspace{0.1cm}
         \rotatebox{90}{\textcolor{gray!50}{\rule{\maxpartialheightsupp}{0.2pt}}}
         &
         \includegraphics[width=0.2\columnwidth,height=1.6cm]{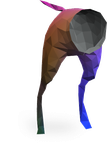}
         &
         \hspace{-0.5cm}
         \includegraphics[width=0.2\columnwidth,height=\maxpartialheightsupp]{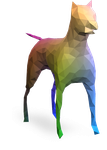}\\
         \hspace{-0.3cm}
        \includegraphics[width=0.2\columnwidth,height=\maxpartialheightsupp]{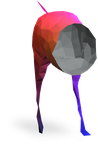}
         &
         \hspace{-0.5cm}
         \includegraphics[width=0.2\columnwidth,height=\maxpartialheightsupp]{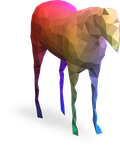}
         &
         \hspace{0.1cm}
         \rotatebox{90}{\textcolor{gray!50}{\rule{\maxpartialheightsupp}{0.2pt}}}
         &
         \includegraphics[width=0.2\columnwidth,height=1.5cm]{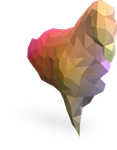}
         &
         \hspace{-0.6cm}
         \includegraphics[width=0.2\columnwidth,height=\maxpartialheightsupp]{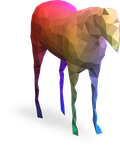}
         &
         \hspace{0.1cm}
         \rotatebox{90}{\textcolor{gray!50}{\rule{\maxpartialheightsupp}{0.2pt}}}
         &
         \includegraphics[width=0.2\columnwidth,height=\maxpartialheightsupp]{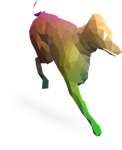}
         &
         \hspace{-0.5cm}
         \includegraphics[width=0.2\columnwidth,height=\maxpartialheightsupp]{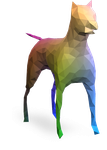}\\
    \end{tabular}
    \caption{More qualitative results computed with our method for \textbf{partial-to-full} shape matching on test set shapes~\cite{ehm2024geometrically} from SHREC'16~\cite{cosmo2016shrec}.}
    \label{fig:supp-partial-sm}
\end{figure}

\begin{table}[]
    \centering
    \hspace{0.2cm}
    {\footnotesize
    \begin{tabular}{l|ccc}
        \toprule
        \textbf{Metric}                                 & \textbf{ULRSSM} & \textbf{SpiderMatch} & \textbf{Ours}  
        \\\midrule
        Mean geodesic errors  ($\downarrow$)   & 0.071  &      0.066  & \textbf{0.062} \\
        Mean Dirichlet energies ($\downarrow$) & 1.262  &      0.613  & \textbf{0.444} \\
        \bottomrule
    \end{tabular}
    }
    \caption{ {Quantitative results for \textbf{partial-to-full} shape matching on the SHREC'16 cuts~\cite{cosmo2016shrec} dataset for ULRSSM~\cite{cao2023unsupervised}, SpiderMatch~\cite{roetzer2024spidermatch} and our method.}}
    \label{tab:partial-quant}
\end{table}

\subsubsection{Planar Graph Matching}\label{sec:supp:planar-graph-matching-results}
\begin{figure}
    \centering
    \hspace{-0.2cm}
    \begin{tabular}{c}
        \includegraphics[width=0.95\columnwidth]{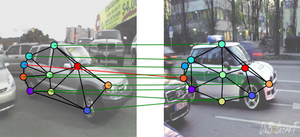}\\
        \includegraphics[width=0.95\columnwidth]{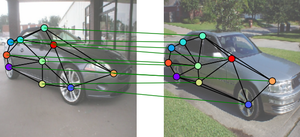}\\
         \includegraphics[width=0.95\columnwidth]{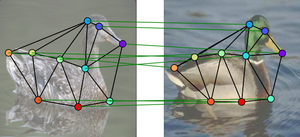}\\
         \includegraphics[width=0.95\columnwidth]{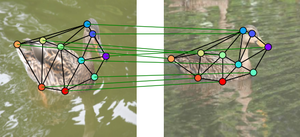}\\
    \end{tabular}
    \caption{More \textbf{planar graph matching} results using our method. For the cars in the top row we can see matching artefacts stemming from graph connectivity differences and enforced geometric consistency (see vertices connected with red line).}
    \label{fig:supp:more-graph-sm}
\end{figure}

In \cref{fig:supp:more-graph-sm}, we show more graph matching examples of instances of WILLOW~\cite{cho2013learning} dataset.
We furthermore note that we only consider the graph matching problem as a proof-of-concept experiment and that we use pixel-coordinates of vertices of respective graphs as input features.
We leave the integration of more elaborate features, when our method is applied to graph matching, to future works.

\end{document}